\documentclass[aps,twocolumn,superscriptaddress]{revtex4-1}%\setkeys{acs}{biblabel=brackets}
\usepackage{ulem}
\usepackage{graphicx}
\usepackage{xcolor}
\usepackage[english]{babel}
\usepackage{amsmath}
\usepackage{amssymb}
\usepackage{bm}
\usepackage{makecell}
\usepackage{array}
\usepackage{float}

\usepackage{algorithm,algpseudocode}
\usepackage{wrapfig}

\definecolor{darkgreen}{RGB}{0, 150, 0}

\def\1{{\mathds{1}}}

\begin{document}

\author{Alice E. A. Allen}
\affiliation%[Center for Nonlinear Studies]
{Center for Nonlinear Studies, Los Alamos National Laboratory, Los Alamos, 87545, NM, United States}
\affiliation%[Theoretical Division]
{Theoretical Division, Los Alamos National Laboratory, Los Alamos, 87545, NM, United States}
\affiliation%[Max Planck Institute for Polymer Research]
{Max Planck Institute for Polymer Research, Ackermannweg 10, 55128 Mainz, Germany,}
\email{allena@mpip-mainz.mpg.de}
\author{Rui Li}
\affiliation%[Center for Nonlinear Studies]
{Division of Chemistry and Chemical Engineering, California Institute of Technology, Pasadena, 91125, CA, United States}
\author{Sakib Matin}
\affiliation%[Center for Nonlinear Studies]
{Center for Nonlinear Studies, Los Alamos National Laboratory, Los Alamos, 87545, NM, United States}
\affiliation%[Theoretical Division]
{Theoretical Division, Los Alamos National Laboratory, Los Alamos, 87545, NM, United States}
\author{Xing Zhang}
\affiliation%[Center for Nonlinear Studies]
{Division of Chemistry and Chemical Engineering, California Institute of Technology, Pasadena, 91125, CA, United States}
\author{Benjamin Nebgen}
\affiliation%[Theoretical Division]
{Theoretical Division, Los Alamos National Laboratory, Los Alamos, 87545, NM, United States}
\author{Nicholas Lubbers}
\affiliation%[CCS]
{Computer, Computational, and Statistical Sciences Division, Los Alamos National Laboratory, Los Alamos, 87545, NM, United States}
\author{Justin S. Smith}
\affiliation%[NVIDIA]
{Nvidia Corporation, 2788 San Tomas Expressway,
Santa Clara, 95051, CA, United States}
\author{Richard Messerly}
\affiliation%[Theoretical Division]
{Theoretical Division, Los Alamos National Laboratory, Los Alamos, 87545, NM, United States}
\affiliation%[OR]
{National Center for Computational Sciences Division, Oak Ridge National Laboratory, Oak Ridge, 37830, TN, United States}
\author{Sergei Tretiak}
\affiliation%[Theoretical Division]
{Theoretical Division, Los Alamos National Laboratory, Los Alamos, 87545, NM, United States}
\affiliation%[Center for Nonlinear Studies]
{Center for Integrated Nanotechnologies, Los Alamos National Laboratory, Los Alamos, 87545,  NM, United States}
\author{Garnet Kin-Lic Chan}
\affiliation%[Center for Nonlinear Studies]
{Division of Chemistry and Chemical Engineering, California Institute of Technology, Pasadena, 91125, CA, United States}
\author{Kipton Barros}
\affiliation%[Center for Nonlinear Studies]
{Center for Nonlinear Studies, Los Alamos National Laboratory, Los Alamos, 87545, NM, United States}
\affiliation%[Theoretical Division]
{Theoretical Division, Los Alamos National Laboratory, Los Alamos, 87545, NM, United States}

%\iffalse

\title[]{
Reactive Chemistry at Unrestricted Coupled Cluster Level: High-throughput Calculations for Training Machine Learning Potentials}

\begin{abstract}
Accurately modeling chemical reactions at the atomistic level requires high-level electronic structure theory due to the presence of unpaired electrons and the need to properly describe bond breaking and making energetics. 
Commonly used approaches such as Density Functional Theory (DFT) frequently fail for this task due to deficiencies that are well recognized. However, for high-fidelity approaches, creating large datasets of energies and forces for reactive processes to train machine learning interatomic potentials or force fields is daunting. For example, the use of the unrestricted coupled cluster level of theory has previously been seen as unfeasible due to high computational costs,
the lack of analytical gradients in many computational codes, and additional challenges such as constructing suitable basis set corrections for forces. 
In this work, we develop new methods and workflows to overcome the challenges inherent to automating unrestricted coupled cluster calculations. Using these advancements, we create a dataset of gas-phase reactions containing energies and forces for 3119 different organic molecules configurations calculated at the gold-standard level of unrestricted CCSD(T) (coupled cluster singles doubles and perturbative triples). 
With this dataset, we provide an analysis of the differences between the density functional and unrestricted CCSD(T) descriptions. 
We develop a transferable machine learning interatomic potential for gas-phase reactions, trained on unrestricted CCSD(T) data, and demonstrate the advantages of transitioning away from DFT data. Transitioning from training to DFT to training to UCCSD(T) datasets yields an improvement of more than 0.1~eV/\AA~in force accuracy and over 0.1~eV in activation energy reproduction.

\end{abstract}

\maketitle

%\clearpage
\section{Introduction}

\begin{figure*}[htb]
    \centering
    \includegraphics[width=6.25in]{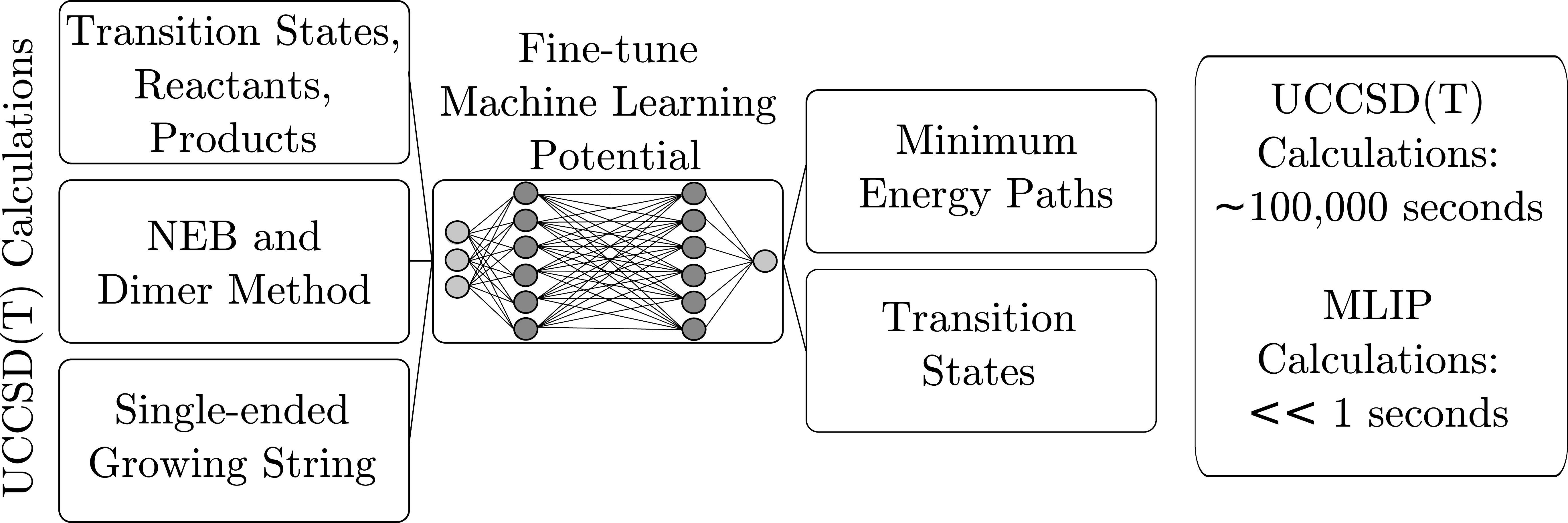}
    \caption{Unrestricted CCSD(T) calculations sampled with a variety of methods are used to fine-tune a MLIP pre-trained to DFT data. The subsequent MLIP can then predict minimum energy paths and transition states of reactions beyond the training data.  }
    \label{fig:Summary}
\end{figure*}
Atomistic simulations of molecular systems underpin the understanding of a wide variety of biological and chemical processes~\cite{Jorgensen1988, Weiner1984,Smith2019, Smith2017, Chmiela2018}.
Simulating chemical reactions is consistently more challenging than handling non-reactive systems, due to multiple factors. Reactive systems often involve modeling configurations with unrestricted formalisms to describe bond breaking and formation processes. Additionally, reactions typically feature highly diverse conformations across the potential energy surface (PES), with pathways and transition state barriers often unknown beforehand. Conquering these challenges is important, as precise knowledge of the transition state energetics is necessary for understanding the probability and efficiency of reactions.
Common quantum mechanical (QM) approximations can fail to describe the reaction paths sufficiently accurately, with significantly more expensive techniques beyond density functional theory (DFT) often required~\cite{Zhao2009, Goerigk2010, Zhao2011, Johnson2008, Chamkin2020,Vermeeren2022,Cohen2008,Li2024}. 
Alternatively, machine learning interatomic potentials (MLIPs) can provide predictions of QM properties (such as energies and forces) that are accurate and orders of magnitude faster than high-level QM techniques~\cite{Smith2017, Schutt2017, bartok2010gaussian, Behler:2011it, ChemRev-MLFF,Mueller2020,Bartok:2018ih}. However, MLIPs must be derived from large datasets of QM calculations and therefore the issues associated with using QM calculations for reactive systems are still present when creating reactive MLIPs. In this work, we develop an automated workflow to create an organic molecule gas-phase dataset using high-level QM data from the unrestricted coupled cluster singles doubles and perturbative triples (CCSD(T)) approximation. We perform analysis on this dataset to gain insight into the differences between unrestricted CCSD(T) (UCCSD(T)) and DFT approximations across a range of chemical reaction space expressed in the training data. Finally, by training an MLIP to the UCCSD(T) dataset and subsequent evaluation of its performance, we demonstrate that transferable MLIPs can be created for reactive chemistry that reproduce high accuracy training data.

%Problems with using DFT for reactions UCCSD(T)
Coupled cluster theory is commonly used for accurate ground-state calculations on small molecules at equilibrium geometries where it is referred to as the gold-standard of quantum chemistry~\cite{Bartlett2007}. In its unrestricted formulation, it also provides reasonably accurate results for typical chemical bond breaking, e.g. those involving single bonds~\cite{Visscher1996, Bartlett2007}. Since the goal of MLIPs is often to replicate a system's ground-state energy, unrestricted calculations offer a practical route to accurate data for reactive systems. The discrepancies between the predictions of DFT and CCSD(T) for reactive systems have been investigated in a number of previous works~\cite{Zhao2009, Goerigk2010, Zhao2011, Johnson2008, Chamkin2020,Vermeeren2022,Cohen2008,Li2024, Korth2009}. These studies looked at a small subset of reaction space, consisting of a few chosen reactions, and investigated differences in the estimated activation energy and transition states between DFT functionals and CCSD(T)~\cite{Zhao2011,Li2024, Hait2018}. 
In this work, we are interested in exploring the difference between DFT (for three commonly used functionals) and UCCSD(T) for a broader range of reactive chemistry across hundreds of different reactions, and investigating the consequences of the level of theory used for MLIP development.  The resulting UCCSD(T) dataset contains the energies and forces for thousands of molecules and
forms the starting point in this work for creating a {transferable} and high-accuracy machine learning model capable of modeling a wide range of reaction space.

% UCCSD(T) details 
Running coupled cluster calculations at scale requires new considerations beyond those typical in studies of a limited number of molecules. For example, UCCSD(T) calculations should use an appropriate Hartree-Fock reference (i.e. one that breaks spin symmetry) if such an unrestricted determinant exists. To enable high-throughput data collection, we automate the construction and selection of the unrestricted Hartree-Fock reference using stability analysis~\cite{Seeger1977}.
Because CCSD(T) requires large basis sets for chemical accuracy, we also employ a basis set correction for both the energy and the forces~\cite{fan2006approximating}. 
We provide evidence of the effectiveness of the basis set correction for our dataset.
Finally, the UCCSD(T) data can be unreliable at points close to the boundary of the Hartree-Fock symmetry breaking point. Thus, we also develop an automatic filtering protocol to remove structures where the UCCSD(T) result is suspected to be inaccurate.

%TL, Al and Sampling, And Workflow
The choice of sampling method for exploring chemical space plays a crucial role in constructing a high-quality QM dataset for machine learning interatomic potentials. Here we focus on building a dataset for molecules containing C, H, N, O atoms and use active learning to select relevant structures in the chemical reaction space~\cite{Smith2018-less}. We start from
a subset of reactants, products, and transition states of molecules in existing datasets in the literature~\cite{Grambow2020}, and subsequently employ several automated structure sampling techniques to generate new structures~\cite{Henkelman1999,Jonsson1998,Zimmerman2015}. By using an ensemble of exploratory MLIP, we detect high uncertainty points amongst these structures to identify points in chemical space to add to the dataset~\cite{Smith2018-less}. Then by using transfer-learning from a MLIP trained to DFT energies and forces on these actively learned points, we build a final MLIP on high quality UCCSD(T) energies and forces on a smaller actively learned set of structures~\cite{Smith2019}. Our final UCCSD(T) data set consists of the energies and forces of over three thousand sampled configurations for molecules with up to 16 atoms.

%What have been done before
To put our MLIP work in context, machine learning interatomic potentials have previously been fit to large, diverse datasets of organic molecules to create transferable potentials capable of modeling a wide range of chemical and biological systems~\cite{Smith2018-less,Smith2017, Kovacs2025,Zubatyuk2019}. Furthermore, this has been extended to reactive systems in Ref.~\citenum{zhang_2023} where high temperature molecular dynamics was performed to sample reaction space in a ``nanoreactor''. Additionally, in Ref.~\citenum{Schreiner2022}  gas-phase reactive datasets were created and fit to model reactive systems. However, these potentials have relied on DFT training data which fundamentally limits the accuracy of the potentials. In Ref.~\citenum{Batatia2024} a foundation model fit to the Materials Project dataset was created~\cite{Jain2013},  which successfully recreated hydrogen combustion reactions, ammonia and borane thermal decomposition, and additional reactive processes. However, given that the Materials Project dataset is also generated at the DFT level, the QM level of theory is a fundamental limitation in the accuracy obtainable with this model~\cite{Batatia2024}. For single molecule MLIPs, a number of works have used CCSD(T) data, or even data from more expensive quantum chemistry methods, but these models have been designed for a single or small number of applications and are not generally transferable~\cite{Braams2009-wi, Qu:2018gj, Huang:2005, Hu2023}. In this work, our intention is to provide a large organic molecule dataset of UCCSD(T) energy and forces for reactive processes, and create a transferable MLIP that can be used on many different organic reactions. This is a significant step towards creating transferable MLIP that can describe even broader classes of reactions, for example, both in the gas phase and condensed phase, at the coupled cluster level of accuracy.

\section{Methods}

\subsection{Machine Learning Interatomic Potential Framework with HIP-NN}
Machine learning interatomic potentials predict the energy and forces of a molecule from atomic coordinates~\cite{Smith2018-less,Smith2017, Kovacs2025,Zubatyuk2019}. MLIPs are fit to quantum mechanical calculations and can offer highly accurate and very fast models of interactions between atoms and molecules.  In this work, we used the model  HIP-NN in two architectures~\cite{Chigaev2023, Lubbers2018}. The first was used for exploring reaction space using active learning.
For this, we used  
%The MLIP used for sampling in this work is 
HIP-NN-TS, a messaging passing neural network with tensor sensitivity~\cite{Chigaev2023, Lubbers2018}. The hyperparameters are set to the same values used in Ref.~\citenum{Chigaev2023} used for training the HIP-NN-TS potential to the ANI-1x dataset. To calculate the uncertainty, an ensemble of four HIP-NN-TS potentials was trained and the standard deviation of the energy (per atom) was used as the uncertainty for the active learning process. 

The second type of MLIP was the one trained to the final datasets of DFT and UCCSD(T) energies and forces. For this, 
the newest version of the HIP-NN class of models was employed. This model, HIP-HOP-NN, is introduced in Ref.~\citenum{hiphop}, and incorporates many-body tensor information directly into the message passing components of the model. An ensemble of 8 HIP-HOP-NN models was used to improve the accuracy of the model when predicting energies and forces. 
We used transfer learning~\cite{Smith2019, Pan2010} by first fitting the MLIP to the (larger) DFT dataset, before fine-tuning to the smaller UCCSD(T) dataset.

\subsection{Existing Datasets Used for Initialization}
Quantum mechanical datasets of molecular energies and forces are necessary for training MLIPs. For our training data, we started from subsets of structures in existing datasets from the literature, which we then augmented with new structures that sample reaction space generated by the active learning sampling protocol. In the following section, we describe the existing datasets used to initialize the sampling process. Note that although we used structural data from these datasets, we always recomputed the energies and forces using the QM methods of this work.

\subsubsection{Reaction Paths from Ref.~\citenum{Grambow2020}}
In Ref.~\citenum{Grambow2020}, $\approx$12,000 reaction pathways were discovered using the single-ended growing string (SEGS) method to suggest products of a given reactant~\cite{Zimmerman2015}. The molecules chosen contained C, H, N, O elements. The minimum energy path and transition state (TS) were subsequently found and the reactants, products and transition state located using the $\omega$B97X-D3/def2-TZVP level of theory (all in the singlet state). 

\subsubsection{Transition-1x}

In the Transition-1x~\cite{Schreiner2022} dataset, the dataset of Ref.~\citenum{Grambow2020} was used to initialize points along the minimum energy paths for $\approx$10,000 reactions. Again, the molecules chosen contained C, H, N, O elements.  The Transition-1x dataset was constructed from the structures encountered as NEB attempted to find a converged reaction path from the Ref.~\citenum{Grambow2020} reaction paths~\cite{Jonsson1998}. The level of theory used was DFT with $\omega$B97X/6-31G(d) and the dataset consisted of 9.6 million configurations.

\subsubsection{QM9}

The QM9 dataset was first introduced in Ref.~\citenum{Ramakrishnan2014}. The QM9 dataset contains 133,885 organic molecules with up to nine heavy atoms and contains the elements C, H, O, N, F. The organic molecules contained are equilibrium configurations calculated at the B3LYP/6-31G(2df,p) level of theory~\cite{Chai2008, Adamo1999, Becke1993, Stephens1994}.

\subsection{QM Methods}

Additional quantum mechanical calculations were then performed at both DFT and coupled cluster levels of theory. 
We now describe the QM methods used to generate energy and force data. 
To facilitate the extension of these datasets, the Python code to carry out the calculations used in this work is available as part of the ALF code at https://github.com/lanl/ALF.

\subsubsection{DFT Calculations}

The DFT calculations were performed on structures produced using the existing datasets above as well as new structures produced by active learning.
The calculations used the ORCA package at the $\omega$B97X/6-31G(d) level of theory~\cite{Neese2020}. All DFT calculations were carried out in the singlet state using restricted Kohn-Sham theory, as commonly performed for reactive datasets constructed for MLIPs~\cite{Grambow2020, Schreiner2022}.

\subsubsection{UCCSD(T) Calculations}

The UCCSD(T) data was computed with PySCF~\cite{Sun2020,Sun2018} with the cc-pVDZ (DZ) basis set alongside a basis set correction obtained from calculations performed at the UMP2/QZ, UMP2/TZ, UCCSD/TZ and UCCSD/DZ levels. This data was combined to estimate the UCCSD(T) QZ energies and forces. The basis set corrected UCCSD(T) is referred to as UCCSD(T)/QZ* throughout this work.
We used a composite scheme with the following form for the energies and forces: %for forces:
\begin{multline}
E_\text{UCCSD(T)/QZ*}=E_\text{UCCSD(T)/DZ} + \\ (E_\text{UCCSD/TZ} - E_\text{UCCSD/DZ})  +  \\ (E_\text{UMP2/QZ} - E_\text{UMP2/TZ}) 
\end{multline}
\begin{multline}
F_\text{UCCSD(T)/QZ*}=F_\text{UCCSD(T)/DZ} +  \\ (F_\text{UCCSD/TZ} - F_\text{UCCSD/DZ})  +  \\ (F_\text{UMP2/QZ} - F_\text{UMP2/TZ}) 
\end{multline}

The error of the correction compared to the higher level UCCSD(T)/QZ forces is 0.073 eV/\AA~per force component which is smaller than the model errors in this work.
Other aspects of this correction are analyzed in detail in Sec.~S1. 

\begin{figure}[htb]
    \centering
    \includegraphics[width=3.25in]{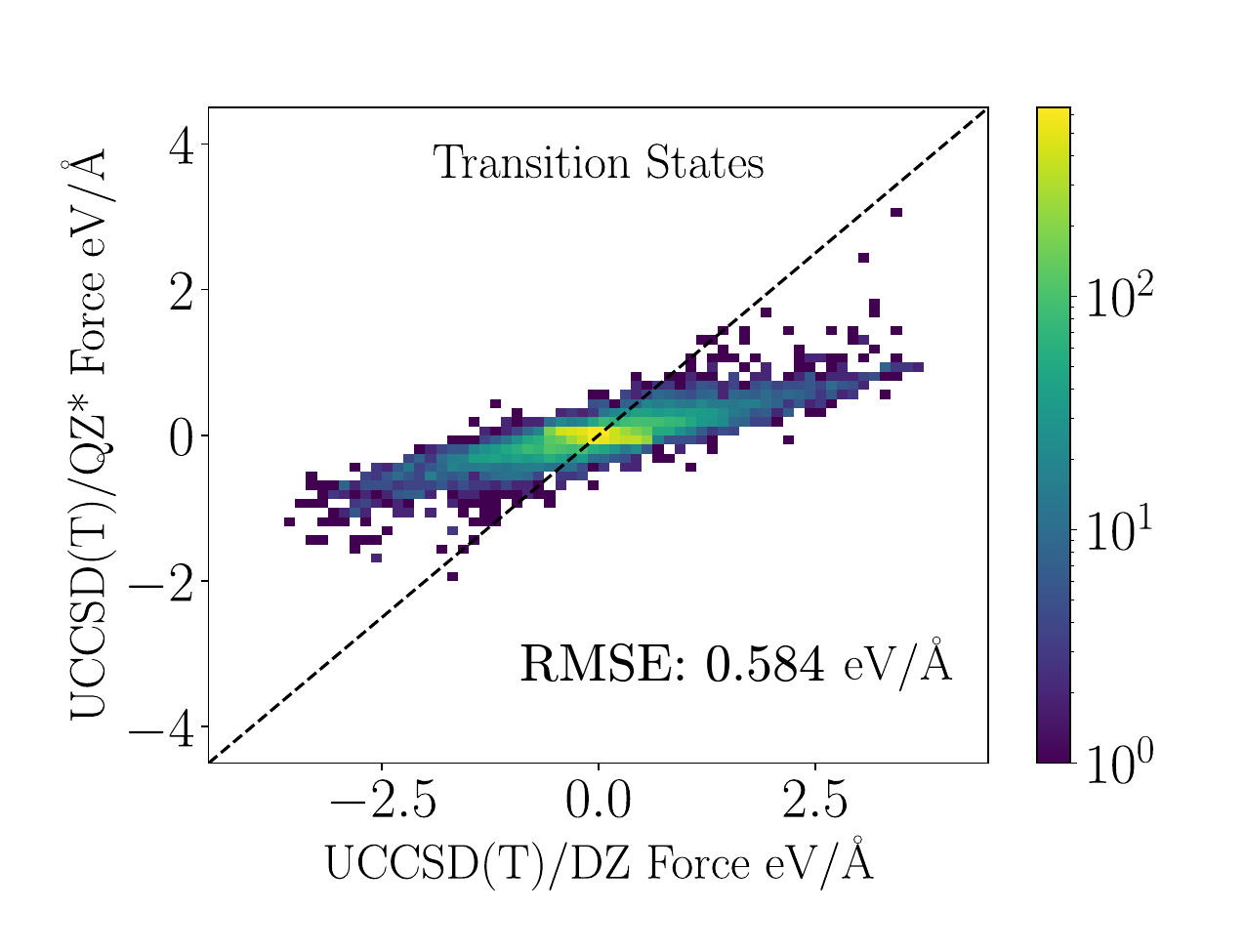}
    \caption{ The forces for UCCSD(T)/DZ compared to the UCCSD(T) forces with a basis set correction up to the QZ level. A systematic error in DZ forces can be seen, demonstrating the importance of basis set corrections and the insufficiency of using UCCSD(T)/DZ for force calculations or geometry optimization.  }
    \label{fig:Forces_TS_uccsdt_dz_uccsdt}
\end{figure}

The effect of the basis set correction used is visualized in Fig.~\ref{fig:Forces_TS_uccsdt_dz_uccsdt} with a comparison of UCCSD(T)/QZ* and UCCSD(T)/DZ forces. UCCSD(T)/DZ forces show a large deviation from the UCCSD(T)/QZ* with systematically larger forces for the DZ basis than the QZ* basis. The correlation of the components used in the basis set correction are shown in Fig.~S1 and Fig.~S2.

To perform the UCCSD(T) calculations, we should start from an appropriate Hartree-Fock reference. Near the equilibrium geometry, for an even number of electrons, the restricted Hartree-Fock reference with a singlet ground-state, where opposite spin electrons occupy the same spatial orbitals, is commonly the lowest energy solution.
However, at some geometries (typically stretched ones) the restricted Hartree-Fock solutions can develop an instability towards an unrestricted solution, where opposite spin electrons occupy different spatial orbitals. We used stability analysis and following of the Hartree-Fock (HF) wavefunction, as presented in Ref.~\citenum{Seeger1977}, to detect instabilities in the HF solution and find stable unrestricted HF (UHF) solutions. Finding the stable ground-state solution for the HF calculation ensures that the electronic structure remains as single reference as possible (by being in the ground state) even at stretched geometries, which is a prerequisite for the accuracy of subsequent UCCSD (T) calculations~\cite{Margraf2017}.

However, there are some complications using unrestricted coupled cluster energies. 
One issue is that the unrestricted wavefunction is no longer guaranteed to be an eigenvalue of the $S^2$ operator~\cite{Schlegel2002}. In practice, this spin contamination is not necessarily a concern for the quality of the predicted energy. However, a more problematic issue is that at the boundary between geometries with unrestricted Hartree-Fock and restricted Hartree-Fock solutions (the Coulson-Fischer points~\cite{Coulson1949}) the energy is continuous but not differentiable, thus the force is undefined. This means that forces are inaccurate near this boundary of the Hartree-Fock instability. This issue is exacerbated when using the basis set correction, as the boundary can manifest at slightly different geometries depending on the basis set, making calculations with different basis sets inconsistent.

\begin{figure}[htb]
    \centering
    \includegraphics[width=3.0in]{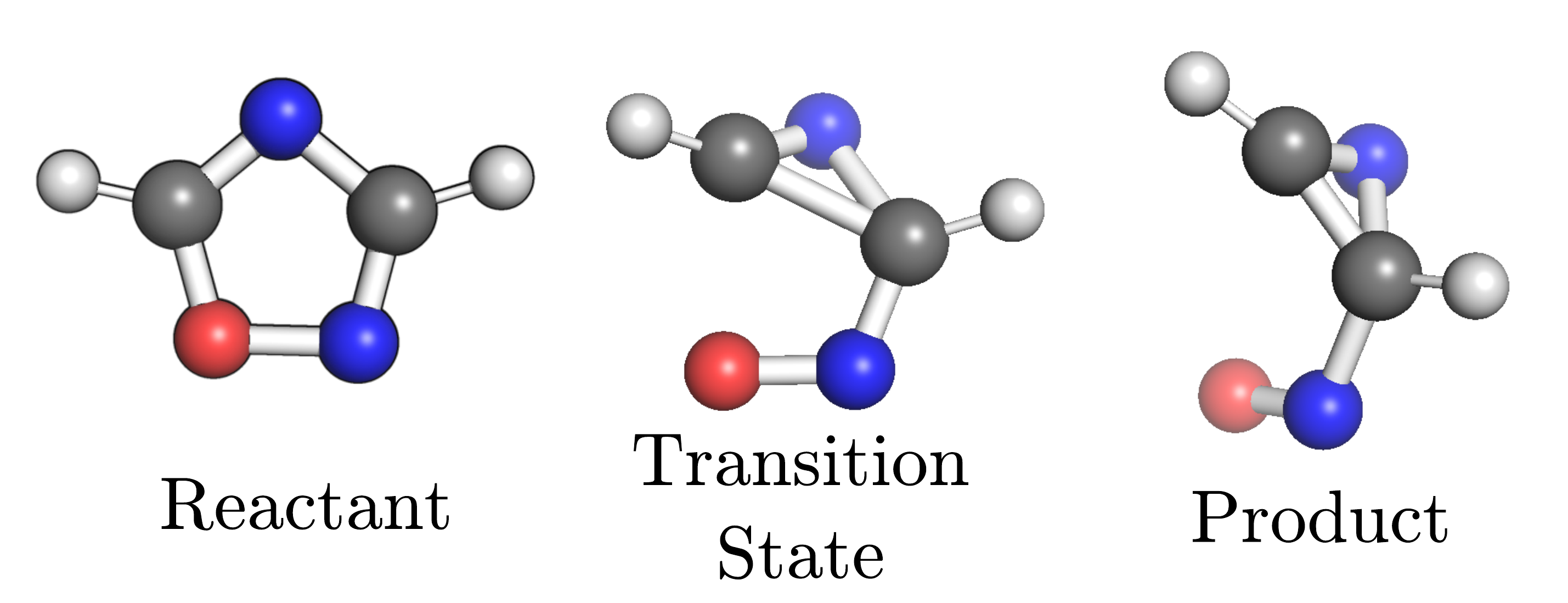}
    \caption{The reactants, products and transition states for an example of an excluded transition state structure in the dataset. The forces for the MP2 and UCCSD/TZ calculations are very high for this nominal transition state (up to 11.12~eV/\AA~for one force component).  }
    \label{fig:BD_BSC}
\end{figure}

To remove these complications, we use a filtering procedure to detect when the geometry is too close to the Coulson-Fischer point and to remove such structures.
A detailed description of the procedure is in Section S1. 
To demonstrate the type of configuration removed in this process,
an example structure removed from the dataset is shown by the transition state in Ref.~\ref{fig:BD_BSC} along with the reactants and products. This reaction is an example of the unusual reactions created with automated approaches for reaction paths. Although the central structure in Fig.~\ref{fig:BD_BSC} is suggested to be a transition state in the DFT simulations, for UCCSD/TZ much higher forces are present, up to 11.12~eV/\AA. Similarly, the MP2 calculation exhibits large forces. For the UHF references using the
DZ/TZ/QZ basis sets, $\langle S^2\rangle = 0.51/0.22/0.13$ demonstrating that the $\langle S^2\rangle$ values are basis set dependent. The TS structure occurs near the point where the 5-membered ring breaks, which likely produces a diradical. The lowest eigenvalue of the orbital Hessian of this structure is  $< 10^{-3}$ Ha, indicating closeness to the Hartree-Fock instability, which is leading to the inaccurate forces.

\subsection{Sampling Methods for Dataset Construction}
When generating molecular datasets to train MLIPs, there are two considerations. The quantum mechanical approximation used and the sampling method employed. The sampling method is the technique used to generate molecular structures. 
To construct our training dataset we used different sampling methods, initialized from structures in the existing datasets above, to efficiently explore reactive space. We first created a larger dataset with energies and forces computed at the DFT level and then used transfer learning to generate  UCCSD(T) data. 

We have provided the sampling techniques for active learning with the dimer method and NEB within the MLIP active learning framework (ALF) available at https://github.com/lanl/ALF~\cite{ALF, zhang_2023}. The SEGS method is freely available in Ref.~\citenum{Zimmerman2015} with code additions to automate sampling presented in Ref.~\citenum{Grambow2020}. In this work, we used the existing SEGS software with minor additions to enable MLIPs to be incorporated.

\subsubsection{AL for the DFT Dataset}
Active learning uses the uncertainty of an ML model to select molecular structures to add to the training data. This reduces the computational cost of dataset generation. The active learning strategy for the DFT dataset is as follows. 

To carry out the sampling, an exploratory MLIP was first trained to DFT data computed for ten percent of the Transition-1x structures. 
Then, from the transition states in Ref.~\cite{Grambow2020},
%proposed in Ref.~\cite{Grambow2020},
the dimer method was used with the exploratory 
MLIP to find the MLIP optimized transition state~\cite{Henkelman1999}. If the proposed transition state had an uncertainty greater than a given threshold the structure was calculated with DFT and added to the new dataset. The number of steps allowed for the dimer method was limited to maintain realistic structures that did not differ too greatly from the starting point. 

Further, using the reactants and products from Ref.~\citenum{Grambow2020}, approximate minimum energy paths were calculated using the NEB method using the exploratory MLIP~\cite{Jonsson1998}. If the structures proposed in the final set of structures had an uncertainty above a set threshold they were calculated with DFT and added to a new dataset. The NEB was not necessarily converged at this final stage and this means that both the true minimum energy path and regions around it were sampled. 

Finally, the SEGS method together with the exploratory MLIP was also used to sample reactive pathways~\cite{Zimmerman2015} beyond those in Ref.~\citenum{Grambow2020}. This method relies on only a reactant and driving coordinates which state the new bonds to be formed and broken. The existing QM9 dataset of equilibrium molecules from Ref.~\citenum{Ramakrishnan2014} was used to suggest reactants and the driving coordinates were randomly chosen using the code provided in Ref.~\citenum{Grambow2020}. The QM9 reactants and driving coordinates were then fed into the SEGS software. Again, QM calculations were performed if the uncertainty was above a set threshold. 

The total size of the DFT dataset sampled in this way consisted of 270720 structures, with a breakdown of structures from each of the sampling techniques given in Sec.~S4. The final HIP-HOP DFT was trained to this dataset in addition to 10\% of Transition-1x.

\begin{figure*}[htb!]
    \centering
    \includegraphics[width=6.5in]{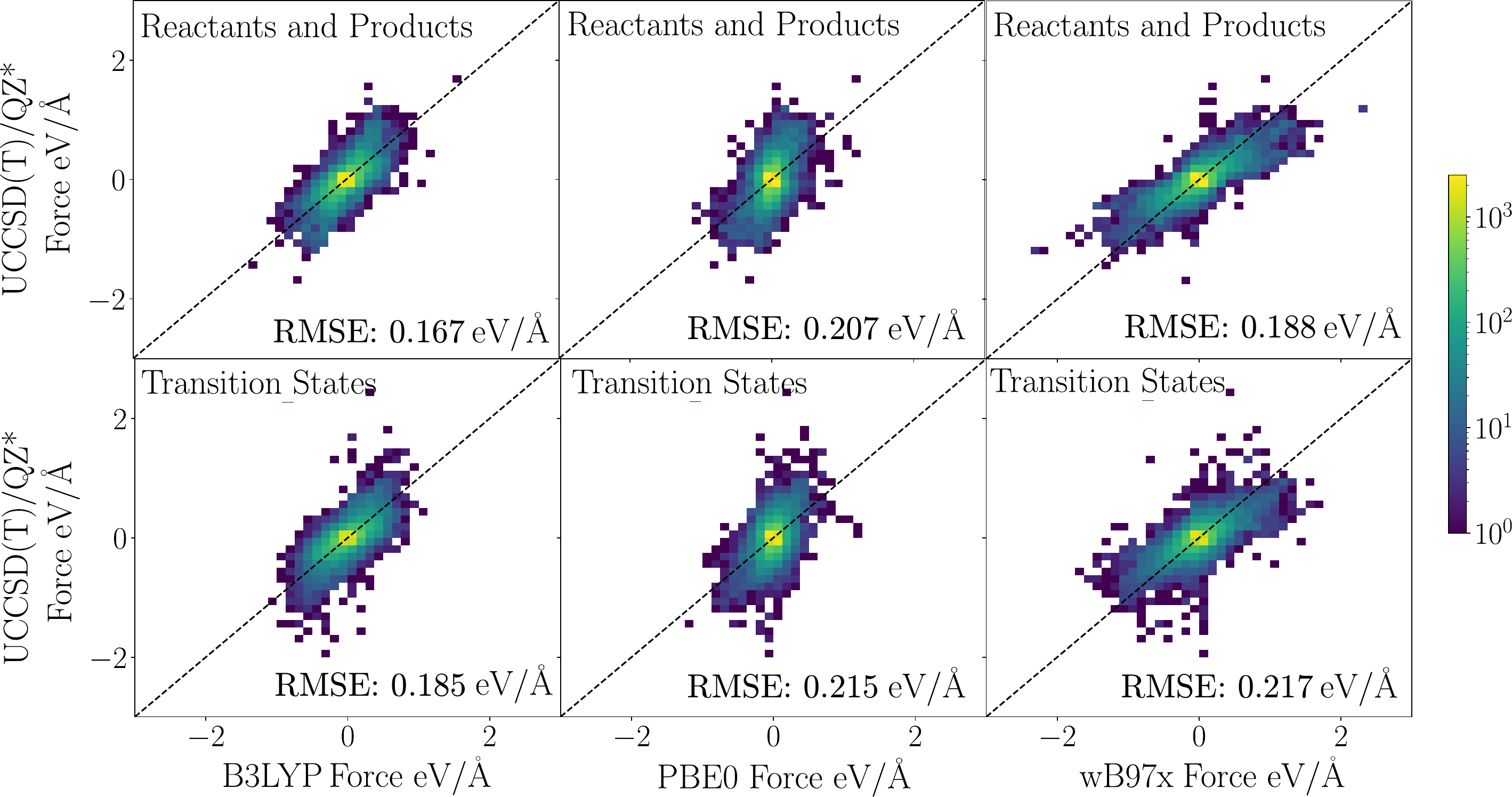}
    \caption{ A comparison of forces for DFT functionals  calculated in the singlet state and UCCSD(T)/QZ* for the transition states and products and reactants. Consistently higher RMSE is seen in the transition states over the reactants and products.}
    \label{fig:TS_P_R_comparison_dft_cc}
\end{figure*}

\subsubsection{AL for the UCCSD(T) Dataset}

The active learning strategy for the UCCSD(T) dataset is now described. In the present work, the molecules from Ref.~\citenum{Grambow2020} were used to build an initial dataset of reactants, products and transition states at the UCCSD(T) level for 950 molecules. Transfer learning was used to retrain the exploratory DFT MLIP to the dataset of transition states, reactants and products to create an exploratory coupled cluster MLIP. 

With this exploratory coupled cluster MLIP, sampling with the dimer method, NEB sampling and the SEGS method was again performed. This provided additional samples %at UCCSD(T) 
to extend the UCCSD(T) dataset to points off the minimum energy path.

An additional set of molecules was also added with bonds stretched or compressed to large and small values by randomly sampling the bond length. These molecules were added to improve bond dissociation curves and repulsive regions at small distances. The molecules chosen for this were primarily very small molecules with less than 7 atoms.

From the active learning process, 167  configurations were found from the dimer sampling method and  49  configurations were found with the SEGS method. Additionally, 1953 small molecules structures ($<$ 9 atoms) were added with bond lengths stretched - as described above. The final UCCSD(T) dataset consisted of 3119 configurations. A summary of the dataset created is given in Sec.~S4. The final HIP-HOP UCCSD(T) MLIP was trained to this dataset by transfer learning from the DFT HIP-HOP potential.

\section{Results}

The creation of a UCCSD(T) dataset offers the unique opportunity to both analyze the differences between QM levels of theory and demonstrate the improved performance of MLIPs fit to UCCSD(T) data. We begin by comparing UCCSD(T) data with DFT data before examining an MLIP fit to UCCSD(T) data. 

\subsection{Analysing the UCCSD(T) Dataset}

The UCCSD(T)/QZ* dataset enables analysis of the differences between UCCSD(T) and DFT quality data across a broad range of chemical reactions. We first focus on the reactants, products, and transition states, before considering the other sampled geometries. There are 950 structures that are reactants, products, and transition states.

\begin{figure*}[htb]
    \centering
    \includegraphics[width=6.75in]{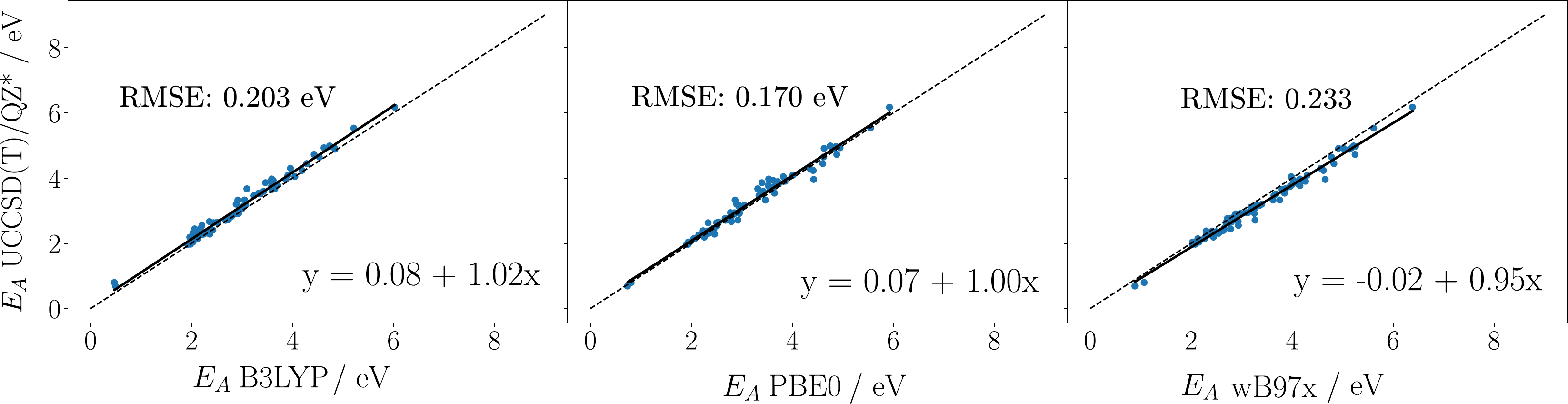}
    \caption{  The activation energy for DFT functionals in comparison to UCCSD(T)/QZ* activation energies. The structures used for the transition state, reactants and products are optimized at the $\omega$B97X-D3/def2-TZVP level and calculated in the singlet state. The dashed black line shows $y=x$ and the solid black line gives the line of best fit through the points with the corresponding equation shown on each figure.}
    \label{fig:EA_DFT}
\end{figure*}

\subsubsection{Reactants, Products, and Transition States}

 The forces for three DFT functionals, $\omega$B97x/6-31G(d), PBE0 and B3LYP, compared to those from UCCSD(T), are shown in Fig.~\ref{fig:TS_P_R_comparison_dft_cc}. Note that these are all evaluated at the structures taken from existing datasets, which were optimized with a different level of theory ($\omega$B97X-D3/def2-TZVP), thus we do not expect the forces to vanish. Nevertheless, all the forces are highly concentrated around zero.
 Different levels of theory are often used for geometry optimization and the final energy and force computation during dataset construction, and this analysis supports that protocol.

However, non-vanishing forces do occur, and the deviations between these forces for the various DFT functionals and UCCSD(T) has some spread. Figure S6 demonstrates this with the distribution of force errors shown for each DFT functional.

The activation energies of the reactions for three DFT functionals are compared to UCCSD(T)/QZ* in Fig.~\ref{fig:EA_DFT} (note, the activation energies are calculated using the transition state from original dataset optimized at DFT/$\omega$B97X-D3/def2-TZVP level).  We see some systematic trends: B3LYP overestimate the barrier height (versus UCCSD(T)/QZ*), PBE0 has the best correlation but still has a systematic overestimation, and $\omega$B97X/6-31G(d) underpredicts the barrier height, as reflected by the equation of the line of best fit through the data points. The difference in activation energy between CCSD(T) and DFT has been highlighted in Ref.~\citenum{Spiekermann2022}, and will cause significant differences in calculated properties such as reaction rates.

\begin{figure*}[htb]
    \centering
    \includegraphics[width=6.5in]{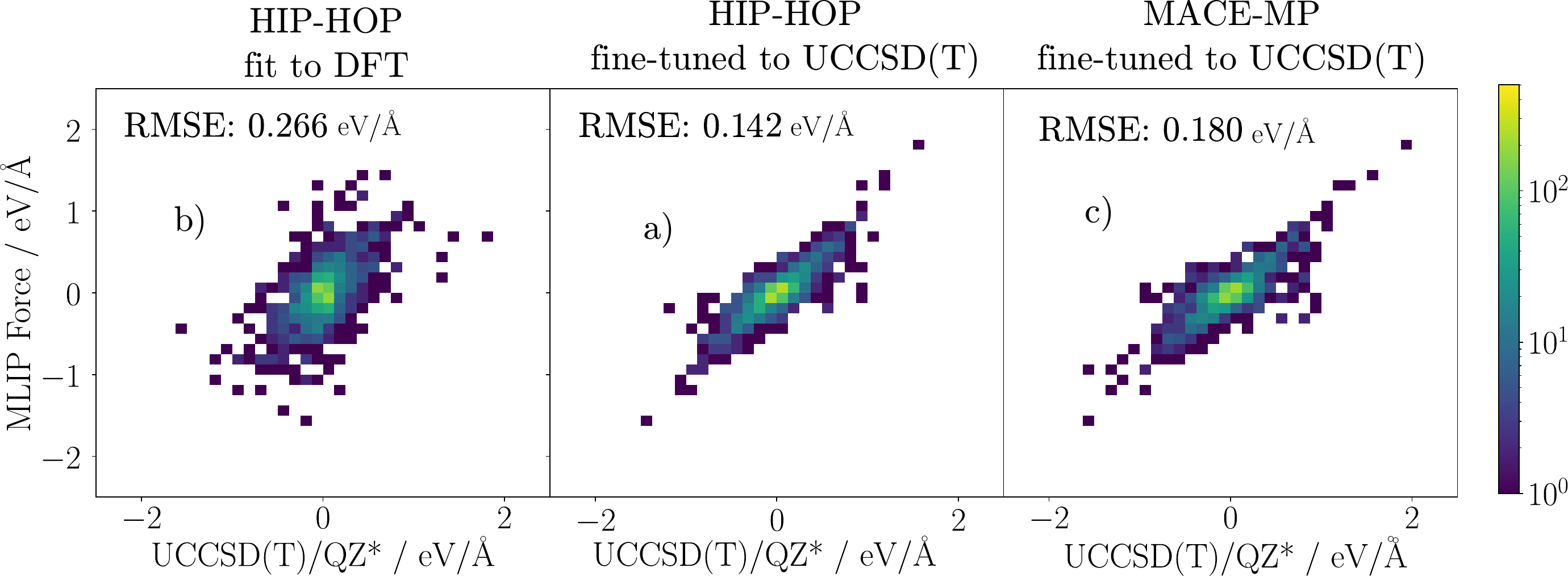}
    \caption{ A comparison of forces calculated with the machine learning interatomic potential HIP-HOP-NN and UCCSD(T)/QZ* for testing datasets containing the transition states and products and reactants (TS, P, R). Part a) shows HIP-HOP-NN model trained to UCCSD(T), b) HIP-HOP-NN trained to DFT data only and c) is the foundation model MACE-MP fine-tuned to UCCSD(T) data. When a MLIP is trained to the UCCSD(T) dataset, comparable (or better) error than the DFT functionals previously shown is achievable.  }
    \label{fig:TS_P_R_comparison_mlip_cc}
\end{figure*}

\subsubsection{Dataset Constructed Along Reaction Path}

Next, we analyze the data for structures generated by sampling along the reaction path. There are 216 of these structures in the dataset. A comparison between the DFT and UCCSD(T) forces is shown in 
% The forces of the structures sampled using NEB and the dimer method are shown in 
Fig.~S12 and the distribution of UCCSD(T)/QZ* forces is shown in Fig.~S11. For the transition states, reactants, and products, the distribution is centered around zero as expected. For the structures sampled with the NEB and dimer methods and the SEGS method, a much larger range of forces are sampled and we can conclude that the diversity of chemical space is increased by the addition of these structures.

For structures sampled by the NEB dimer method, the force component RMSE is 0.470/0.466/0.498~eV/\AA~for the B3LYP/PBE0/$\omega$B97X functional. For the SEGS method, the force component RMSE is 0.746/0.734/0.812~eV/\AA~for the B3LYP/PBE0/$\omega$B97X functional. In addition, the DFT data has different slopes compared to the reference UCCSD(T) data: the fitted slopes are between 0.85 and 0.92 and larger deviations between DFT and UCCSD(T) forces occur as the magnitude of the force component increases. Overall, the force error for the DFT functionals for structures far from equilibrium is greater than for the equilibrium structures. This highlights the importance of high-quality data for off-equilibrium calculations when studying reaction paths, mechanisms, and performing free-energy calculations. 

\subsection{Fitting Transferable MLIPs to UCCSD(T) Data}

We now report the performance of the MLIP trained on the DFT and UCCSD(T) datasets % were used to train the 
using the latest version of the HIP-NN model, HIP-HOP-NN~\cite{hiphop}. The computational cost of a force evaluation for a small molecule (less than 20 atoms) for the HIP-HOP-NN model is approximately 15~milliseconds on an A100 GPU.

Figure \ref{fig:TS_P_R_comparison_mlip_cc} demonstrates the level of accuracy that is achievable with MLIPs. In Fig.~\ref{fig:TS_P_R_comparison_mlip_cc}~a) HIP-HOP-NN is trained to UCCSD(T) data and for this model, the force recreation for the transition states, products and reactants is excellent, significantly improving on the correlation between QM DFT forces and those of UCCSD(T) reported above in Fig.~\ref{fig:TS_P_R_comparison_dft_cc}. 
In contrast, as depicted in Fig.~\ref{fig:TS_P_R_comparison_mlip_cc}~b), the HIP-HOP-NN model trained on $\omega$B97x DFT data has nearly double the error on the UCCSD(T) test set. %, emphasizing the limitations imposed by training on DFT data.

\begin{figure}[htb]
    \centering
    \includegraphics[width=3.35in]{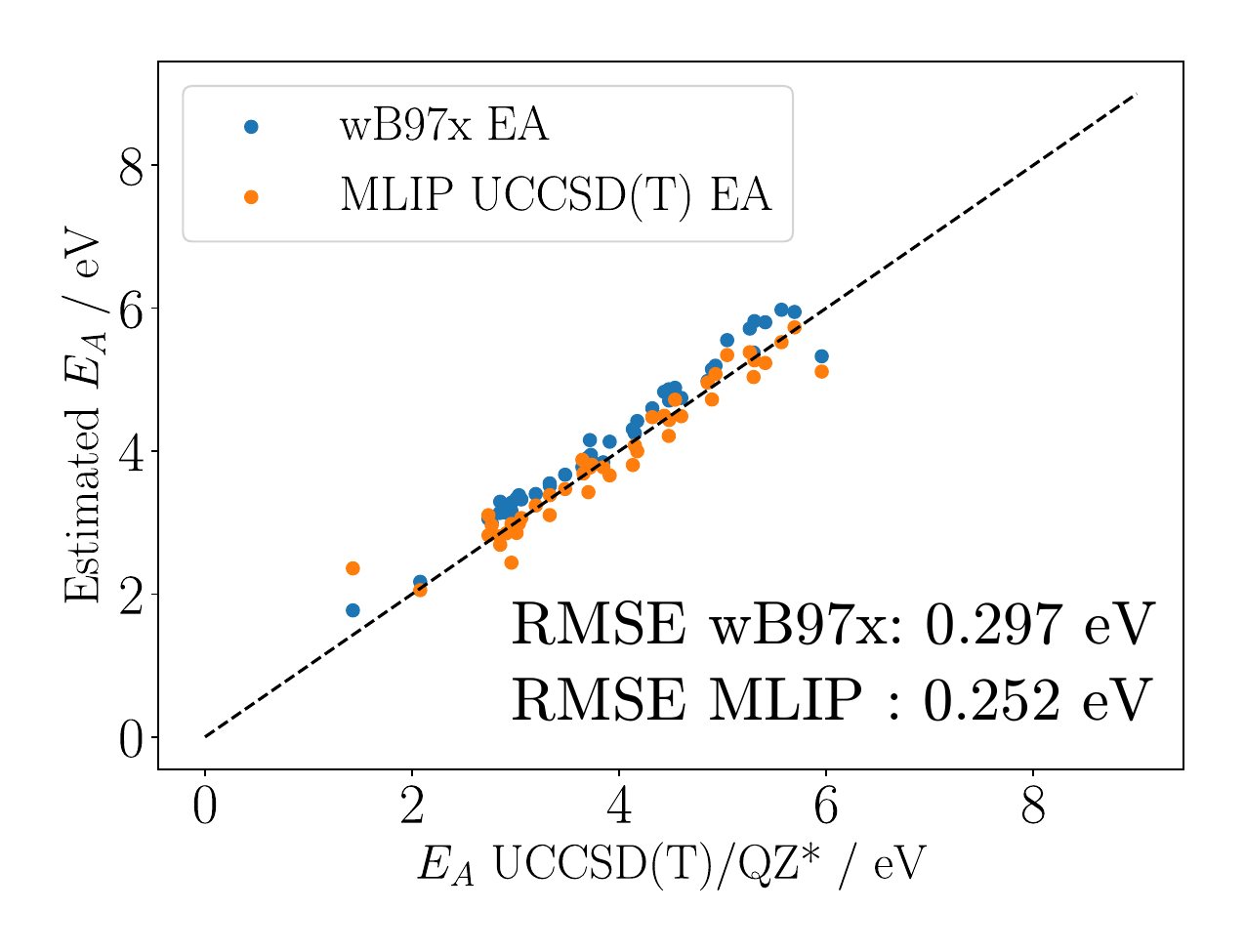}
    \caption{The activation energies for the HIP-HOP-NN model trained to UCCSD(T) data compared to the $\omega$B97x DFT functional. Reactions are taken from Ref.~\citenum{Grambow2020} and are not included in the UCCSD(T) training dataset.  }
    \label{fig:TS_ML}
\end{figure}

The energy error for the HIP-HOP-NN model also provides a measure of accuracy of the potentials. The test set energy errors  for the dataset containing the TS, P, R and AL structures is  0.0134~eV/atom for HIP-HOP-NN trained to UCCSD(T). The error in the activation energy recreation also provides an informative measure of the utility of the potential - see Fig.~\ref{fig:TS_ML}. The reaction paths chosen for the activation energies are not included in the UCCSD(T) training data used to train the model. The MLIP is able to recreate the activation energies of the reactions with a RMSE of 0.252~eV. This exceeds the performance of the more expensive QM DFT calculation shown. The accuracy of the recreating activation energies for the HIP-HOP-NN model trained to DFT data with the $\omega$B97X functional is 0.377~eV. This highlights that for both recreation of forces and activation energies, training to UCCSD(T) data can significantly improve performance.

Foundation models have been developed by a number of groups to provide a pre-trained MLIP that is transferable to multiple systems and can be fine-tuned to specific systems of interest. The MACE-MP model is pre-trained to the Materials Project dataset of DFT data. The error of this foundation model without fine-tuning has a large RMSE in the forces of 0.770~eV/\AA. We can fine-tune this model to the 
UCCSD(T) dataset (see Sec.~S5 for training details) to illustrate the performance of a different training method and model. We find that the fine-tuned model achieves a similarly low force error to the HIP-HOP-NN model trained with transfer learning, of 0.180 eV/\AA.

We further used the HIP-HOP-NN MLIP to explore the minimum energy reaction paths with the NEB and dimer methods previously discussed. For this, we chose reaction paths which did not have any UCCSD(T) data in the training set. The dimer methods initiated from the DFT TS structures consistently converged as did the NEB method over the reaction paths. Example transition states found with HIP-HOP-NN UCCSD(T) model are given in Fig.~S14. The MLIP consistently finds TS nearby to those proposed by the DFT method used in the Transition-1x dataset and of similar quality with respect to the true UCCSD(T) TS. 
Furthermore, the UCCSD(T) forces of the TS found with the MLIPs were calculated and are shown in Fig.S15. The MLIP trained to the UCCSD(T) outperforms the MLIP trained to the DFT data, but the TS found with DFT still outperform the MLIPs. This could be a consequence of limited UCCSD(T) data.

As an additional test of extrapolation capabilities, the bond dissociation curves for hydrogen abstraction for 11 randomly chosen molecules from QM9 were calculated with the MLIP. All the molecules chosen had more than 16 atoms and were larger than the molecules included in the UCCSD(T) training data. The curves and bond dissociation energies are shown in Fig.~S13.  The UCCSD(T) bond dissociation energies are also shown and compared to the MLIP values and singlet DFT values. Relatively smooth bond dissociation curves were generated, with an average RMSE of 0.485~eV in the bond dissociation energy  achieved. The large discrepancy between singlet DFT and UCCSD(T) is also clearly shown. Given that bond dissociation energies were not the focus of this work, the results are promising and further fragmentation data will be added in future iterations of the dataset to improve performance. 

\begin{figure}[htb]
    \centering
    \includegraphics[width=3.25in]{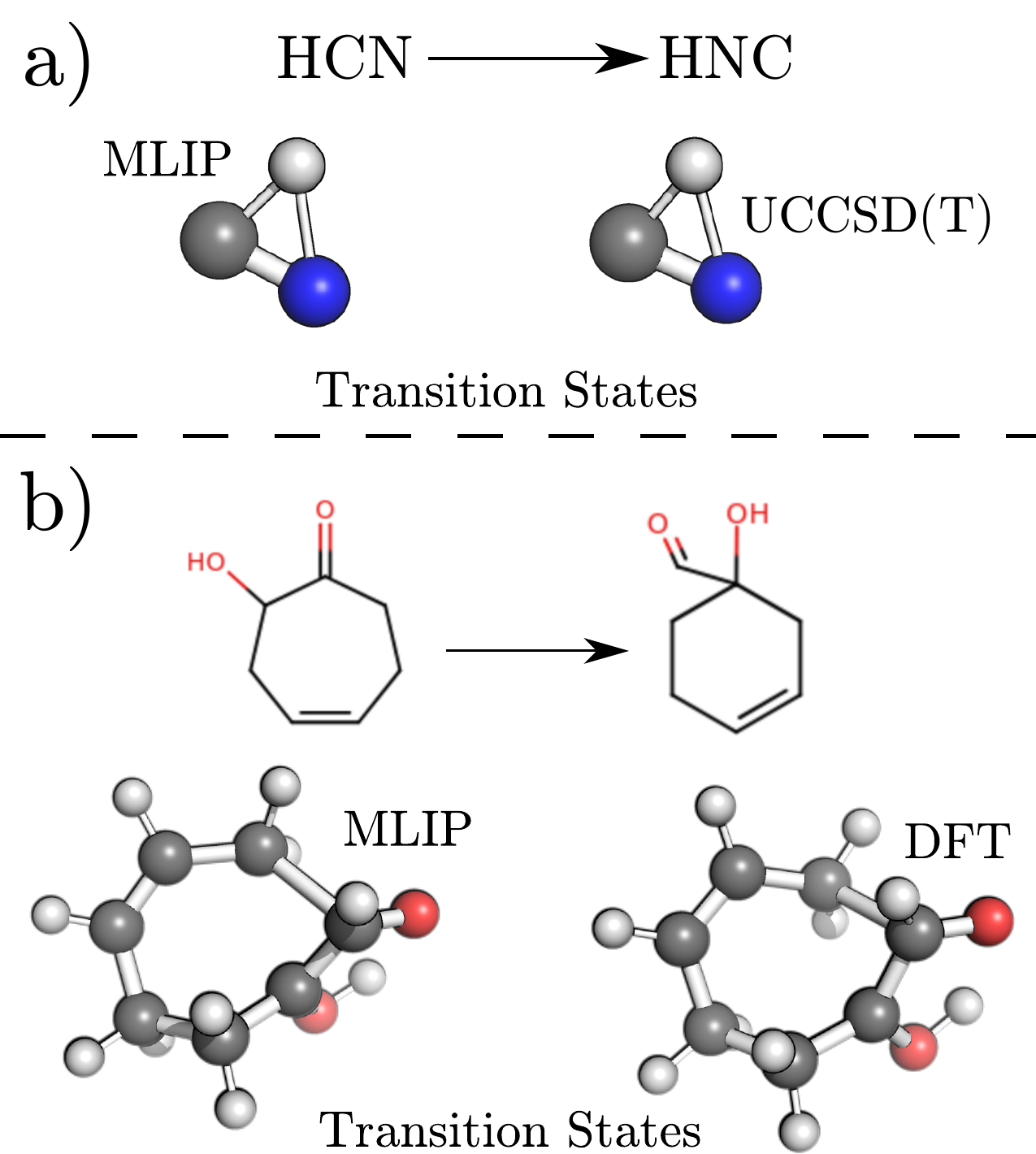}
    \caption{The transition states with the MLIP trained to UCCSD(T) data are compared to the a) UCCSD(T) TS for HCN isomerization and b) a DFT TS for the cyclic isomerization reaction from Ref.~\citenum{Chen2018}.  }
    \label{fig:Cyclic_MEP}
\end{figure}

As a final demonstration of the MLIP, minimum energy paths and TS states for two isomerization reactions are given in Fig.~S16, Fig.~S17 and Fig.~\ref{fig:Cyclic_MEP}. The HCN to HNC isomerization in Fig.~S16 shows a smooth reaction path with NEB readily converging. The MLIP barrier height is  above the UCCSD(T) and DFT energies (0.467 eV greater), but the product energy is very close to the UCCSD(T) energy (0.024 eV). This highlights the need for further energy calculations for transition states, including for small systems, in future work. The reaction path in Fig.~S17 demonstrates a cyclic isomerization reaction from Ref.~\citenum{Chen2018}. The molecules involved are larger than those included in the training dataset. Again, the MLIP-calculated minimum energy path follows a similar shape to the DFT calculation, with some overestimation of the reaction barrier height. 
For the cyclic isomerization reaction from Ref.~\citenum{Chen2018}, the MLIP also 
recreates the reaction mechanism with the proton transfer occurring after the TS. Note that the NEB and dimer method using a MLIP take minutes to perform for this system, but are prohibitively expensive with UCCSD(T)/QZ*.

\section{Conclusion}
Modeling reactive processes with atomistic simulation poses an ongoing challenge for the computational chemistry community. Whilst MLIPs provide a fast and accurate model, creating datasets of highly accurate quantum mechanical calculations for training transferable MLIPs has remained an open challenge. 
We have demonstrated that analytical force and energy calculations at the UCCSD(T) level can be performed at a large-scale and in an automated way in order to generate datasets for machine learning. 
Some practical adaptations we developed for calculations at this scale include a basis corrections for forces and strategies to identify `problematic' UCCSD(T) structures, particularly those arising from spin symmetry breaking.
We have further shown how UCCSD(T) datasets of thousands of molecules together with transfer learning allows for the development of MLIPs for chemical reactions which show improved performance over MLIPs trained to DFT calculations and in a number of cases, improvements over DFT calculations themselves. The dataset produced will be provided to the community upon publication of this work. 

%QM software and BSC/BSE
In the process of carrying out this work, a number of important considerations for the future of quantum chemistry and MLIP arose. Firstly, it became clear that improving both the robustness and performance of software for QM calculations remains a vital pursuit for the advancement of MLIPs. This is necessary for the continued creation of highly accurate datasets at the CCSD(T) level. 
The development of automated approaches to multi-reference calculations could provide a future direction for the automated construction of data sets for reactive processes~\cite{Wardzala2024, King2022}. 
Furthermore, the recent development of analytical force methods for local coupled cluster calculations paves the way for MLIPs to be able to recreate properties of larger molecules, including in condensed phase simulations~\cite{Zhang2024}.

The expansion of the techniques developed in this work to the condensed phase is an open challenge which would greatly expand the applicability of the MLIP. This would require improvements in the sampling techniques used to construct the datasets, for example, so that small systems can be extracted and calculated at a higher level of theory~\cite{Herbold2023}, as well as further improvements in the high-level quantum chemistry methods applicable to condensed phase systems.

Additionally, analysis of problematic structures in the dataset revealed how automated sampling techniques produce reaction paths and transition states that are very unlikely to be produced by a chemist. There are benefits to this, as chemical space is not limited to known and predictable reactions.
However, it does mean that the structures produced stretch quantum chemistry methods to new, unseen limits. As a practical way forward, excluding structures that are unreliably described, as we have done, makes building organic molecule datasets simpler than carefully tailoring methods to an individual system.

We have taken the first steps to building a transferable reactive machine learning potential that is capable of outperforming DFT and offers orders of magnitude better computational performance. By advancing toward UCCSD(T)-level accuracy, MLIPs have the potential to revolutionize reactive simulations across diverse applications and offer unique insights into chemical reactivity at scale.

\acknowledgments
This work was supported by the United States Department of Energy (US DOE), Office of Science, Basic Energy Sciences, Chemical Sciences, Geosciences, and Biosciences Division under Triad National Security, LLC (‘Triad’) contract grant no. 89233218CNA000001 (FWP: LANLE3F2 and LANLE83B). A. E. A. Allen and S. Matin also acknowledge the Center for Nonlinear Studies. Computer time was provided by the CCS-7 Darwin cluster and Institutional Computing clusters at LANL.

\bibliography{refs}

\providecommand{\latin}[1]{#1}
\makeatletter
\providecommand{\doi}
  {\begingroup\let\do\@makeother\dospecials
  \catcode`\{=1 \catcode`\}=2 \doi@aux}
\providecommand{\doi@aux}[1]{\endgroup\texttt{#1}}
\makeatother
\providecommand*\mcitethebibliography{\thebibliography}
\csname @ifundefined\endcsname{endmcitethebibliography}  {\let\endmcitethebibliography\endthebibliography}{}
\begin{mcitethebibliography}{3}
\providecommand*\natexlab[1]{#1}
\providecommand*\mciteSetBstSublistMode[1]{}
\providecommand*\mciteSetBstMaxWidthForm[2]{}
\providecommand*\mciteBstWouldAddEndPuncttrue
  {\def\EndOfBibitem{\unskip.}}
\providecommand*\mciteBstWouldAddEndPunctfalse
  {\let\EndOfBibitem\relax}
\providecommand*\mciteSetBstMidEndSepPunct[3]{}
\providecommand*\mciteSetBstSublistLabelBeginEnd[3]{}
\providecommand*\EndOfBibitem{}
\mciteSetBstSublistMode{f}
\mciteSetBstMaxWidthForm{subitem}{(\alph{mcitesubitemcount})}
\mciteSetBstSublistLabelBeginEnd
  {\mcitemaxwidthsubitemform\space}
  {\relax}
  {\relax}

\bibitem[Sun \latin{et~al.}(2020)Sun, Zhang, Banerjee, Bao, Barbry, Blunt, Bogdanov, Booth, Chen, Cui, Eriksen, Gao, Guo, Hermann, Hermes, Koh, Koval, Lehtola, Li, Liu, Mardirossian, McClain, Motta, Mussard, Pham, Pulkin, Purwanto, Robinson, Ronca, Sayfutyarova, Scheurer, Schurkus, Smith, Sun, Sun, Upadhyay, Wagner, Wang, White, Whitfield, Williamson, Wouters, Yang, Yu, Zhu, Berkelbach, Sharma, Sokolov, and Chan]{Sun2020}
Sun,~Q.; Zhang,~X.; Banerjee,~S.; Bao,~P.; Barbry,~M.; Blunt,~N.~S.; Bogdanov,~N.~A.; Booth,~G.~H.; Chen,~J.; Cui,~Z.-H.; Eriksen,~J.~J.; Gao,~Y.; Guo,~S.; Hermann,~J.; Hermes,~M.~R.; Koh,~K.; Koval,~P.; Lehtola,~S.; Li,~Z.; Liu,~J.; Mardirossian,~N.; McClain,~J.~D.; Motta,~M.; Mussard,~B.; Pham,~H.~Q.; Pulkin,~A.; Purwanto,~W.; Robinson,~P.~J.; Ronca,~E.; Sayfutyarova,~E.~R.; Scheurer,~M.; Schurkus,~H.~F.; Smith,~J. E.~T.; Sun,~C.; Sun,~S.-N.; Upadhyay,~S.; Wagner,~L.~K.; Wang,~X.; White,~A.; Whitfield,~J.~D.; Williamson,~M.~J.; Wouters,~S.; Yang,~J.; Yu,~J.~M.; Zhu,~T.; Berkelbach,~T.~C.; Sharma,~S.; Sokolov,~A.~Y.; Chan,~G. K.-L. {Recent developments in the PySCF program package}. \emph{J. Chem. Phys.} \textbf{2020}, \emph{153}, 024109\relax
\mciteBstWouldAddEndPuncttrue
\mciteSetBstMidEndSepPunct{\mcitedefaultmidpunct}
{\mcitedefaultendpunct}{\mcitedefaultseppunct}\relax
\EndOfBibitem
\bibitem[Sun \latin{et~al.}(2018)Sun, Berkelbach, Blunt, Booth, Guo, Li, Liu, McClain, Sayfutyarova, Sharma, Wouters, and Chan]{Sun2018}
Sun,~Q.; Berkelbach,~T.~C.; Blunt,~N.~S.; Booth,~G.~H.; Guo,~S.; Li,~Z.; Liu,~J.; McClain,~J.~D.; Sayfutyarova,~E.~R.; Sharma,~S.; Wouters,~S.; Chan,~G. K.-L. PySCF: the Python-based simulations of chemistry framework. \emph{WIREs Comput. Mol. Sci.} \textbf{2018}, \emph{8}, e1340\relax
\mciteBstWouldAddEndPuncttrue
\mciteSetBstMidEndSepPunct{\mcitedefaultmidpunct}
{\mcitedefaultendpunct}{\mcitedefaultseppunct}\relax
\EndOfBibitem
\end{mcitethebibliography}


\begin{thebibliography}{62}%
\makeatletter
\providecommand \@ifxundefined [1]{%
 \@ifx{#1\undefined}
}%
\providecommand \@ifnum [1]{%
 \ifnum #1\expandafter \@firstoftwo
 \else \expandafter \@secondoftwo
 \fi
}%
\providecommand \@ifx [1]{%
 \ifx #1\expandafter \@firstoftwo
 \else \expandafter \@secondoftwo
 \fi
}%
\providecommand \natexlab [1]{#1}%
\providecommand \enquote  [1]{``#1''}%
\providecommand \bibnamefont  [1]{#1}%
\providecommand \bibfnamefont [1]{#1}%
\providecommand \citenamefont [1]{#1}%
\providecommand \href@noop [0]{\@secondoftwo}%
\providecommand \href [0]{\begingroup \@sanitize@url \@href}%
\providecommand \@href[1]{\@@startlink{#1}\@@href}%
\providecommand \@@href[1]{\endgroup#1\@@endlink}%
\providecommand \@sanitize@url [0]{\catcode `\\12\catcode `\$12\catcode `\&12\catcode `\#12\catcode `\^12\catcode `\_12\catcode `\%12\relax}%
\providecommand \@@startlink[1]{}%
\providecommand \@@endlink[0]{}%
\providecommand \url  [0]{\begingroup\@sanitize@url \@url }%
\providecommand \@url [1]{\endgroup\@href {#1}{\urlprefix }}%
\providecommand \urlprefix  [0]{URL }%
\providecommand \Eprint [0]{\href }%
\providecommand \doibase [0]{http://dx.doi.org/}%
\providecommand \selectlanguage [0]{\@gobble}%
\providecommand \bibinfo  [0]{\@secondoftwo}%
\providecommand \bibfield  [0]{\@secondoftwo}%
\providecommand \translation [1]{[#1]}%
\providecommand \BibitemOpen [0]{}%
\providecommand \bibitemStop [0]{}%
\providecommand \bibitemNoStop [0]{.\EOS\space}%
\providecommand \EOS [0]{\spacefactor3000\relax}%
\providecommand \BibitemShut  [1]{\csname bibitem#1\endcsname}%
\let\auto@bib@innerbib\@empty
%</preamble>
\bibitem [{\citenamefont {Jorgensen}\ and\ \citenamefont {Tirado-Rives}(1988)}]{Jorgensen1988}%
  \BibitemOpen
  \bibfield  {author} {\bibinfo {author} {\bibfnamefont {W.~L.}\ \bibnamefont {Jorgensen}}\ and\ \bibinfo {author} {\bibfnamefont {J.}~\bibnamefont {Tirado-Rives}},\ }\href@noop {} {\bibfield  {journal} {\bibinfo  {journal} {J. Am. Chem. Soc.}\ }\textbf {\bibinfo {volume} {110}},\ \bibinfo {pages} {1657} (\bibinfo {year} {1988})}\BibitemShut {NoStop}%
\bibitem [{\citenamefont {Weiner}\ \emph {et~al.}(1984)\citenamefont {Weiner}, \citenamefont {Kollman}, \citenamefont {Case}, \citenamefont {Singh}, \citenamefont {Ghio}, \citenamefont {Alagona}, \citenamefont {Profeta},\ and\ \citenamefont {Weiner}}]{Weiner1984}%
  \BibitemOpen
  \bibfield  {author} {\bibinfo {author} {\bibfnamefont {S.~J.}\ \bibnamefont {Weiner}}, \bibinfo {author} {\bibfnamefont {P.~A.}\ \bibnamefont {Kollman}}, \bibinfo {author} {\bibfnamefont {D.~A.}\ \bibnamefont {Case}}, \bibinfo {author} {\bibfnamefont {U.~C.}\ \bibnamefont {Singh}}, \bibinfo {author} {\bibfnamefont {C.}~\bibnamefont {Ghio}}, \bibinfo {author} {\bibfnamefont {G.}~\bibnamefont {Alagona}}, \bibinfo {author} {\bibfnamefont {S.}~\bibnamefont {Profeta}}, \ and\ \bibinfo {author} {\bibfnamefont {P.}~\bibnamefont {Weiner}},\ }\href@noop {} {\bibfield  {journal} {\bibinfo  {journal} {J. Am. Chem. Soc.}\ }\textbf {\bibinfo {volume} {106}},\ \bibinfo {pages} {765} (\bibinfo {year} {1984})}\BibitemShut {NoStop}%
\bibitem [{\citenamefont {Smith}\ \emph {et~al.}(2019)\citenamefont {Smith}, \citenamefont {Nebgen}, \citenamefont {Zubatyuk}, \citenamefont {Lubbers}, \citenamefont {Devereux}, \citenamefont {Barros}, \citenamefont {Tretiak}, \citenamefont {Isayev},\ and\ \citenamefont {Roitberg}}]{Smith2019}%
  \BibitemOpen
  \bibfield  {author} {\bibinfo {author} {\bibfnamefont {J.~S.}\ \bibnamefont {Smith}}, \bibinfo {author} {\bibfnamefont {B.~T.}\ \bibnamefont {Nebgen}}, \bibinfo {author} {\bibfnamefont {R.}~\bibnamefont {Zubatyuk}}, \bibinfo {author} {\bibfnamefont {N.}~\bibnamefont {Lubbers}}, \bibinfo {author} {\bibfnamefont {C.}~\bibnamefont {Devereux}}, \bibinfo {author} {\bibfnamefont {K.}~\bibnamefont {Barros}}, \bibinfo {author} {\bibfnamefont {S.}~\bibnamefont {Tretiak}}, \bibinfo {author} {\bibfnamefont {O.}~\bibnamefont {Isayev}}, \ and\ \bibinfo {author} {\bibfnamefont {A.~E.}\ \bibnamefont {Roitberg}},\ }\href@noop {} {\bibfield  {journal} {\bibinfo  {journal} {Nat. Commun.}\ }\textbf {\bibinfo {volume} {10}},\ \bibinfo {pages} {2903} (\bibinfo {year} {2019})}\BibitemShut {NoStop}%
\bibitem [{\citenamefont {Smith}\ \emph {et~al.}(2017)\citenamefont {Smith}, \citenamefont {Isayev},\ and\ \citenamefont {Roitberg}}]{Smith2017}%
  \BibitemOpen
  \bibfield  {author} {\bibinfo {author} {\bibfnamefont {J.~S.}\ \bibnamefont {Smith}}, \bibinfo {author} {\bibfnamefont {O.}~\bibnamefont {Isayev}}, \ and\ \bibinfo {author} {\bibfnamefont {A.~E.}\ \bibnamefont {Roitberg}},\ }\href@noop {} {\bibfield  {journal} {\bibinfo  {journal} {Chem. Sci.}\ }\textbf {\bibinfo {volume} {8}},\ \bibinfo {pages} {3192} (\bibinfo {year} {2017})}\BibitemShut {NoStop}%
\bibitem [{\citenamefont {Chmiela}\ \emph {et~al.}(2018)\citenamefont {Chmiela}, \citenamefont {Sauceda},\ and\ \citenamefont {Tkatchenko}}]{Chmiela2018}%
  \BibitemOpen
  \bibfield  {author} {\bibinfo {author} {\bibfnamefont {S.}~\bibnamefont {Chmiela}}, \bibinfo {author} {\bibfnamefont {K.-R.}\ \bibnamefont {Sauceda}, \bibfnamefont {Huziel E.and~M{\"u}ller}}, \ and\ \bibinfo {author} {\bibfnamefont {A.}~\bibnamefont {Tkatchenko}},\ }\href@noop {} {\bibfield  {journal} {\bibinfo  {journal} {Nat. Commun.}\ }\textbf {\bibinfo {volume} {9}},\ \bibinfo {pages} {3887} (\bibinfo {year} {2018})}\BibitemShut {NoStop}%
\bibitem [{\citenamefont {Zhao}\ \emph {et~al.}(2009)\citenamefont {Zhao}, \citenamefont {Tishchenko}, \citenamefont {Gour}, \citenamefont {Li}, \citenamefont {Lutz}, \citenamefont {Piecuch},\ and\ \citenamefont {Truhlar}}]{Zhao2009}%
  \BibitemOpen
  \bibfield  {author} {\bibinfo {author} {\bibfnamefont {Y.}~\bibnamefont {Zhao}}, \bibinfo {author} {\bibfnamefont {O.}~\bibnamefont {Tishchenko}}, \bibinfo {author} {\bibfnamefont {J.~R.}\ \bibnamefont {Gour}}, \bibinfo {author} {\bibfnamefont {W.}~\bibnamefont {Li}}, \bibinfo {author} {\bibfnamefont {J.~J.}\ \bibnamefont {Lutz}}, \bibinfo {author} {\bibfnamefont {P.}~\bibnamefont {Piecuch}}, \ and\ \bibinfo {author} {\bibfnamefont {D.~G.}\ \bibnamefont {Truhlar}},\ }\href {\doibase 10.1021/jp811054n} {\bibfield  {journal} {\bibinfo  {journal} {J. Phys. Chem. A}\ }\textbf {\bibinfo {volume} {113}},\ \bibinfo {pages} {5786} (\bibinfo {year} {2009})},\ \bibinfo {note} {pMID: 19374412}\BibitemShut {NoStop}%
\bibitem [{\citenamefont {Goerigk}\ and\ \citenamefont {Grimme}(2010)}]{Goerigk2010}%
  \BibitemOpen
  \bibfield  {author} {\bibinfo {author} {\bibfnamefont {L.}~\bibnamefont {Goerigk}}\ and\ \bibinfo {author} {\bibfnamefont {S.}~\bibnamefont {Grimme}},\ }\href {\doibase 10.1021/ct900489g} {\bibfield  {journal} {\bibinfo  {journal} {J. Chem. Theory Comput.}\ }\textbf {\bibinfo {volume} {6}},\ \bibinfo {pages} {107} (\bibinfo {year} {2010})}\BibitemShut {NoStop}%
\bibitem [{\citenamefont {Zhao}\ and\ \citenamefont {Truhlar}(2011)}]{Zhao2011}%
  \BibitemOpen
  \bibfield  {author} {\bibinfo {author} {\bibfnamefont {Y.}~\bibnamefont {Zhao}}\ and\ \bibinfo {author} {\bibfnamefont {D.~G.}\ \bibnamefont {Truhlar}},\ }\href {\doibase 10.1021/ct1006604} {\bibfield  {journal} {\bibinfo  {journal} {J. Chem. Theory Comput.}\ }\textbf {\bibinfo {volume} {7}},\ \bibinfo {pages} {669} (\bibinfo {year} {2011})}\BibitemShut {NoStop}%
\bibitem [{\citenamefont {Johnson}\ \emph {et~al.}(2008)\citenamefont {Johnson}, \citenamefont {Mori-Sánchez}, \citenamefont {Cohen},\ and\ \citenamefont {Yang}}]{Johnson2008}%
  \BibitemOpen
  \bibfield  {author} {\bibinfo {author} {\bibfnamefont {E.~R.}\ \bibnamefont {Johnson}}, \bibinfo {author} {\bibfnamefont {P.}~\bibnamefont {Mori-Sánchez}}, \bibinfo {author} {\bibfnamefont {A.~J.}\ \bibnamefont {Cohen}}, \ and\ \bibinfo {author} {\bibfnamefont {W.}~\bibnamefont {Yang}},\ }\href {\doibase 10.1063/1.3021474} {\bibfield  {journal} {\bibinfo  {journal} {J. Chem. Phys.}\ }\textbf {\bibinfo {volume} {129}},\ \bibinfo {pages} {204112} (\bibinfo {year} {2008})}\BibitemShut {NoStop}%
\bibitem [{\citenamefont {Chamkin}\ and\ \citenamefont {Serkova}(2020)}]{Chamkin2020}%
  \BibitemOpen
  \bibfield  {author} {\bibinfo {author} {\bibfnamefont {A.~A.}\ \bibnamefont {Chamkin}}\ and\ \bibinfo {author} {\bibfnamefont {E.~S.}\ \bibnamefont {Serkova}},\ }\href {\doibase https://doi.org/10.1002/jcc.26398} {\bibfield  {journal} {\bibinfo  {journal} {J. Comput. Chem.}\ }\textbf {\bibinfo {volume} {41}},\ \bibinfo {pages} {2388} (\bibinfo {year} {2020})}\BibitemShut {NoStop}%
\bibitem [{\citenamefont {Vermeeren}\ \emph {et~al.}(2022)\citenamefont {Vermeeren}, \citenamefont {Dalla~Tiezza}, \citenamefont {Wolf}, \citenamefont {Lahm}, \citenamefont {Allen}, \citenamefont {Schaefer}, \citenamefont {Hamlin},\ and\ \citenamefont {Bickelhaupt}}]{Vermeeren2022}%
  \BibitemOpen
  \bibfield  {author} {\bibinfo {author} {\bibfnamefont {P.}~\bibnamefont {Vermeeren}}, \bibinfo {author} {\bibfnamefont {M.}~\bibnamefont {Dalla~Tiezza}}, \bibinfo {author} {\bibfnamefont {M.~E.}\ \bibnamefont {Wolf}}, \bibinfo {author} {\bibfnamefont {M.~E.}\ \bibnamefont {Lahm}}, \bibinfo {author} {\bibfnamefont {W.~D.}\ \bibnamefont {Allen}}, \bibinfo {author} {\bibfnamefont {H.~F.}\ \bibnamefont {Schaefer}}, \bibinfo {author} {\bibfnamefont {T.~A.}\ \bibnamefont {Hamlin}}, \ and\ \bibinfo {author} {\bibfnamefont {F.~M.}\ \bibnamefont {Bickelhaupt}},\ }\href {\doibase 10.1039/D2CP02234F} {\bibfield  {journal} {\bibinfo  {journal} {Phys. Chem. Chem. Phys.}\ }\textbf {\bibinfo {volume} {24}},\ \bibinfo {pages} {18028} (\bibinfo {year} {2022})}\BibitemShut {NoStop}%
\bibitem [{\citenamefont {Cohen}\ \emph {et~al.}(2008)\citenamefont {Cohen}, \citenamefont {Mori-Sánchez},\ and\ \citenamefont {Yang}}]{Cohen2008}%
  \BibitemOpen
  \bibfield  {author} {\bibinfo {author} {\bibfnamefont {A.~J.}\ \bibnamefont {Cohen}}, \bibinfo {author} {\bibfnamefont {P.}~\bibnamefont {Mori-Sánchez}}, \ and\ \bibinfo {author} {\bibfnamefont {W.}~\bibnamefont {Yang}},\ }\href {\doibase 10.1126/science.1158722} {\bibfield  {journal} {\bibinfo  {journal} {Science}\ }\textbf {\bibinfo {volume} {321}},\ \bibinfo {pages} {792} (\bibinfo {year} {2008})}\BibitemShut {NoStop}%
\bibitem [{\citenamefont {Li}\ \emph {et~al.}(2024)\citenamefont {Li}, \citenamefont {Mansoori~Kermani}, \citenamefont {Ottochian}, \citenamefont {Crescenzi}, \citenamefont {Janesko}, \citenamefont {Truhlar}, \citenamefont {Scalmani}, \citenamefont {Frisch}, \citenamefont {Ciofini},\ and\ \citenamefont {Adamo}}]{Li2024}%
  \BibitemOpen
  \bibfield  {author} {\bibinfo {author} {\bibfnamefont {H.}~\bibnamefont {Li}}, \bibinfo {author} {\bibfnamefont {M.}~\bibnamefont {Mansoori~Kermani}}, \bibinfo {author} {\bibfnamefont {A.}~\bibnamefont {Ottochian}}, \bibinfo {author} {\bibfnamefont {O.}~\bibnamefont {Crescenzi}}, \bibinfo {author} {\bibfnamefont {B.~G.}\ \bibnamefont {Janesko}}, \bibinfo {author} {\bibfnamefont {D.~G.}\ \bibnamefont {Truhlar}}, \bibinfo {author} {\bibfnamefont {G.}~\bibnamefont {Scalmani}}, \bibinfo {author} {\bibfnamefont {M.~J.}\ \bibnamefont {Frisch}}, \bibinfo {author} {\bibfnamefont {I.}~\bibnamefont {Ciofini}}, \ and\ \bibinfo {author} {\bibfnamefont {C.}~\bibnamefont {Adamo}},\ }\href {\doibase 10.1021/jacs.3c12713} {\bibfield  {journal} {\bibinfo  {journal} {J. Am. Chem. Soc.}\ }\textbf {\bibinfo {volume} {146}},\ \bibinfo {pages} {6721} (\bibinfo {year} {2024})}\BibitemShut {NoStop}%
\bibitem [{\citenamefont {Sch{\"u}tt}\ \emph {et~al.}(2017)\citenamefont {Sch{\"u}tt}, \citenamefont {Arbabzadah}, \citenamefont {Chmiela}, \citenamefont {M{\"u}ller},\ and\ \citenamefont {Tkatchenko}}]{Schutt2017}%
  \BibitemOpen
  \bibfield  {author} {\bibinfo {author} {\bibfnamefont {K.~T.}\ \bibnamefont {Sch{\"u}tt}}, \bibinfo {author} {\bibfnamefont {F.}~\bibnamefont {Arbabzadah}}, \bibinfo {author} {\bibfnamefont {S.}~\bibnamefont {Chmiela}}, \bibinfo {author} {\bibfnamefont {K.-R.}\ \bibnamefont {M{\"u}ller}}, \ and\ \bibinfo {author} {\bibfnamefont {A.}~\bibnamefont {Tkatchenko}},\ }\href@noop {} {\bibfield  {journal} {\bibinfo  {journal} {Nat. Commun.}\ }\textbf {\bibinfo {volume} {8}},\ \bibinfo {pages} {13890} (\bibinfo {year} {2017})}\BibitemShut {NoStop}%
\bibitem [{\citenamefont {Bart{\'o}k}\ \emph {et~al.}(2010)\citenamefont {Bart{\'o}k}, \citenamefont {Payne}, \citenamefont {Kondor},\ and\ \citenamefont {Cs{\'a}nyi}}]{bartok2010gaussian}%
  \BibitemOpen
  \bibfield  {author} {\bibinfo {author} {\bibfnamefont {A.~P.}\ \bibnamefont {Bart{\'o}k}}, \bibinfo {author} {\bibfnamefont {M.~C.}\ \bibnamefont {Payne}}, \bibinfo {author} {\bibfnamefont {R.}~\bibnamefont {Kondor}}, \ and\ \bibinfo {author} {\bibfnamefont {G.}~\bibnamefont {Cs{\'a}nyi}},\ }\href@noop {} {\bibfield  {journal} {\bibinfo  {journal} {Phys. Rev. Lett.}\ }\textbf {\bibinfo {volume} {104}},\ \bibinfo {pages} {136403} (\bibinfo {year} {2010})}\BibitemShut {NoStop}%
\bibitem [{\citenamefont {Behler}(2011)}]{Behler:2011it}%
  \BibitemOpen
  \bibfield  {author} {\bibinfo {author} {\bibfnamefont {J.}~\bibnamefont {Behler}},\ }\href@noop {} {\bibfield  {journal} {\bibinfo  {journal} {J. Chem. Phys}\ }\textbf {\bibinfo {volume} {134}},\ \bibinfo {pages} {074106} (\bibinfo {year} {2011})}\BibitemShut {NoStop}%
\bibitem [{\citenamefont {Unke}\ \emph {et~al.}(2021)\citenamefont {Unke}, \citenamefont {Chmiela}, \citenamefont {Sauceda}, \citenamefont {Gastegger}, \citenamefont {Poltavsky}, \citenamefont {Sch{\"u}tt}, \citenamefont {Tkatchenko},\ and\ \citenamefont {M{\"u}ller}}]{ChemRev-MLFF}%
  \BibitemOpen
  \bibfield  {author} {\bibinfo {author} {\bibfnamefont {O.~T.}\ \bibnamefont {Unke}}, \bibinfo {author} {\bibfnamefont {S.}~\bibnamefont {Chmiela}}, \bibinfo {author} {\bibfnamefont {H.~E.}\ \bibnamefont {Sauceda}}, \bibinfo {author} {\bibfnamefont {M.}~\bibnamefont {Gastegger}}, \bibinfo {author} {\bibfnamefont {I.}~\bibnamefont {Poltavsky}}, \bibinfo {author} {\bibfnamefont {K.~T.}\ \bibnamefont {Sch{\"u}tt}}, \bibinfo {author} {\bibfnamefont {A.}~\bibnamefont {Tkatchenko}}, \ and\ \bibinfo {author} {\bibfnamefont {K.-R.}\ \bibnamefont {M{\"u}ller}},\ }\href@noop {} {\bibfield  {journal} {\bibinfo  {journal} {Chem. Rev.}\ }\textbf {\bibinfo {volume} {121}},\ \bibinfo {pages} {10142} (\bibinfo {year} {2021})}\BibitemShut {NoStop}%
\bibitem [{\citenamefont {Mueller}\ \emph {et~al.}(2020)\citenamefont {Mueller}, \citenamefont {Hernandez},\ and\ \citenamefont {Wang}}]{Mueller2020}%
  \BibitemOpen
  \bibfield  {author} {\bibinfo {author} {\bibfnamefont {T.}~\bibnamefont {Mueller}}, \bibinfo {author} {\bibfnamefont {A.}~\bibnamefont {Hernandez}}, \ and\ \bibinfo {author} {\bibfnamefont {C.}~\bibnamefont {Wang}},\ }\href@noop {} {\bibfield  {journal} {\bibinfo  {journal} {J. Chem. Phys.}\ }\textbf {\bibinfo {volume} {152}},\ \bibinfo {pages} {050902} (\bibinfo {year} {2020})}\BibitemShut {NoStop}%
\bibitem [{\citenamefont {Bart{\'o}k}\ \emph {et~al.}(2018)\citenamefont {Bart{\'o}k}, \citenamefont {Kermode}, \citenamefont {Bernstein},\ and\ \citenamefont {Cs{\'a}nyi}}]{Bartok:2018ih}%
  \BibitemOpen
  \bibfield  {author} {\bibinfo {author} {\bibfnamefont {A.~P.}\ \bibnamefont {Bart{\'o}k}}, \bibinfo {author} {\bibfnamefont {J.}~\bibnamefont {Kermode}}, \bibinfo {author} {\bibfnamefont {N.}~\bibnamefont {Bernstein}}, \ and\ \bibinfo {author} {\bibfnamefont {G.}~\bibnamefont {Cs{\'a}nyi}},\ }\href@noop {} {\bibfield  {journal} {\bibinfo  {journal} {Phys. Rev. X}\ }\textbf {\bibinfo {volume} {8}},\ \bibinfo {pages} {041048} (\bibinfo {year} {2018})}\BibitemShut {NoStop}%
\bibitem [{\citenamefont {Bartlett}\ and\ \citenamefont {Musia\l{}}(2007)}]{Bartlett2007}%
  \BibitemOpen
  \bibfield  {author} {\bibinfo {author} {\bibfnamefont {R.~J.}\ \bibnamefont {Bartlett}}\ and\ \bibinfo {author} {\bibfnamefont {M.}~\bibnamefont {Musia\l{}}},\ }\href {\doibase 10.1103/RevModPhys.79.291} {\bibfield  {journal} {\bibinfo  {journal} {Rev. Mod. Phys.}\ }\textbf {\bibinfo {volume} {79}},\ \bibinfo {pages} {291} (\bibinfo {year} {2007})}\BibitemShut {NoStop}%
\bibitem [{\citenamefont {Visscher}\ \emph {et~al.}(1996)\citenamefont {Visscher}, \citenamefont {Lee},\ and\ \citenamefont {Dyall}}]{Visscher1996}%
  \BibitemOpen
  \bibfield  {author} {\bibinfo {author} {\bibfnamefont {L.}~\bibnamefont {Visscher}}, \bibinfo {author} {\bibfnamefont {T.~J.}\ \bibnamefont {Lee}}, \ and\ \bibinfo {author} {\bibfnamefont {K.~G.}\ \bibnamefont {Dyall}},\ }\href {\doibase 10.1063/1.472655} {\bibfield  {journal} {\bibinfo  {journal} {J. Chem. Phys.}\ }\textbf {\bibinfo {volume} {105}},\ \bibinfo {pages} {8769} (\bibinfo {year} {1996})}\BibitemShut {NoStop}%
\bibitem [{\citenamefont {Korth}\ and\ \citenamefont {Grimme}(2009)}]{Korth2009}%
  \BibitemOpen
  \bibfield  {author} {\bibinfo {author} {\bibfnamefont {M.}~\bibnamefont {Korth}}\ and\ \bibinfo {author} {\bibfnamefont {S.}~\bibnamefont {Grimme}},\ }\href {\doibase 10.1021/ct800511q} {\bibfield  {journal} {\bibinfo  {journal} {J. Chem. Theory Comput.}\ }\textbf {\bibinfo {volume} {5}},\ \bibinfo {pages} {993} (\bibinfo {year} {2009})}\BibitemShut {NoStop}%
\bibitem [{\citenamefont {Hait}\ and\ \citenamefont {Head-Gordon}(2018)}]{Hait2018}%
  \BibitemOpen
  \bibfield  {author} {\bibinfo {author} {\bibfnamefont {D.}~\bibnamefont {Hait}}\ and\ \bibinfo {author} {\bibfnamefont {M.}~\bibnamefont {Head-Gordon}},\ }\href@noop {} {\bibfield  {journal} {\bibinfo  {journal} {J. Phys. Chem. Letters}\ }\textbf {\bibinfo {volume} {9}},\ \bibinfo {pages} {6280} (\bibinfo {year} {2018})}\BibitemShut {NoStop}%
\bibitem [{\citenamefont {Seeger}\ and\ \citenamefont {Pople}(1977)}]{Seeger1977}%
  \BibitemOpen
  \bibfield  {author} {\bibinfo {author} {\bibfnamefont {R.}~\bibnamefont {Seeger}}\ and\ \bibinfo {author} {\bibfnamefont {J.~A.}\ \bibnamefont {Pople}},\ }\href@noop {} {\bibfield  {journal} {\bibinfo  {journal} {J. Chem. Phys.}\ }\textbf {\bibinfo {volume} {66}},\ \bibinfo {pages} {3045} (\bibinfo {year} {1977})}\BibitemShut {NoStop}%
\bibitem [{\citenamefont {Fan}\ \emph {et~al.}(2006)\citenamefont {Fan}, \citenamefont {Ho},\ and\ \citenamefont {Bettens}}]{fan2006approximating}%
  \BibitemOpen
  \bibfield  {author} {\bibinfo {author} {\bibfnamefont {Y.}~\bibnamefont {Fan}}, \bibinfo {author} {\bibfnamefont {J.}~\bibnamefont {Ho}}, \ and\ \bibinfo {author} {\bibfnamefont {R.~P.}\ \bibnamefont {Bettens}},\ }\href@noop {} {\bibfield  {journal} {\bibinfo  {journal} {J. Phys. Chem. A}\ }\textbf {\bibinfo {volume} {110}},\ \bibinfo {pages} {2796} (\bibinfo {year} {2006})}\BibitemShut {NoStop}%
\bibitem [{\citenamefont {Smith}\ \emph {et~al.}(2018)\citenamefont {Smith}, \citenamefont {Nebgen}, \citenamefont {Lubbers}, \citenamefont {Isayev},\ and\ \citenamefont {Roitberg}}]{Smith2018-less}%
  \BibitemOpen
  \bibfield  {author} {\bibinfo {author} {\bibfnamefont {J.~S.}\ \bibnamefont {Smith}}, \bibinfo {author} {\bibfnamefont {B.}~\bibnamefont {Nebgen}}, \bibinfo {author} {\bibfnamefont {N.}~\bibnamefont {Lubbers}}, \bibinfo {author} {\bibfnamefont {O.}~\bibnamefont {Isayev}}, \ and\ \bibinfo {author} {\bibfnamefont {A.~E.}\ \bibnamefont {Roitberg}},\ }\href@noop {} {\bibfield  {journal} {\bibinfo  {journal} {J. Chem. Phys.}\ }\textbf {\bibinfo {volume} {148}},\ \bibinfo {pages} {241733} (\bibinfo {year} {2018})}\BibitemShut {NoStop}%
\bibitem [{\citenamefont {Grambow}\ \emph {et~al.}(2020)\citenamefont {Grambow}, \citenamefont {Pattanaik},\ and\ \citenamefont {Green}}]{Grambow2020}%
  \BibitemOpen
  \bibfield  {author} {\bibinfo {author} {\bibfnamefont {C.~A.}\ \bibnamefont {Grambow}}, \bibinfo {author} {\bibfnamefont {L.}~\bibnamefont {Pattanaik}}, \ and\ \bibinfo {author} {\bibfnamefont {W.~H.}\ \bibnamefont {Green}},\ }\href {\doibase 10.1038/s41597-020-0460-4} {\bibfield  {journal} {\bibinfo  {journal} {Sci. Data}\ }\textbf {\bibinfo {volume} {7}},\ \bibinfo {pages} {137} (\bibinfo {year} {2020})}\BibitemShut {NoStop}%
\bibitem [{\citenamefont {Henkelman}\ and\ \citenamefont {Jónsson}(1999)}]{Henkelman1999}%
  \BibitemOpen
  \bibfield  {author} {\bibinfo {author} {\bibfnamefont {G.}~\bibnamefont {Henkelman}}\ and\ \bibinfo {author} {\bibfnamefont {H.}~\bibnamefont {Jónsson}},\ }\href@noop {} {\bibfield  {journal} {\bibinfo  {journal} {J. Chem. Phys.}\ }\textbf {\bibinfo {volume} {111}},\ \bibinfo {pages} {7010} (\bibinfo {year} {1999})}\BibitemShut {NoStop}%
\bibitem [{\citenamefont {Jonsson}\ \emph {et~al.}(1998)\citenamefont {Jonsson}, \citenamefont {Mills},\ and\ \citenamefont {Jacobsen}}]{Jonsson1998}%
  \BibitemOpen
  \bibfield  {author} {\bibinfo {author} {\bibfnamefont {H.}~\bibnamefont {Jonsson}}, \bibinfo {author} {\bibfnamefont {G.}~\bibnamefont {Mills}}, \ and\ \bibinfo {author} {\bibfnamefont {K.~W.}\ \bibnamefont {Jacobsen}},\ }\href {\doibase 10.1142/9789812839664_0016} {\emph {\bibinfo {title} {Classical and Quantum Dynamics in Condensed Phase}}}\ (\bibinfo {year} {1998})\ pp.\ \bibinfo {pages} {385--404}\BibitemShut {NoStop}%
\bibitem [{\citenamefont {Zimmerman}(2015)}]{Zimmerman2015}%
  \BibitemOpen
  \bibfield  {author} {\bibinfo {author} {\bibfnamefont {P.~M.}\ \bibnamefont {Zimmerman}},\ }\href {\doibase https://doi.org/10.1002/jcc.23833} {\bibfield  {journal} {\bibinfo  {journal} {J. Comput. Chem.}\ }\textbf {\bibinfo {volume} {36}},\ \bibinfo {pages} {601} (\bibinfo {year} {2015})}\BibitemShut {NoStop}%
\bibitem [{\citenamefont {Kovács}\ \emph {et~al.}(2025)\citenamefont {Kovács}, \citenamefont {Moore}, \citenamefont {Browning}, \citenamefont {Batatia}, \citenamefont {Horton}, \citenamefont {Pu}, \citenamefont {Kapil}, \citenamefont {Witt}, \citenamefont {Magdău}, \citenamefont {Cole},\ and\ \citenamefont {Csányi}}]{Kovacs2025}%
  \BibitemOpen
  \bibfield  {author} {\bibinfo {author} {\bibfnamefont {D.~P.}\ \bibnamefont {Kovács}}, \bibinfo {author} {\bibfnamefont {J.~H.}\ \bibnamefont {Moore}}, \bibinfo {author} {\bibfnamefont {N.~J.}\ \bibnamefont {Browning}}, \bibinfo {author} {\bibfnamefont {I.}~\bibnamefont {Batatia}}, \bibinfo {author} {\bibfnamefont {J.~T.}\ \bibnamefont {Horton}}, \bibinfo {author} {\bibfnamefont {Y.}~\bibnamefont {Pu}}, \bibinfo {author} {\bibfnamefont {V.}~\bibnamefont {Kapil}}, \bibinfo {author} {\bibfnamefont {W.~C.}\ \bibnamefont {Witt}}, \bibinfo {author} {\bibfnamefont {I.-B.}\ \bibnamefont {Magdău}}, \bibinfo {author} {\bibfnamefont {D.~J.}\ \bibnamefont {Cole}}, \ and\ \bibinfo {author} {\bibfnamefont {G.}~\bibnamefont {Csányi}},\ }\href {\doibase 10.1021/jacs.4c07099} {\bibfield  {journal} {\bibinfo  {journal} {Journal of the American Chemical Society}\ }\textbf {\bibinfo {volume} {147}},\ \bibinfo {pages} {17598} (\bibinfo {year} {2025})},\ \bibinfo {note} {pMID: 40387214},\ \Eprint
  {http://arxiv.org/abs/https://doi.org/10.1021/jacs.4c07099} {https://doi.org/10.1021/jacs.4c07099} \BibitemShut {NoStop}%
\bibitem [{\citenamefont {Zubatyuk}\ \emph {et~al.}(2019)\citenamefont {Zubatyuk}, \citenamefont {Smith}, \citenamefont {Leszczynski},\ and\ \citenamefont {Isayev}}]{Zubatyuk2019}%
  \BibitemOpen
  \bibfield  {author} {\bibinfo {author} {\bibfnamefont {R.}~\bibnamefont {Zubatyuk}}, \bibinfo {author} {\bibfnamefont {J.~S.}\ \bibnamefont {Smith}}, \bibinfo {author} {\bibfnamefont {J.}~\bibnamefont {Leszczynski}}, \ and\ \bibinfo {author} {\bibfnamefont {O.}~\bibnamefont {Isayev}},\ }\href {\doibase 10.1126/sciadv.aav6490} {\bibfield  {journal} {\bibinfo  {journal} {Science Advances}\ }\textbf {\bibinfo {volume} {5}},\ \bibinfo {pages} {eaav6490} (\bibinfo {year} {2019})}\BibitemShut {NoStop}%
\bibitem [{\citenamefont {Zhang}\ \emph {et~al.}(2024{\natexlab{a}})\citenamefont {Zhang}, \citenamefont {Mako{\'{s}}}, \citenamefont {Jadrich}, \citenamefont {Kraka}, \citenamefont {Barros}, \citenamefont {Nebgen}, \citenamefont {Tretiak}, \citenamefont {Isayev}, \citenamefont {Lubbers}, \citenamefont {Messerly},\ and\ \citenamefont {Smith}}]{zhang_2023}%
  \BibitemOpen
  \bibfield  {author} {\bibinfo {author} {\bibfnamefont {S.}~\bibnamefont {Zhang}}, \bibinfo {author} {\bibfnamefont {M.~Z.}\ \bibnamefont {Mako{\'{s}}}}, \bibinfo {author} {\bibfnamefont {R.~B.}\ \bibnamefont {Jadrich}}, \bibinfo {author} {\bibfnamefont {E.}~\bibnamefont {Kraka}}, \bibinfo {author} {\bibfnamefont {K.}~\bibnamefont {Barros}}, \bibinfo {author} {\bibfnamefont {B.~T.}\ \bibnamefont {Nebgen}}, \bibinfo {author} {\bibfnamefont {S.}~\bibnamefont {Tretiak}}, \bibinfo {author} {\bibfnamefont {O.}~\bibnamefont {Isayev}}, \bibinfo {author} {\bibfnamefont {N.}~\bibnamefont {Lubbers}}, \bibinfo {author} {\bibfnamefont {R.~A.}\ \bibnamefont {Messerly}}, \ and\ \bibinfo {author} {\bibfnamefont {J.~S.}\ \bibnamefont {Smith}},\ }\href@noop {} {\bibfield  {journal} {\bibinfo  {journal} {Nat. Chem}\ }\textbf {\bibinfo {volume} {16}},\ \bibinfo {pages} {727} (\bibinfo {year} {2024}{\natexlab{a}})}\BibitemShut {NoStop}%
\bibitem [{\citenamefont {Schreiner}\ \emph {et~al.}(2022)\citenamefont {Schreiner}, \citenamefont {Bhowmik}, \citenamefont {Vegge}, \citenamefont {Busk},\ and\ \citenamefont {Winther}}]{Schreiner2022}%
  \BibitemOpen
  \bibfield  {author} {\bibinfo {author} {\bibfnamefont {M.}~\bibnamefont {Schreiner}}, \bibinfo {author} {\bibfnamefont {A.}~\bibnamefont {Bhowmik}}, \bibinfo {author} {\bibfnamefont {T.}~\bibnamefont {Vegge}}, \bibinfo {author} {\bibfnamefont {J.}~\bibnamefont {Busk}}, \ and\ \bibinfo {author} {\bibfnamefont {O.}~\bibnamefont {Winther}},\ }\href@noop {} {\bibfield  {journal} {\bibinfo  {journal} {Sci. Data}\ }\textbf {\bibinfo {volume} {9}},\ \bibinfo {pages} {779} (\bibinfo {year} {2022})}\BibitemShut {NoStop}%
\bibitem [{\citenamefont {Batatia}\ \emph {et~al.}(2024)\citenamefont {Batatia}, \citenamefont {Benner}, \citenamefont {Chiang}, \citenamefont {Elena}, \citenamefont {Kovács}, \citenamefont {Riebesell}, \citenamefont {Advincula}, \citenamefont {Asta}, \citenamefont {Avaylon}, \citenamefont {Baldwin}, \citenamefont {Berger}, \citenamefont {Bernstein}, \citenamefont {Bhowmik}, \citenamefont {Blau}, \citenamefont {Cărare}, \citenamefont {Darby}, \citenamefont {De}, \citenamefont {Pia}, \citenamefont {Deringer}, \citenamefont {Elijošius}, \citenamefont {El-Machachi}, \citenamefont {Falcioni}, \citenamefont {Fako}, \citenamefont {Ferrari}, \citenamefont {Genreith-Schriever}, \citenamefont {George}, \citenamefont {Goodall}, \citenamefont {Grey}, \citenamefont {Grigorev}, \citenamefont {Han}, \citenamefont {Handley}, \citenamefont {Heenen}, \citenamefont {Hermansson}, \citenamefont {Holm}, \citenamefont {Jaafar}, \citenamefont {Hofmann}, \citenamefont {Jakob}, \citenamefont {Jung}, \citenamefont {Kapil},
  \citenamefont {Kaplan}, \citenamefont {Karimitari}, \citenamefont {Kermode}, \citenamefont {Kroupa}, \citenamefont {Kullgren}, \citenamefont {Kuner}, \citenamefont {Kuryla}, \citenamefont {Liepuoniute}, \citenamefont {Margraf}, \citenamefont {Magdău}, \citenamefont {Michaelides}, \citenamefont {Moore}, \citenamefont {Naik}, \citenamefont {Niblett}, \citenamefont {Norwood}, \citenamefont {O'Neill}, \citenamefont {Ortner}, \citenamefont {Persson}, \citenamefont {Reuter}, \citenamefont {Rosen}, \citenamefont {Schaaf}, \citenamefont {Schran}, \citenamefont {Shi}, \citenamefont {Sivonxay}, \citenamefont {Stenczel}, \citenamefont {Svahn}, \citenamefont {Sutton}, \citenamefont {Swinburne}, \citenamefont {Tilly}, \citenamefont {van~der Oord}, \citenamefont {Varga-Umbrich}, \citenamefont {Vegge}, \citenamefont {Vondrák}, \citenamefont {Wang}, \citenamefont {Witt}, \citenamefont {Zills},\ and\ \citenamefont {Csányi}}]{Batatia2024}%
  \BibitemOpen
  \bibfield  {author} {\bibinfo {author} {\bibfnamefont {I.}~\bibnamefont {Batatia}}, \bibinfo {author} {\bibfnamefont {P.}~\bibnamefont {Benner}}, \bibinfo {author} {\bibfnamefont {Y.}~\bibnamefont {Chiang}}, \bibinfo {author} {\bibfnamefont {A.~M.}\ \bibnamefont {Elena}}, \bibinfo {author} {\bibfnamefont {D.~P.}\ \bibnamefont {Kovács}}, \bibinfo {author} {\bibfnamefont {J.}~\bibnamefont {Riebesell}}, \bibinfo {author} {\bibfnamefont {X.~R.}\ \bibnamefont {Advincula}}, \bibinfo {author} {\bibfnamefont {M.}~\bibnamefont {Asta}}, \bibinfo {author} {\bibfnamefont {M.}~\bibnamefont {Avaylon}}, \bibinfo {author} {\bibfnamefont {W.~J.}\ \bibnamefont {Baldwin}}, \bibinfo {author} {\bibfnamefont {F.}~\bibnamefont {Berger}}, \bibinfo {author} {\bibfnamefont {N.}~\bibnamefont {Bernstein}}, \bibinfo {author} {\bibfnamefont {A.}~\bibnamefont {Bhowmik}}, \bibinfo {author} {\bibfnamefont {S.~M.}\ \bibnamefont {Blau}}, \bibinfo {author} {\bibfnamefont {V.}~\bibnamefont {Cărare}}, \bibinfo {author} {\bibfnamefont {J.~P.}\
  \bibnamefont {Darby}}, \bibinfo {author} {\bibfnamefont {S.}~\bibnamefont {De}}, \bibinfo {author} {\bibfnamefont {F.~D.}\ \bibnamefont {Pia}}, \bibinfo {author} {\bibfnamefont {V.~L.}\ \bibnamefont {Deringer}}, \bibinfo {author} {\bibfnamefont {R.}~\bibnamefont {Elijošius}}, \bibinfo {author} {\bibfnamefont {Z.}~\bibnamefont {El-Machachi}}, \bibinfo {author} {\bibfnamefont {F.}~\bibnamefont {Falcioni}}, \bibinfo {author} {\bibfnamefont {E.}~\bibnamefont {Fako}}, \bibinfo {author} {\bibfnamefont {A.~C.}\ \bibnamefont {Ferrari}}, \bibinfo {author} {\bibfnamefont {A.}~\bibnamefont {Genreith-Schriever}}, \bibinfo {author} {\bibfnamefont {J.}~\bibnamefont {George}}, \bibinfo {author} {\bibfnamefont {R.~E.~A.}\ \bibnamefont {Goodall}}, \bibinfo {author} {\bibfnamefont {C.~P.}\ \bibnamefont {Grey}}, \bibinfo {author} {\bibfnamefont {P.}~\bibnamefont {Grigorev}}, \bibinfo {author} {\bibfnamefont {S.}~\bibnamefont {Han}}, \bibinfo {author} {\bibfnamefont {W.}~\bibnamefont {Handley}}, \bibinfo {author}
  {\bibfnamefont {H.~H.}\ \bibnamefont {Heenen}}, \bibinfo {author} {\bibfnamefont {K.}~\bibnamefont {Hermansson}}, \bibinfo {author} {\bibfnamefont {C.}~\bibnamefont {Holm}}, \bibinfo {author} {\bibfnamefont {J.}~\bibnamefont {Jaafar}}, \bibinfo {author} {\bibfnamefont {S.}~\bibnamefont {Hofmann}}, \bibinfo {author} {\bibfnamefont {K.~S.}\ \bibnamefont {Jakob}}, \bibinfo {author} {\bibfnamefont {H.}~\bibnamefont {Jung}}, \bibinfo {author} {\bibfnamefont {V.}~\bibnamefont {Kapil}}, \bibinfo {author} {\bibfnamefont {A.~D.}\ \bibnamefont {Kaplan}}, \bibinfo {author} {\bibfnamefont {N.}~\bibnamefont {Karimitari}}, \bibinfo {author} {\bibfnamefont {J.~R.}\ \bibnamefont {Kermode}}, \bibinfo {author} {\bibfnamefont {N.}~\bibnamefont {Kroupa}}, \bibinfo {author} {\bibfnamefont {J.}~\bibnamefont {Kullgren}}, \bibinfo {author} {\bibfnamefont {M.~C.}\ \bibnamefont {Kuner}}, \bibinfo {author} {\bibfnamefont {D.}~\bibnamefont {Kuryla}}, \bibinfo {author} {\bibfnamefont {G.}~\bibnamefont {Liepuoniute}}, \bibinfo {author}
  {\bibfnamefont {J.~T.}\ \bibnamefont {Margraf}}, \bibinfo {author} {\bibfnamefont {I.-B.}\ \bibnamefont {Magdău}}, \bibinfo {author} {\bibfnamefont {A.}~\bibnamefont {Michaelides}}, \bibinfo {author} {\bibfnamefont {J.~H.}\ \bibnamefont {Moore}}, \bibinfo {author} {\bibfnamefont {A.~A.}\ \bibnamefont {Naik}}, \bibinfo {author} {\bibfnamefont {S.~P.}\ \bibnamefont {Niblett}}, \bibinfo {author} {\bibfnamefont {S.~W.}\ \bibnamefont {Norwood}}, \bibinfo {author} {\bibfnamefont {N.}~\bibnamefont {O'Neill}}, \bibinfo {author} {\bibfnamefont {C.}~\bibnamefont {Ortner}}, \bibinfo {author} {\bibfnamefont {K.~A.}\ \bibnamefont {Persson}}, \bibinfo {author} {\bibfnamefont {K.}~\bibnamefont {Reuter}}, \bibinfo {author} {\bibfnamefont {A.~S.}\ \bibnamefont {Rosen}}, \bibinfo {author} {\bibfnamefont {L.~L.}\ \bibnamefont {Schaaf}}, \bibinfo {author} {\bibfnamefont {C.}~\bibnamefont {Schran}}, \bibinfo {author} {\bibfnamefont {B.~X.}\ \bibnamefont {Shi}}, \bibinfo {author} {\bibfnamefont {E.}~\bibnamefont {Sivonxay}},
  \bibinfo {author} {\bibfnamefont {T.~K.}\ \bibnamefont {Stenczel}}, \bibinfo {author} {\bibfnamefont {V.}~\bibnamefont {Svahn}}, \bibinfo {author} {\bibfnamefont {C.}~\bibnamefont {Sutton}}, \bibinfo {author} {\bibfnamefont {T.~D.}\ \bibnamefont {Swinburne}}, \bibinfo {author} {\bibfnamefont {J.}~\bibnamefont {Tilly}}, \bibinfo {author} {\bibfnamefont {C.}~\bibnamefont {van~der Oord}}, \bibinfo {author} {\bibfnamefont {E.}~\bibnamefont {Varga-Umbrich}}, \bibinfo {author} {\bibfnamefont {T.}~\bibnamefont {Vegge}}, \bibinfo {author} {\bibfnamefont {M.}~\bibnamefont {Vondrák}}, \bibinfo {author} {\bibfnamefont {Y.}~\bibnamefont {Wang}}, \bibinfo {author} {\bibfnamefont {W.~C.}\ \bibnamefont {Witt}}, \bibinfo {author} {\bibfnamefont {F.}~\bibnamefont {Zills}}, \ and\ \bibinfo {author} {\bibfnamefont {G.}~\bibnamefont {Csányi}},\ }\href {https://arxiv.org/abs/2401.00096} {\enquote {\bibinfo {title} {A foundation model for atomistic materials chemistry},}\ } (\bibinfo {year} {2024}),\ \Eprint
  {http://arxiv.org/abs/2401.00096} {arXiv:2401.00096 [physics.chem-ph]} \BibitemShut {NoStop}%
\bibitem [{\citenamefont {Jain}\ \emph {et~al.}(2013)\citenamefont {Jain}, \citenamefont {Ong}, \citenamefont {Hautier}, \citenamefont {Chen}, \citenamefont {Richards}, \citenamefont {Dacek}, \citenamefont {Cholia}, \citenamefont {Gunter}, \citenamefont {Skinner}, \citenamefont {Ceder},\ and\ \citenamefont {Persson}}]{Jain2013}%
  \BibitemOpen
  \bibfield  {author} {\bibinfo {author} {\bibfnamefont {A.}~\bibnamefont {Jain}}, \bibinfo {author} {\bibfnamefont {S.~P.}\ \bibnamefont {Ong}}, \bibinfo {author} {\bibfnamefont {G.}~\bibnamefont {Hautier}}, \bibinfo {author} {\bibfnamefont {W.}~\bibnamefont {Chen}}, \bibinfo {author} {\bibfnamefont {W.~D.}\ \bibnamefont {Richards}}, \bibinfo {author} {\bibfnamefont {S.}~\bibnamefont {Dacek}}, \bibinfo {author} {\bibfnamefont {S.}~\bibnamefont {Cholia}}, \bibinfo {author} {\bibfnamefont {D.}~\bibnamefont {Gunter}}, \bibinfo {author} {\bibfnamefont {D.}~\bibnamefont {Skinner}}, \bibinfo {author} {\bibfnamefont {G.}~\bibnamefont {Ceder}}, \ and\ \bibinfo {author} {\bibfnamefont {K.~A.}\ \bibnamefont {Persson}},\ }\href {\doibase 10.1063/1.4812323} {\bibfield  {journal} {\bibinfo  {journal} {APL Materials}\ }\textbf {\bibinfo {volume} {1}},\ \bibinfo {pages} {011002} (\bibinfo {year} {2013})}\BibitemShut {NoStop}%
\bibitem [{\citenamefont {Braams}\ and\ \citenamefont {Bowman}(2009)}]{Braams2009-wi}%
  \BibitemOpen
  \bibfield  {author} {\bibinfo {author} {\bibfnamefont {B.~J.}\ \bibnamefont {Braams}}\ and\ \bibinfo {author} {\bibfnamefont {J.~M.}\ \bibnamefont {Bowman}},\ }\href@noop {} {\bibfield  {journal} {\bibinfo  {journal} {Int. Rev. Phys. Chem.}\ }\textbf {\bibinfo {volume} {28}},\ \bibinfo {pages} {577} (\bibinfo {year} {2009})}\BibitemShut {NoStop}%
\bibitem [{\citenamefont {Qu}\ \emph {et~al.}(2018)\citenamefont {Qu}, \citenamefont {Yu},\ and\ \citenamefont {Bowman}}]{Qu:2018gj}%
  \BibitemOpen
  \bibfield  {author} {\bibinfo {author} {\bibfnamefont {C.}~\bibnamefont {Qu}}, \bibinfo {author} {\bibfnamefont {Q.}~\bibnamefont {Yu}}, \ and\ \bibinfo {author} {\bibfnamefont {J.~M.}\ \bibnamefont {Bowman}},\ }\href@noop {} {\bibfield  {journal} {\bibinfo  {journal} {Annu. Rev. Phys. Chem.}\ }\textbf {\bibinfo {volume} {69}},\ \bibinfo {pages} {151} (\bibinfo {year} {2018})}\BibitemShut {NoStop}%
\bibitem [{\citenamefont {Huang}\ \emph {et~al.}(2005)\citenamefont {Huang}, \citenamefont {Braams},\ and\ \citenamefont {Bowman}}]{Huang:2005}%
  \BibitemOpen
  \bibfield  {author} {\bibinfo {author} {\bibfnamefont {X.}~\bibnamefont {Huang}}, \bibinfo {author} {\bibfnamefont {B.~J.}\ \bibnamefont {Braams}}, \ and\ \bibinfo {author} {\bibfnamefont {J.~M.}\ \bibnamefont {Bowman}},\ }\href@noop {} {\bibfield  {journal} {\bibinfo  {journal} {J. Chem. Phys}\ }\textbf {\bibinfo {volume} {122}},\ \bibinfo {pages} {044308} (\bibinfo {year} {2005})}\BibitemShut {NoStop}%
\bibitem [{\citenamefont {Hu}\ \emph {et~al.}(2023)\citenamefont {Hu}, \citenamefont {Johannesen}, \citenamefont {Graham},\ and\ \citenamefont {Goodpaster}}]{Hu2023}%
  \BibitemOpen
  \bibfield  {author} {\bibinfo {author} {\bibfnamefont {Q.~H.}\ \bibnamefont {Hu}}, \bibinfo {author} {\bibfnamefont {A.~M.}\ \bibnamefont {Johannesen}}, \bibinfo {author} {\bibfnamefont {D.~S.}\ \bibnamefont {Graham}}, \ and\ \bibinfo {author} {\bibfnamefont {J.~D.}\ \bibnamefont {Goodpaster}},\ }\href@noop {} {\bibfield  {journal} {\bibinfo  {journal} {Digit. Discov.}\ }\textbf {\bibinfo {volume} {2}},\ \bibinfo {pages} {1058} (\bibinfo {year} {2023})}\BibitemShut {NoStop}%
\bibitem [{\citenamefont {Chigaev}\ \emph {et~al.}(2023)\citenamefont {Chigaev}, \citenamefont {Smith}, \citenamefont {Anaya}, \citenamefont {Nebgen}, \citenamefont {Bettencourt}, \citenamefont {Barros},\ and\ \citenamefont {Lubbers}}]{Chigaev2023}%
  \BibitemOpen
  \bibfield  {author} {\bibinfo {author} {\bibfnamefont {M.}~\bibnamefont {Chigaev}}, \bibinfo {author} {\bibfnamefont {J.~S.}\ \bibnamefont {Smith}}, \bibinfo {author} {\bibfnamefont {S.}~\bibnamefont {Anaya}}, \bibinfo {author} {\bibfnamefont {B.}~\bibnamefont {Nebgen}}, \bibinfo {author} {\bibfnamefont {M.}~\bibnamefont {Bettencourt}}, \bibinfo {author} {\bibfnamefont {K.}~\bibnamefont {Barros}}, \ and\ \bibinfo {author} {\bibfnamefont {N.}~\bibnamefont {Lubbers}},\ }\href {\doibase 10.1063/5.0142127} {\bibfield  {journal} {\bibinfo  {journal} {J. Chem. Phys.}\ }\textbf {\bibinfo {volume} {158}},\ \bibinfo {pages} {184108} (\bibinfo {year} {2023})}\BibitemShut {NoStop}%
\bibitem [{\citenamefont {Lubbers}\ \emph {et~al.}(2018)\citenamefont {Lubbers}, \citenamefont {Smith},\ and\ \citenamefont {Barros}}]{Lubbers2018}%
  \BibitemOpen
  \bibfield  {author} {\bibinfo {author} {\bibfnamefont {N.}~\bibnamefont {Lubbers}}, \bibinfo {author} {\bibfnamefont {J.~S.}\ \bibnamefont {Smith}}, \ and\ \bibinfo {author} {\bibfnamefont {K.}~\bibnamefont {Barros}},\ }\href@noop {} {\bibfield  {journal} {\bibinfo  {journal} {J. Chem. Phys.}\ }\textbf {\bibinfo {volume} {148}} (\bibinfo {year} {2018})}\BibitemShut {NoStop}%
\bibitem [{\citenamefont {Allen}\ \emph {et~al.}(2025)\citenamefont {Allen}, \citenamefont {Shinkle}, \citenamefont {Bujack},\ and\ \citenamefont {Lubbers}}]{hiphop}%
  \BibitemOpen
  \bibfield  {author} {\bibinfo {author} {\bibfnamefont {A.~E.~A.}\ \bibnamefont {Allen}}, \bibinfo {author} {\bibfnamefont {E.}~\bibnamefont {Shinkle}}, \bibinfo {author} {\bibfnamefont {R.}~\bibnamefont {Bujack}}, \ and\ \bibinfo {author} {\bibfnamefont {N.}~\bibnamefont {Lubbers}},\ }\href {https://arxiv.org/abs/2503.23515} {\enquote {\bibinfo {title} {Optimal invariant bases for atomistic machine learning},}\ } (\bibinfo {year} {2025}),\ \Eprint {http://arxiv.org/abs/2503.23515} {arXiv:2503.23515 [physics.chem-ph]} \BibitemShut {NoStop}%
\bibitem [{\citenamefont {Pan}\ and\ \citenamefont {Yang}(2010)}]{Pan2010}%
  \BibitemOpen
  \bibfield  {author} {\bibinfo {author} {\bibfnamefont {S.~J.}\ \bibnamefont {Pan}}\ and\ \bibinfo {author} {\bibfnamefont {Q.}~\bibnamefont {Yang}},\ }\href@noop {} {\bibfield  {journal} {\bibinfo  {journal} {IEEE Trans. Knowl. Data Eng.}\ }\textbf {\bibinfo {volume} {22}},\ \bibinfo {pages} {1345} (\bibinfo {year} {2010})}\BibitemShut {NoStop}%
\bibitem [{\citenamefont {Ramakrishnan}\ \emph {et~al.}(2014)\citenamefont {Ramakrishnan}, \citenamefont {Dral}, \citenamefont {Rupp},\ and\ \citenamefont {von Lilienfeld}}]{Ramakrishnan2014}%
  \BibitemOpen
  \bibfield  {author} {\bibinfo {author} {\bibfnamefont {R.}~\bibnamefont {Ramakrishnan}}, \bibinfo {author} {\bibfnamefont {P.~O.}\ \bibnamefont {Dral}}, \bibinfo {author} {\bibfnamefont {M.}~\bibnamefont {Rupp}}, \ and\ \bibinfo {author} {\bibfnamefont {O.~A.}\ \bibnamefont {von Lilienfeld}},\ }\href@noop {} {\bibfield  {journal} {\bibinfo  {journal} {Sci. Data}\ }\textbf {\bibinfo {volume} {1}},\ \bibinfo {pages} {140022} (\bibinfo {year} {2014})}\BibitemShut {NoStop}%
\bibitem [{\citenamefont {Chai}\ and\ \citenamefont {Head-Gordon}(2008)}]{Chai2008}%
  \BibitemOpen
  \bibfield  {author} {\bibinfo {author} {\bibfnamefont {J.-D.}\ \bibnamefont {Chai}}\ and\ \bibinfo {author} {\bibfnamefont {M.}~\bibnamefont {Head-Gordon}},\ }\href {\doibase 10.1063/1.2834918} {\bibfield  {journal} {\bibinfo  {journal} {J. Chem. Phys.}\ }\textbf {\bibinfo {volume} {128}},\ \bibinfo {pages} {084106} (\bibinfo {year} {2008})}\BibitemShut {NoStop}%
\bibitem [{\citenamefont {Adamo}\ and\ \citenamefont {Barone}(1999)}]{Adamo1999}%
  \BibitemOpen
  \bibfield  {author} {\bibinfo {author} {\bibfnamefont {C.}~\bibnamefont {Adamo}}\ and\ \bibinfo {author} {\bibfnamefont {V.}~\bibnamefont {Barone}},\ }\href {\doibase 10.1063/1.478522} {\bibfield  {journal} {\bibinfo  {journal} {J. Chem. Phys.}\ }\textbf {\bibinfo {volume} {110}},\ \bibinfo {pages} {6158} (\bibinfo {year} {1999})}\BibitemShut {NoStop}%
\bibitem [{\citenamefont {Becke}(1993)}]{Becke1993}%
  \BibitemOpen
  \bibfield  {author} {\bibinfo {author} {\bibfnamefont {A.~D.}\ \bibnamefont {Becke}},\ }\href@noop {} {\bibfield  {journal} {\bibinfo  {journal} {J. Chem. Phys.}\ }\textbf {\bibinfo {volume} {98}},\ \bibinfo {pages} {5648} (\bibinfo {year} {1993})}\BibitemShut {NoStop}%
\bibitem [{\citenamefont {Stephens}\ \emph {et~al.}(1994)\citenamefont {Stephens}, \citenamefont {Devlin}, \citenamefont {Chabalowski},\ and\ \citenamefont {Frisch}}]{Stephens1994}%
  \BibitemOpen
  \bibfield  {author} {\bibinfo {author} {\bibfnamefont {P.~J.}\ \bibnamefont {Stephens}}, \bibinfo {author} {\bibfnamefont {F.~J.}\ \bibnamefont {Devlin}}, \bibinfo {author} {\bibfnamefont {C.~F.}\ \bibnamefont {Chabalowski}}, \ and\ \bibinfo {author} {\bibfnamefont {M.~J.}\ \bibnamefont {Frisch}},\ }\href@noop {} {\bibfield  {journal} {\bibinfo  {journal} {J. Phys. Chem.}\ }\textbf {\bibinfo {volume} {98}},\ \bibinfo {pages} {11623} (\bibinfo {year} {1994})}\BibitemShut {NoStop}%
\bibitem [{\citenamefont {Neese}\ \emph {et~al.}(2020)\citenamefont {Neese}, \citenamefont {Wennmohs}, \citenamefont {Becker},\ and\ \citenamefont {Riplinger}}]{Neese2020}%
  \BibitemOpen
  \bibfield  {author} {\bibinfo {author} {\bibfnamefont {F.}~\bibnamefont {Neese}}, \bibinfo {author} {\bibfnamefont {F.}~\bibnamefont {Wennmohs}}, \bibinfo {author} {\bibfnamefont {U.}~\bibnamefont {Becker}}, \ and\ \bibinfo {author} {\bibfnamefont {C.}~\bibnamefont {Riplinger}},\ }\href@noop {} {\bibfield  {journal} {\bibinfo  {journal} {J. Chem. Phys.}\ }\textbf {\bibinfo {volume} {152}} (\bibinfo {year} {2020})}\BibitemShut {NoStop}%
\bibitem [{\citenamefont {Sun}\ \emph {et~al.}(2020)\citenamefont {Sun}, \citenamefont {Zhang}, \citenamefont {Banerjee}, \citenamefont {Bao}, \citenamefont {Barbry}, \citenamefont {Blunt}, \citenamefont {Bogdanov}, \citenamefont {Booth}, \citenamefont {Chen}, \citenamefont {Cui}, \citenamefont {Eriksen}, \citenamefont {Gao}, \citenamefont {Guo}, \citenamefont {Hermann}, \citenamefont {Hermes}, \citenamefont {Koh}, \citenamefont {Koval}, \citenamefont {Lehtola}, \citenamefont {Li}, \citenamefont {Liu}, \citenamefont {Mardirossian}, \citenamefont {McClain}, \citenamefont {Motta}, \citenamefont {Mussard}, \citenamefont {Pham}, \citenamefont {Pulkin}, \citenamefont {Purwanto}, \citenamefont {Robinson}, \citenamefont {Ronca}, \citenamefont {Sayfutyarova}, \citenamefont {Scheurer}, \citenamefont {Schurkus}, \citenamefont {Smith}, \citenamefont {Sun}, \citenamefont {Sun}, \citenamefont {Upadhyay}, \citenamefont {Wagner}, \citenamefont {Wang}, \citenamefont {White}, \citenamefont {Whitfield}, \citenamefont
  {Williamson}, \citenamefont {Wouters}, \citenamefont {Yang}, \citenamefont {Yu}, \citenamefont {Zhu}, \citenamefont {Berkelbach}, \citenamefont {Sharma}, \citenamefont {Sokolov},\ and\ \citenamefont {Chan}}]{Sun2020}%
  \BibitemOpen
  \bibfield  {author} {\bibinfo {author} {\bibfnamefont {Q.}~\bibnamefont {Sun}}, \bibinfo {author} {\bibfnamefont {X.}~\bibnamefont {Zhang}}, \bibinfo {author} {\bibfnamefont {S.}~\bibnamefont {Banerjee}}, \bibinfo {author} {\bibfnamefont {P.}~\bibnamefont {Bao}}, \bibinfo {author} {\bibfnamefont {M.}~\bibnamefont {Barbry}}, \bibinfo {author} {\bibfnamefont {N.~S.}\ \bibnamefont {Blunt}}, \bibinfo {author} {\bibfnamefont {N.~A.}\ \bibnamefont {Bogdanov}}, \bibinfo {author} {\bibfnamefont {G.~H.}\ \bibnamefont {Booth}}, \bibinfo {author} {\bibfnamefont {J.}~\bibnamefont {Chen}}, \bibinfo {author} {\bibfnamefont {Z.-H.}\ \bibnamefont {Cui}}, \bibinfo {author} {\bibfnamefont {J.~J.}\ \bibnamefont {Eriksen}}, \bibinfo {author} {\bibfnamefont {Y.}~\bibnamefont {Gao}}, \bibinfo {author} {\bibfnamefont {S.}~\bibnamefont {Guo}}, \bibinfo {author} {\bibfnamefont {J.}~\bibnamefont {Hermann}}, \bibinfo {author} {\bibfnamefont {M.~R.}\ \bibnamefont {Hermes}}, \bibinfo {author} {\bibfnamefont {K.}~\bibnamefont {Koh}},
  \bibinfo {author} {\bibfnamefont {P.}~\bibnamefont {Koval}}, \bibinfo {author} {\bibfnamefont {S.}~\bibnamefont {Lehtola}}, \bibinfo {author} {\bibfnamefont {Z.}~\bibnamefont {Li}}, \bibinfo {author} {\bibfnamefont {J.}~\bibnamefont {Liu}}, \bibinfo {author} {\bibfnamefont {N.}~\bibnamefont {Mardirossian}}, \bibinfo {author} {\bibfnamefont {J.~D.}\ \bibnamefont {McClain}}, \bibinfo {author} {\bibfnamefont {M.}~\bibnamefont {Motta}}, \bibinfo {author} {\bibfnamefont {B.}~\bibnamefont {Mussard}}, \bibinfo {author} {\bibfnamefont {H.~Q.}\ \bibnamefont {Pham}}, \bibinfo {author} {\bibfnamefont {A.}~\bibnamefont {Pulkin}}, \bibinfo {author} {\bibfnamefont {W.}~\bibnamefont {Purwanto}}, \bibinfo {author} {\bibfnamefont {P.~J.}\ \bibnamefont {Robinson}}, \bibinfo {author} {\bibfnamefont {E.}~\bibnamefont {Ronca}}, \bibinfo {author} {\bibfnamefont {E.~R.}\ \bibnamefont {Sayfutyarova}}, \bibinfo {author} {\bibfnamefont {M.}~\bibnamefont {Scheurer}}, \bibinfo {author} {\bibfnamefont {H.~F.}\ \bibnamefont {Schurkus}},
  \bibinfo {author} {\bibfnamefont {J.~E.~T.}\ \bibnamefont {Smith}}, \bibinfo {author} {\bibfnamefont {C.}~\bibnamefont {Sun}}, \bibinfo {author} {\bibfnamefont {S.-N.}\ \bibnamefont {Sun}}, \bibinfo {author} {\bibfnamefont {S.}~\bibnamefont {Upadhyay}}, \bibinfo {author} {\bibfnamefont {L.~K.}\ \bibnamefont {Wagner}}, \bibinfo {author} {\bibfnamefont {X.}~\bibnamefont {Wang}}, \bibinfo {author} {\bibfnamefont {A.}~\bibnamefont {White}}, \bibinfo {author} {\bibfnamefont {J.~D.}\ \bibnamefont {Whitfield}}, \bibinfo {author} {\bibfnamefont {M.~J.}\ \bibnamefont {Williamson}}, \bibinfo {author} {\bibfnamefont {S.}~\bibnamefont {Wouters}}, \bibinfo {author} {\bibfnamefont {J.}~\bibnamefont {Yang}}, \bibinfo {author} {\bibfnamefont {J.~M.}\ \bibnamefont {Yu}}, \bibinfo {author} {\bibfnamefont {T.}~\bibnamefont {Zhu}}, \bibinfo {author} {\bibfnamefont {T.~C.}\ \bibnamefont {Berkelbach}}, \bibinfo {author} {\bibfnamefont {S.}~\bibnamefont {Sharma}}, \bibinfo {author} {\bibfnamefont {A.~Y.}\ \bibnamefont {Sokolov}},
  \ and\ \bibinfo {author} {\bibfnamefont {G.~K.-L.}\ \bibnamefont {Chan}},\ }\href@noop {} {\bibfield  {journal} {\bibinfo  {journal} {J. Chem. Phys.}\ }\textbf {\bibinfo {volume} {153}},\ \bibinfo {pages} {024109} (\bibinfo {year} {2020})}\BibitemShut {NoStop}%
\bibitem [{\citenamefont {Sun}\ \emph {et~al.}(2018)\citenamefont {Sun}, \citenamefont {Berkelbach}, \citenamefont {Blunt}, \citenamefont {Booth}, \citenamefont {Guo}, \citenamefont {Li}, \citenamefont {Liu}, \citenamefont {McClain}, \citenamefont {Sayfutyarova}, \citenamefont {Sharma}, \citenamefont {Wouters},\ and\ \citenamefont {Chan}}]{Sun2018}%
  \BibitemOpen
  \bibfield  {author} {\bibinfo {author} {\bibfnamefont {Q.}~\bibnamefont {Sun}}, \bibinfo {author} {\bibfnamefont {T.~C.}\ \bibnamefont {Berkelbach}}, \bibinfo {author} {\bibfnamefont {N.~S.}\ \bibnamefont {Blunt}}, \bibinfo {author} {\bibfnamefont {G.~H.}\ \bibnamefont {Booth}}, \bibinfo {author} {\bibfnamefont {S.}~\bibnamefont {Guo}}, \bibinfo {author} {\bibfnamefont {Z.}~\bibnamefont {Li}}, \bibinfo {author} {\bibfnamefont {J.}~\bibnamefont {Liu}}, \bibinfo {author} {\bibfnamefont {J.~D.}\ \bibnamefont {McClain}}, \bibinfo {author} {\bibfnamefont {E.~R.}\ \bibnamefont {Sayfutyarova}}, \bibinfo {author} {\bibfnamefont {S.}~\bibnamefont {Sharma}}, \bibinfo {author} {\bibfnamefont {S.}~\bibnamefont {Wouters}}, \ and\ \bibinfo {author} {\bibfnamefont {G.~K.-L.}\ \bibnamefont {Chan}},\ }\href@noop {} {\bibfield  {journal} {\bibinfo  {journal} {WIREs Comput. Mol. Sci.}\ }\textbf {\bibinfo {volume} {8}},\ \bibinfo {pages} {e1340} (\bibinfo {year} {2018})}\BibitemShut {NoStop}%
\bibitem [{\citenamefont {Margraf}\ \emph {et~al.}(2017)\citenamefont {Margraf}, \citenamefont {Perera}, \citenamefont {Lutz},\ and\ \citenamefont {Bartlett}}]{Margraf2017}%
  \BibitemOpen
  \bibfield  {author} {\bibinfo {author} {\bibfnamefont {J.~T.}\ \bibnamefont {Margraf}}, \bibinfo {author} {\bibfnamefont {A.}~\bibnamefont {Perera}}, \bibinfo {author} {\bibfnamefont {J.~J.}\ \bibnamefont {Lutz}}, \ and\ \bibinfo {author} {\bibfnamefont {R.~J.}\ \bibnamefont {Bartlett}},\ }\href {\doibase 10.1063/1.5003128} {\bibfield  {journal} {\bibinfo  {journal} {J. Chem. Phys.}\ }\textbf {\bibinfo {volume} {147}},\ \bibinfo {pages} {184101} (\bibinfo {year} {2017})}\BibitemShut {NoStop}%
\bibitem [{\citenamefont {Schlegel}(2002)}]{Schlegel2002}%
  \BibitemOpen
  \bibfield  {author} {\bibinfo {author} {\bibfnamefont {H.~B.}\ \bibnamefont {Schlegel}},\ }\href@noop {} {\bibfield  {journal} {\bibinfo  {journal} {Encyclopedia of Computational Chemistry}\ } (\bibinfo {year} {2002})}\BibitemShut {NoStop}%
\bibitem [{\citenamefont {Coulson}\ and\ \citenamefont {Fischer}(1949)}]{Coulson1949}%
  \BibitemOpen
  \bibfield  {author} {\bibinfo {author} {\bibfnamefont {C.}~\bibnamefont {Coulson}}\ and\ \bibinfo {author} {\bibfnamefont {I.}~\bibnamefont {Fischer}},\ }\href@noop {} {\bibfield  {journal} {\bibinfo  {journal} {Lond. Edinb. Dubl. Phil. Mag.}\ }\textbf {\bibinfo {volume} {40}},\ \bibinfo {pages} {386} (\bibinfo {year} {1949})}\BibitemShut {NoStop}%
\bibitem [{\citenamefont {Grizzi}\ \emph {et~al.}()\citenamefont {Grizzi}, \citenamefont {Fedik}, \citenamefont {Anaya}, \citenamefont {Gomes}, \citenamefont {Li}, \citenamefont {Barros}, \citenamefont {Lubbers}, \citenamefont {Messerly}, \citenamefont {Smith}, \citenamefont {Nebgen},\ and\ \citenamefont {Tretiak}}]{ALF}%
  \BibitemOpen
  \bibfield  {author} {\bibinfo {author} {\bibfnamefont {V.~F.}\ \bibnamefont {Grizzi}}, \bibinfo {author} {\bibfnamefont {N.}~\bibnamefont {Fedik}}, \bibinfo {author} {\bibfnamefont {S.}~\bibnamefont {Anaya}}, \bibinfo {author} {\bibfnamefont {J.}~\bibnamefont {Gomes}}, \bibinfo {author} {\bibfnamefont {Y.~W.}\ \bibnamefont {Li}}, \bibinfo {author} {\bibfnamefont {K.}~\bibnamefont {Barros}}, \bibinfo {author} {\bibfnamefont {N.}~\bibnamefont {Lubbers}}, \bibinfo {author} {\bibfnamefont {R.}~\bibnamefont {Messerly}}, \bibinfo {author} {\bibfnamefont {J.}~\bibnamefont {Smith}}, \bibinfo {author} {\bibfnamefont {B.}~\bibnamefont {Nebgen}}, \ and\ \bibinfo {author} {\bibfnamefont {S.}~\bibnamefont {Tretiak}},\ }\href@noop {} {\enquote {\bibinfo {title} {{ALF}, an active learning framework for chemical data automation: open-source code release and accurate molten salts simulations},}\ }\BibitemShut {NoStop}%
\bibitem [{\citenamefont {Spiekermann}\ \emph {et~al.}(2022)\citenamefont {Spiekermann}, \citenamefont {Pattanaik},\ and\ \citenamefont {Green}}]{Spiekermann2022}%
  \BibitemOpen
  \bibfield  {author} {\bibinfo {author} {\bibfnamefont {K.}~\bibnamefont {Spiekermann}}, \bibinfo {author} {\bibfnamefont {L.}~\bibnamefont {Pattanaik}}, \ and\ \bibinfo {author} {\bibfnamefont {W.~H.}\ \bibnamefont {Green}},\ }\href@noop {} {\bibfield  {journal} {\bibinfo  {journal} {Scient. Data}\ }\textbf {\bibinfo {volume} {9}},\ \bibinfo {pages} {417} (\bibinfo {year} {2022})}\BibitemShut {NoStop}%
\bibitem [{\citenamefont {Chen}\ \emph {et~al.}(2018)\citenamefont {Chen}, \citenamefont {Yu},\ and\ \citenamefont {Houk}}]{Chen2018}%
  \BibitemOpen
  \bibfield  {author} {\bibinfo {author} {\bibfnamefont {S.}~\bibnamefont {Chen}}, \bibinfo {author} {\bibfnamefont {P.}~\bibnamefont {Yu}}, \ and\ \bibinfo {author} {\bibfnamefont {K.~N.}\ \bibnamefont {Houk}},\ }\href@noop {} {\bibfield  {journal} {\bibinfo  {journal} {J. Am. Chem. Soc.}\ }\textbf {\bibinfo {volume} {140}},\ \bibinfo {pages} {18124} (\bibinfo {year} {2018})}\BibitemShut {NoStop}%
\bibitem [{\citenamefont {Wardzala}\ \emph {et~al.}(2024)\citenamefont {Wardzala}, \citenamefont {King}, \citenamefont {Ogunfowora}, \citenamefont {Savoie},\ and\ \citenamefont {Gagliardi}}]{Wardzala2024}%
  \BibitemOpen
  \bibfield  {author} {\bibinfo {author} {\bibfnamefont {J.~J.}\ \bibnamefont {Wardzala}}, \bibinfo {author} {\bibfnamefont {D.~S.}\ \bibnamefont {King}}, \bibinfo {author} {\bibfnamefont {L.}~\bibnamefont {Ogunfowora}}, \bibinfo {author} {\bibfnamefont {B.}~\bibnamefont {Savoie}}, \ and\ \bibinfo {author} {\bibfnamefont {L.}~\bibnamefont {Gagliardi}},\ }\href {\doibase 10.1021/acscentsci.3c01559} {\bibfield  {journal} {\bibinfo  {journal} {ACS Central Science}\ }\textbf {\bibinfo {volume} {10}},\ \bibinfo {pages} {833} (\bibinfo {year} {2024})}\BibitemShut {NoStop}%
\bibitem [{\citenamefont {King}\ \emph {et~al.}(2022)\citenamefont {King}, \citenamefont {Hermes}, \citenamefont {Truhlar},\ and\ \citenamefont {Gagliardi}}]{King2022}%
  \BibitemOpen
  \bibfield  {author} {\bibinfo {author} {\bibfnamefont {D.~S.}\ \bibnamefont {King}}, \bibinfo {author} {\bibfnamefont {M.~R.}\ \bibnamefont {Hermes}}, \bibinfo {author} {\bibfnamefont {D.~G.}\ \bibnamefont {Truhlar}}, \ and\ \bibinfo {author} {\bibfnamefont {L.}~\bibnamefont {Gagliardi}},\ }\href {\doibase 10.1021/acs.jctc.2c00630} {\bibfield  {journal} {\bibinfo  {journal} {J. Chem. Theory Comput.}\ }\textbf {\bibinfo {volume} {18}},\ \bibinfo {pages} {6065} (\bibinfo {year} {2022})},\ \bibinfo {note} {pMID: 36112354}\BibitemShut {NoStop}%
\bibitem [{\citenamefont {Zhang}\ \emph {et~al.}(2024{\natexlab{b}})\citenamefont {Zhang}, \citenamefont {Li}, \citenamefont {Ye}, \citenamefont {Berkelbach},\ and\ \citenamefont {Chan}}]{Zhang2024}%
  \BibitemOpen
  \bibfield  {author} {\bibinfo {author} {\bibfnamefont {X.}~\bibnamefont {Zhang}}, \bibinfo {author} {\bibfnamefont {C.}~\bibnamefont {Li}}, \bibinfo {author} {\bibfnamefont {H.-Z.}\ \bibnamefont {Ye}}, \bibinfo {author} {\bibfnamefont {T.~C.}\ \bibnamefont {Berkelbach}}, \ and\ \bibinfo {author} {\bibfnamefont {G.~K.-L.}\ \bibnamefont {Chan}},\ }\href@noop {} {\bibfield  {journal} {\bibinfo  {journal} {J. Chem. Phys.}\ }\textbf {\bibinfo {volume} {161}},\ \bibinfo {pages} {014109} (\bibinfo {year} {2024}{\natexlab{b}})}\BibitemShut {NoStop}%
\bibitem [{\citenamefont {Herbold}\ and\ \citenamefont {Behler}(2023)}]{Herbold2023}%
  \BibitemOpen
  \bibfield  {author} {\bibinfo {author} {\bibfnamefont {M.}~\bibnamefont {Herbold}}\ and\ \bibinfo {author} {\bibfnamefont {J.}~\bibnamefont {Behler}},\ }\href@noop {} {\bibfield  {journal} {\bibinfo  {journal} {Phys. Chem. Chem. Phys.}\ }\textbf {\bibinfo {volume} {25}},\ \bibinfo {pages} {12979} (\bibinfo {year} {2023})}\BibitemShut {NoStop}%
\end{thebibliography}%

\end{document}

\maketitle
\tableofcontents

\section{Supplementary Methods}

\subsection{UCCSD Data}
Whilst building the workflow for the AL process, UCCSD calculations were performed as an interim step between the DFT and UCCSD(T) calculations. The UCCSD/DZ calculations used the PySCF software with the cc-pVDZ basis set~\cite{Sun2020,Sun2018}. 
%It was hypothesized that these structures could improve performance for the final UCCSD(T) model, however this was not found to be the case. The AL sampling techniques previously mentioned were used. 
For completeness, the UCCSD data is included.  

\subsection{Basis Set Corrections}

We began by investigating the accuracy of basis set corrections for forces at the unrestricted CCSD(T) level. The computational cost of obtaining UCCSD(T)/TZ and UCCD(T)/QZ  forces to benchmark against becomes extremely expensive for larger basis sets and memory requirements also become prohibitive. Given this, we begin by looking at the smallest molecules present in the dataset with just 6 atoms. Both reactants, products and transition states are included. 

\begin{figure*}[htb]
    \centering
    \includegraphics[width=6.85in]{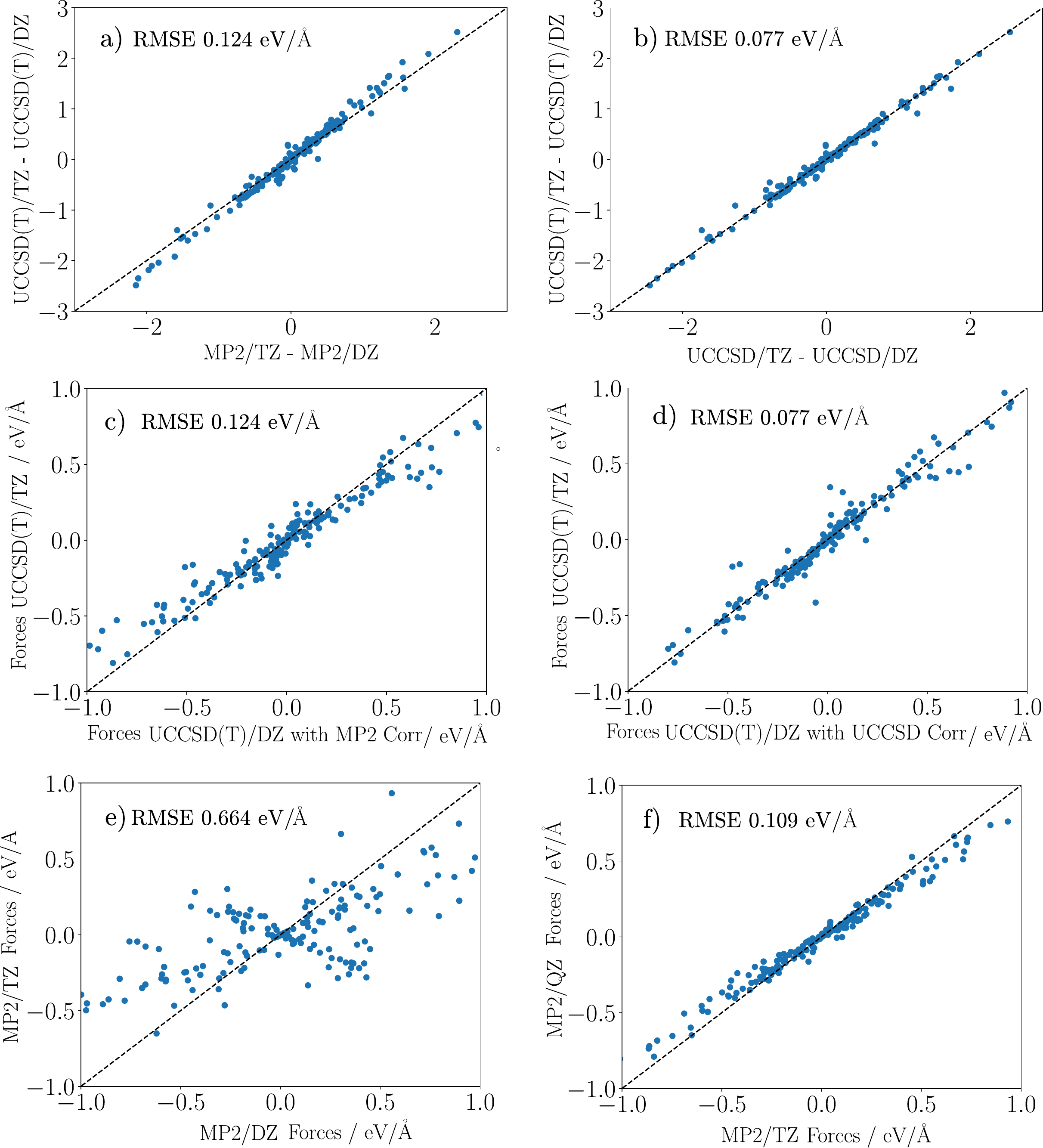}
    \caption{The difference between UCCSD(T) forces with a TZ and DZ basis set compared to the difference  between the TZ and DZ forces with a) MP2 and b) UCCSD respectively. The accuracy of basis set corrections to UCCSD(T)/DZ with c) MP2 and d) UCCSD compared to the UCCSD(T)/TZ forces is then shown. A comparison of the MP2 TZ forces with DZ is shown in part d) and MP2 QZ forces with TZ is show in part e). The dataset used contains both reactant, products and transition states.}
    \label{fig:Basis_set_forces}
\end{figure*}

The fundamental assumption that we make in the basis set constructed is that the difference in forces between two basis sets (ie. between QZ and TZ) is correlated between MP2, UCCSD and UCCSD(T). We first demonstrate this is true. The difference between the TZ and DZ basis set forces for MP2(UCCSD) compared to the difference for UCCSD(T) is shown in Fig.~\ref{fig:Basis_set_forces} a) (Fig.~\ref{fig:Basis_set_forces} b)). The differences between the basis sets are highly correlated between the different levels of theory.  A correction to UCCSD(T)/DZ using MP2/TZ and UCCSD/TZ forces is then shown in Fig.~\ref{fig:Basis_set_forces} c) and d). From this, we can see that using a UCCSD/TZ correction to the UCCSD(T)/DZ basis is approximately 50\% more accurate than an MP2 correction. Additionally, even for the larger forces present the UCCSD correction remains accurate, whereas for the MP2 correction larger deviation appear at larger forces. We can also see from Fig.~\ref{fig:Basis_set_forces} a) and b) that the basis set correction from DZ to TZ is quite large, with a maximum value of 2.31 and 2.55 for MP2 and UCCSD respectively - this is also shown for MP2 in Fig.~\ref{fig:Basis_set_forces} e). This shows the importance of basis set corrections.

\begin{figure}[htb]
    \centering
    \includegraphics[width=2.65in]{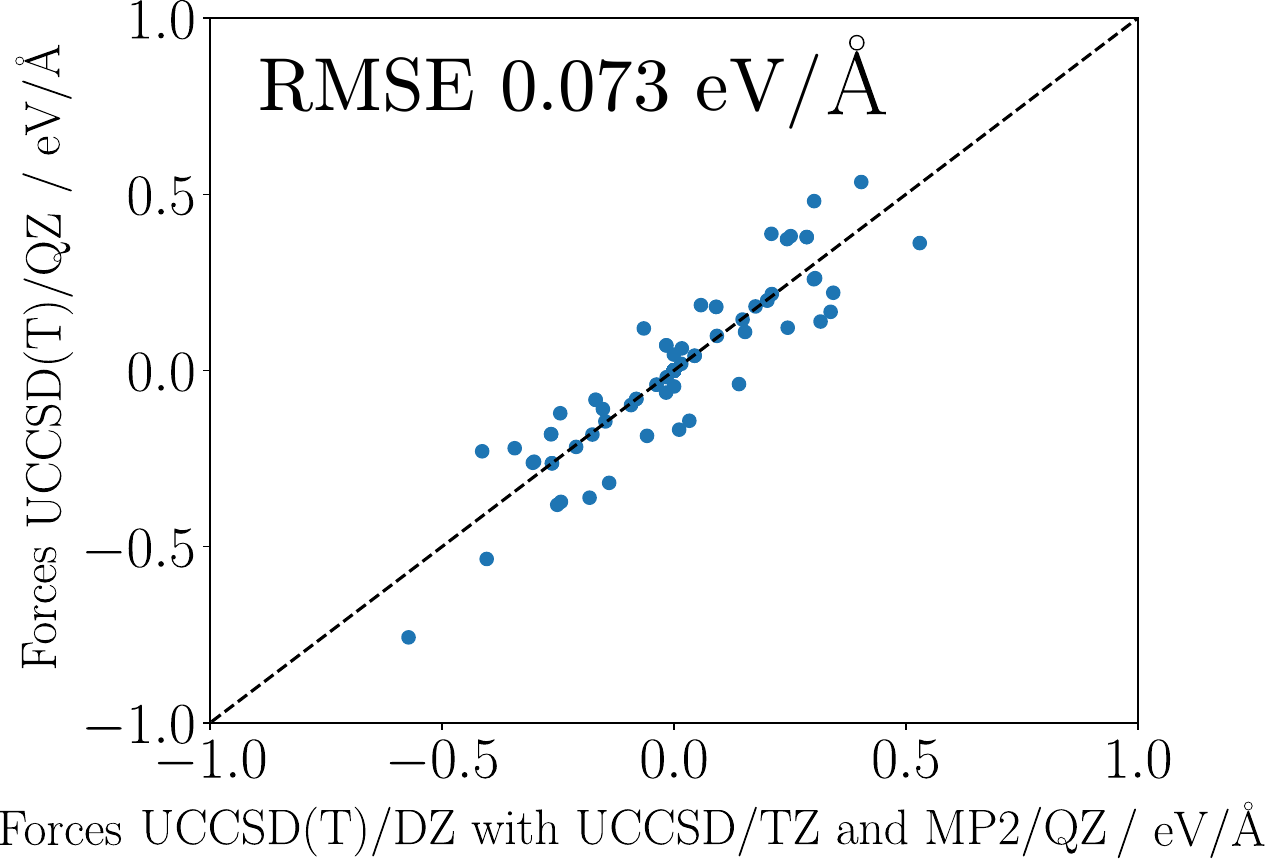}
    \caption{  The difference between UCCSD(T)/QZ forces and UCCSD(T)/QZ* forces. The UCCSD(T)/QZ* forces contain a basis set correction  with a UCCSD/TZ and MP2/QZ calculation performed.}
    \label{fig:Basis_set_forces_QZ}
\end{figure}

Even for six atom molecules, analytical UCCSD(T)/QZ forces cannot be readily computed with PySCF. However, the forces can be compared for MP2/QZ and MP2/TZ as shown in Fig.~\ref{fig:Basis_set_forces} d). This figure demonstrates that the MP2/QZ and MP2/TZ forces are much more highly correlated than MP2/TZ and MP2/DZ forces. 

To correct to the QZ basis set limit, we propose a new method and use the MP2/QZ forces in addition to the UCCSD/TZ correction. The forces are:
\\
$F_{UCCSD(T)/QZ*}=F_{UCCSD(T)/DZ} + (F_{UCCSD/TZ} - F_{UCCSD/DZ})  + (F_{UMP2/QZ} - F_{UMP2/TZ}) $.
\\
The accuracy of this correction was tested on a set of molecules containing configurations of $\mathrm{CO_2}$, $\mathrm{H_2O}$, $\mathrm{C_2H_2}$ and $\mathrm{H_2CO}$ generated using 300K MD. The results are shown in Fig.~\ref{fig:Basis_set_forces_QZ} with the proposed correction of UCCSD and MP2 offering a higher accuracy than correcting just to the TZ level with UCSSD or using MP2 alone to correct the force to the QZ basis set - see Fig.~\ref{fig:Basis_set_forces_QZ}. The corrected basis set is described as UCCSD(T)/QZ* throughout this work. The UCCSD calculations are corrected from the DZ basis set using MP2 alone and is described as UCCSD/QZ*.

\iffalse
\begin{figure*}[htb]
    \centering
    \includegraphics[width=6.85in]{./SI_Graphs/Comparison_corrections.pdf}
    \caption{  A comparison between UCCSD(T) with a QZ basis set and corrections using a) only UCCSD/TZ, b) UMP2/QZ and c) UCCSD/TZ and UMP2/QZ corrections together.   }
    \label{fig:Basis_set_qz}
\end{figure*}
\fi

\begin{figure*}[htb]
    \centering
    \includegraphics[width=3.0in]{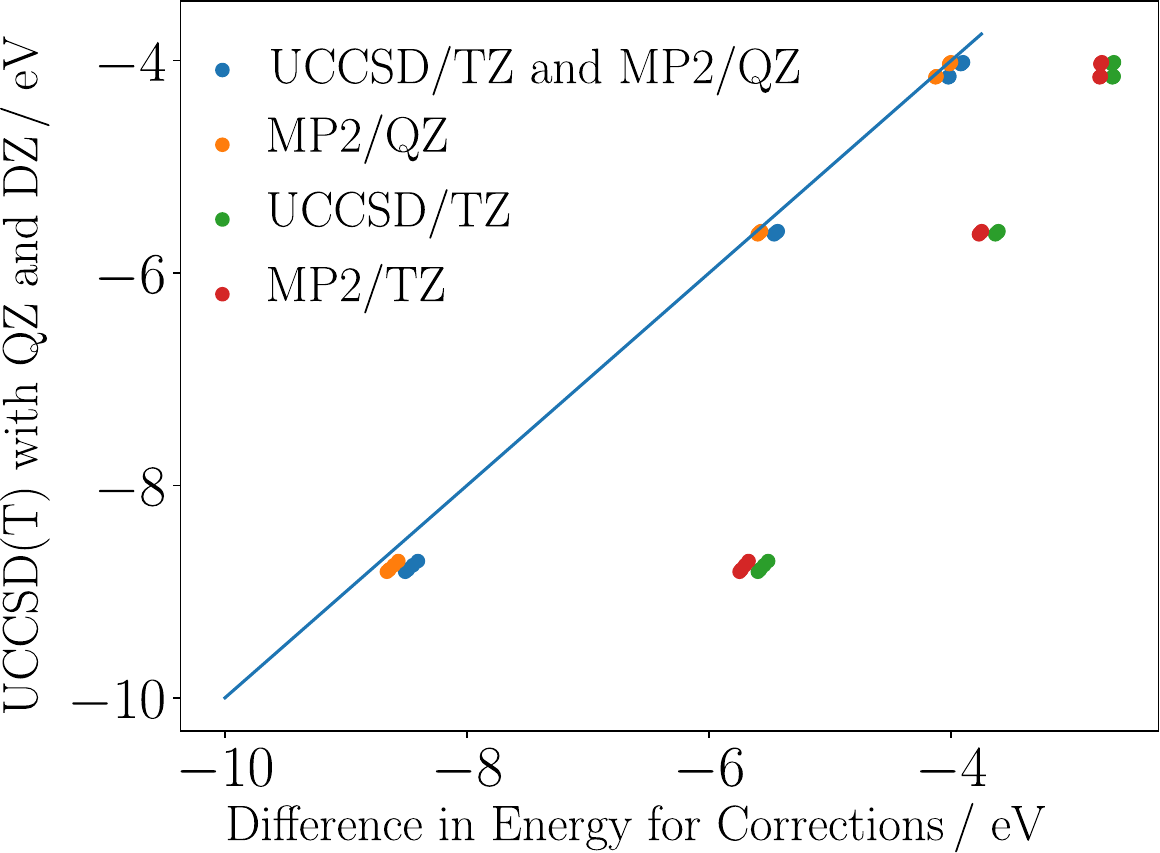}
    \caption{  The difference in energy between a QZ and DZ basis set at the UCCSD(T) level is compared to the difference in energy with various approximations. The UMP2(UCCSD)/TZ data points correspond to the difference between UMP2(UCCSD)/TZ and UMP2(UCCSD)/DZ with the UCCSD(T) data. The UMP2/QZ data points correspond to the difference between UMP2/QZ and UMP2/DZ with the UCCSD(T) data. The UCCSD/TZ and UMP2/QZ data points correspond to the difference between (UCCSD/TZ - UCCSD/DZ + UMP2/QZ - UMP2/TZ) with the UCCSD(T) data.      }
    \label{fig:Basis_set_energy}
\end{figure*}

\clearpage
\subsection{Procedure for Identifying Possibly Inaccurate Structures}
\subsubsection{Transition States, Reactants and Products}

As discussed, cases were found where extremely large forces were present for the calculation to be performed. To overcome this, a procedure was developed to identify structures whose forces may be particularly large. Alongside the unrestricted calculations, if the spin state was non-zero then a restricted calculation was also performed. If the restricted calculation was lower energy than the unrestricted calculation, or the difference between the two was less than 0.1eV, the structure was identified as possibly being near an intersection point of spin states. For the transition states, reactants and products, if the difference in forces was greater than 1eV/\AA, these points were then removed. Additionally, for all structures if the force basis set correction was greater that 1eV/\AA~  for the MP2 component then the structure was also not included. For the dataset containing products, reactants and transition states, this resulted in 114 calculations being removed from the main dataset due to this procedure.  A comparison of the filtered and unfiltered forces for this dataset are shown in Fig.~\ref{fig:U_R_forces_2}. There are a number of possible causes for the behaviour seen. For example, spin-contaminated solution from HF may distort the PES obtained from CCSD(T), and hence cause the atomic gradient to become erroneous.   

\subsubsection{The Dimer, NEB and Single-Ended String Structures}
The structures sampled along the reaction path are further from the intersection point and the coupled cluster calculations are less problematic. Therefore, for these cases, the structures were removed if the basis set correction for the forces contained a component greater than 5~eV/\AA~ for the UCCSD component and greater than 1~eV/\AA~ for the UMP2 component. 

\begin{figure}[htb]
    \centering
    \includegraphics[width=6.85in]{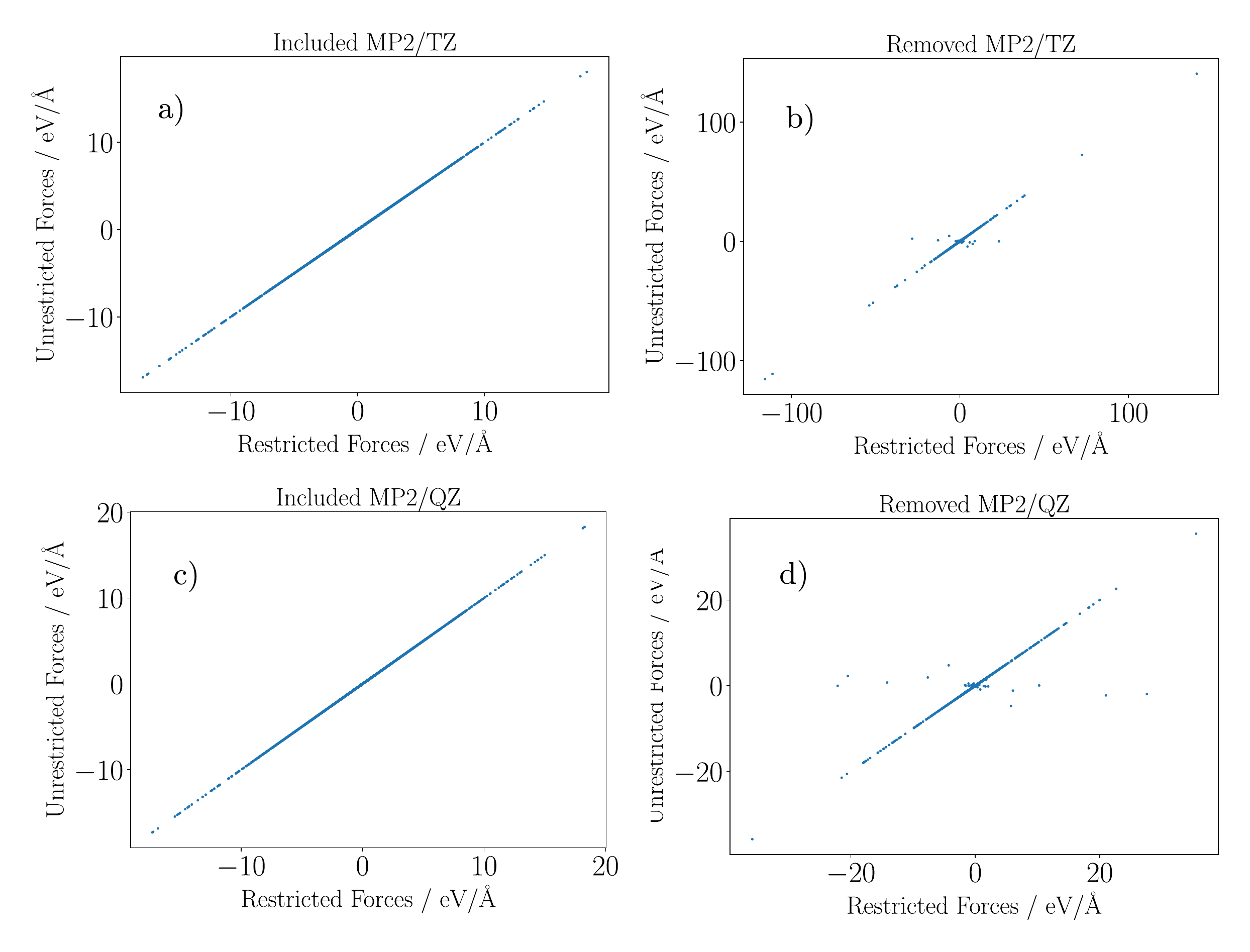}
    \caption*{  A comparison of the unrestricted and restricted forces for the included and removed structures for MP2 level of theory.  }
    \label{fig:U_R_forces_1}
\end{figure}

\begin{figure*}[htb]
    \centering
    \includegraphics[width=6.85in]{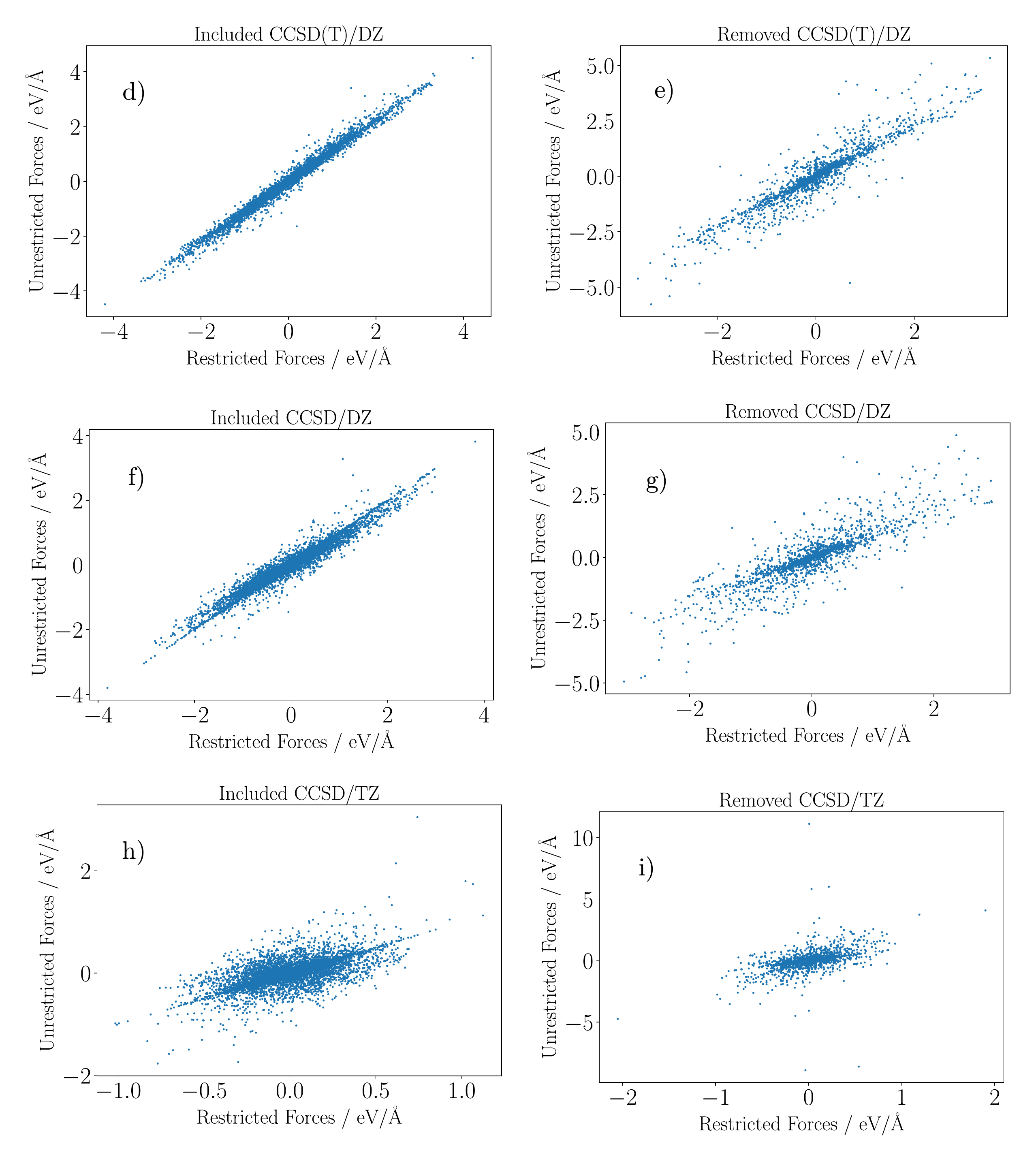}
    \caption{  A comparison of the unrestricted and restricted forces for the included and removed structures for CCSD and CCSD(T) level of theory.  }
    \label{fig:U_R_forces_2}
\end{figure*}

\clearpage

\subsection{Inconsistent Spin States for DZ/TZ/QZ}

We also identified that inconsistent spin states  for the DZ/TZ/QZ basis states found with the automated searching method resulted in incorrect calculations. To analyze the disparities present, we show a comparison of the ${\langle S^2 \rangle}$ values for the different basis sets in Fig.~\ref{fig:U_R_forces_2}. Structures with $|{\langle S^2 \rangle}_{QZ} - {\langle S^2 \rangle}_{TZ} | > 0.1 $ or $|{\langle S^2 \rangle}_{QZ} - {\langle S^2 \rangle}_{DZ} | > 0.1$  were removed.

\begin{figure*}[htb]
    \centering
    \includegraphics[width=6.65in]{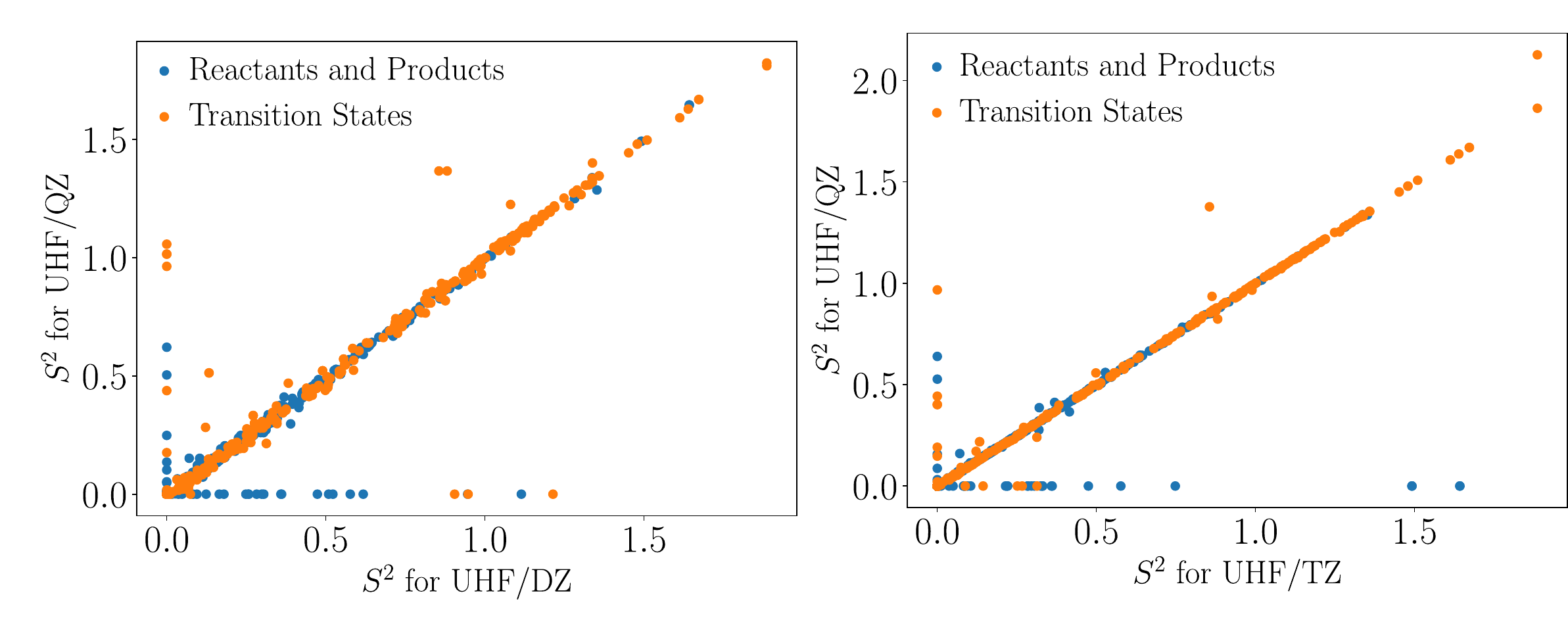}
    \caption{  A comparison between the UHF/QZ spin states with the UHF/TZ and UHF/DZ spin states for the reactants, products and transition states.}
    \label{fig:U_R_forces_2}
\end{figure*}

\clearpage
\section{Transition States, Products, Reactants}

Further analysis of the transition states, products and reactants is given in this section for various levels of theory. 

\subsection{DFT Datasets}

For the DFT datasets, additional analysis of the error distribution for the forces was performed. 

\begin{figure*}[htb]
    \centering
    \includegraphics[width=7.0in]{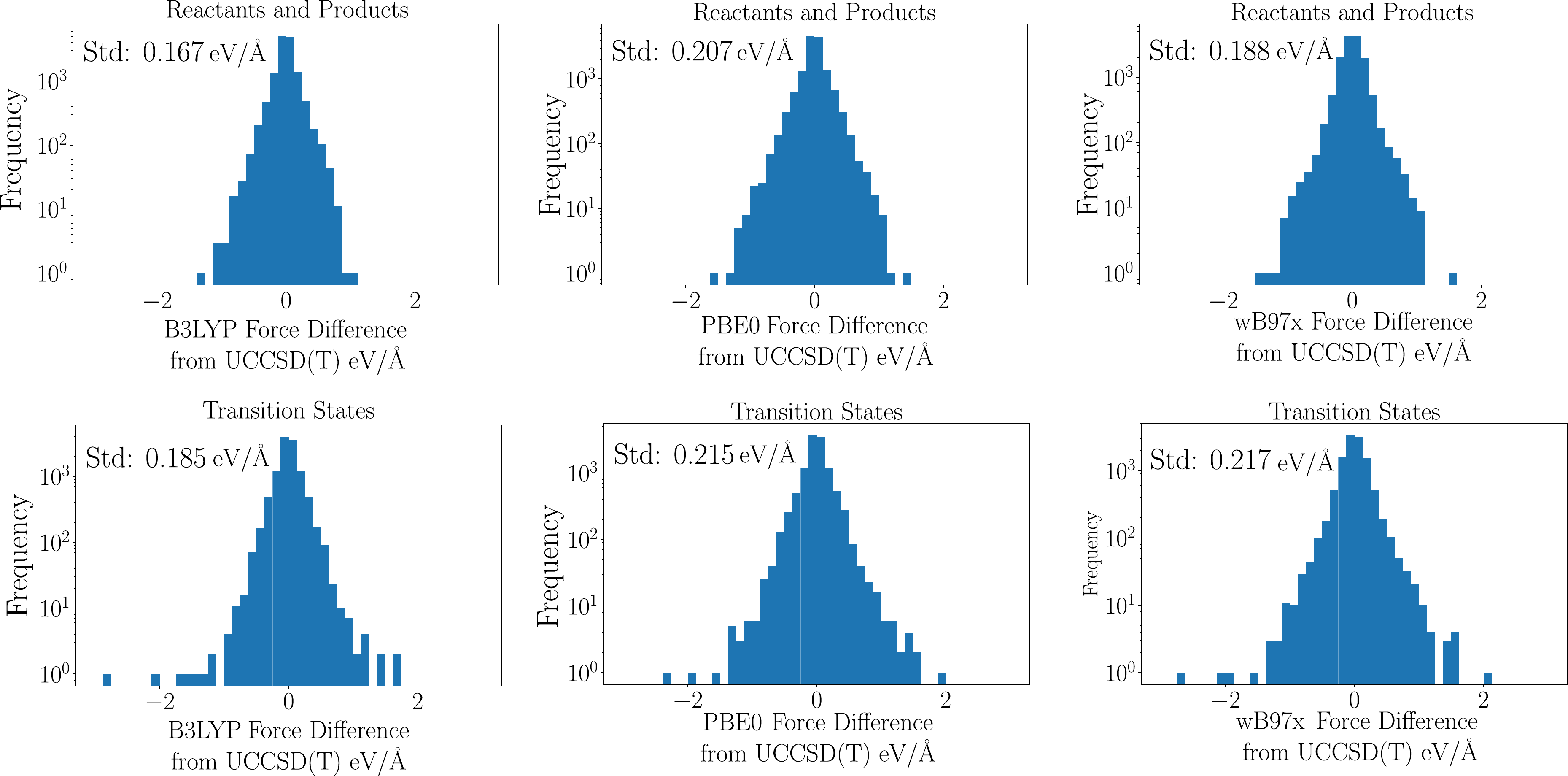}
    \caption{  The distributions of force differences between the unrestricted CCSD(T)/QZ* level of theory and various levels of DFT. The mean value of all the distributions is 0.00.  }
    \label{fig:UCCSDT_force}
\end{figure*}

\subsection{UCCSD Dataset}

\begin{figure*}[htb]
    \centering
    \includegraphics[width=3.10in]{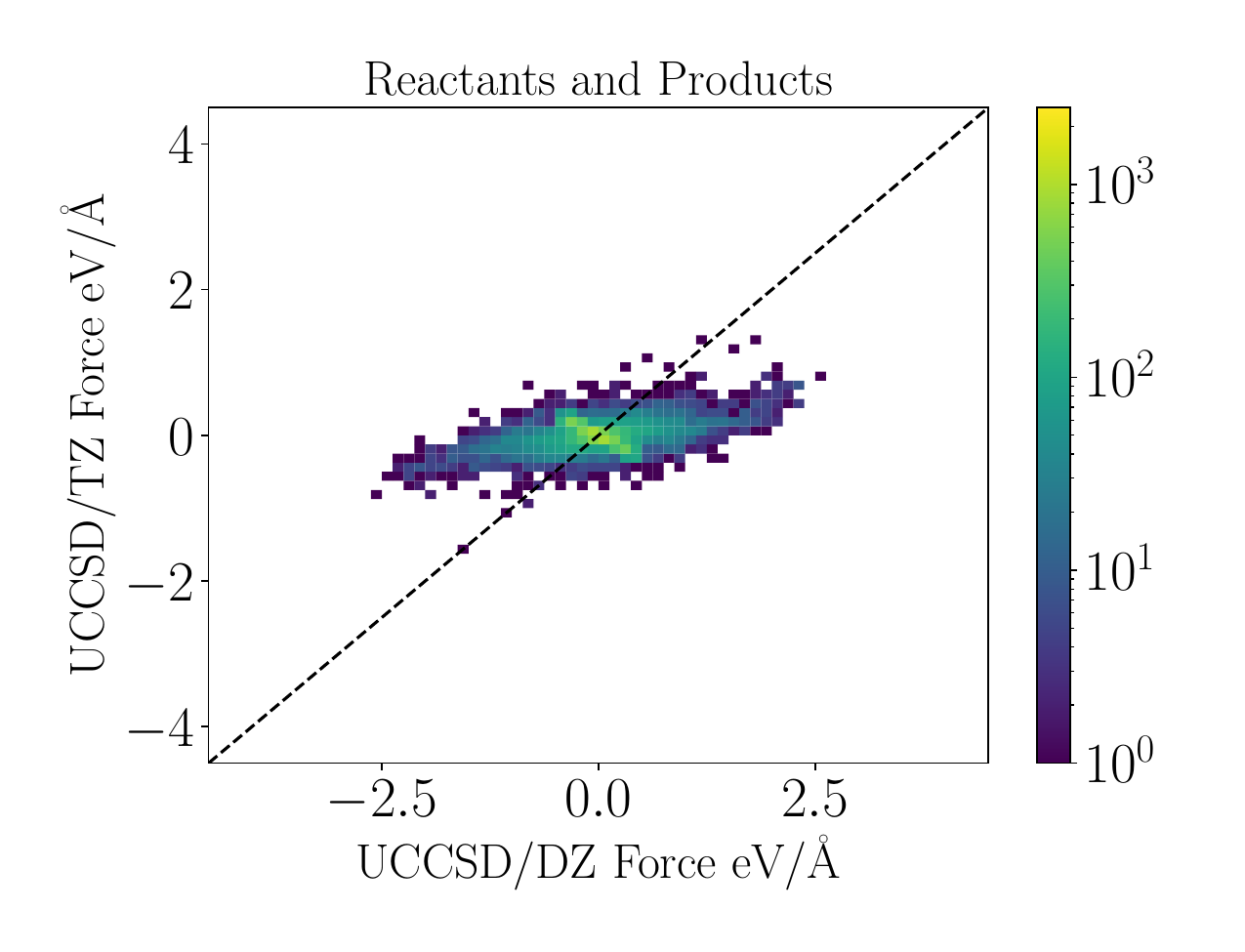}
    \includegraphics[width=3.10in]{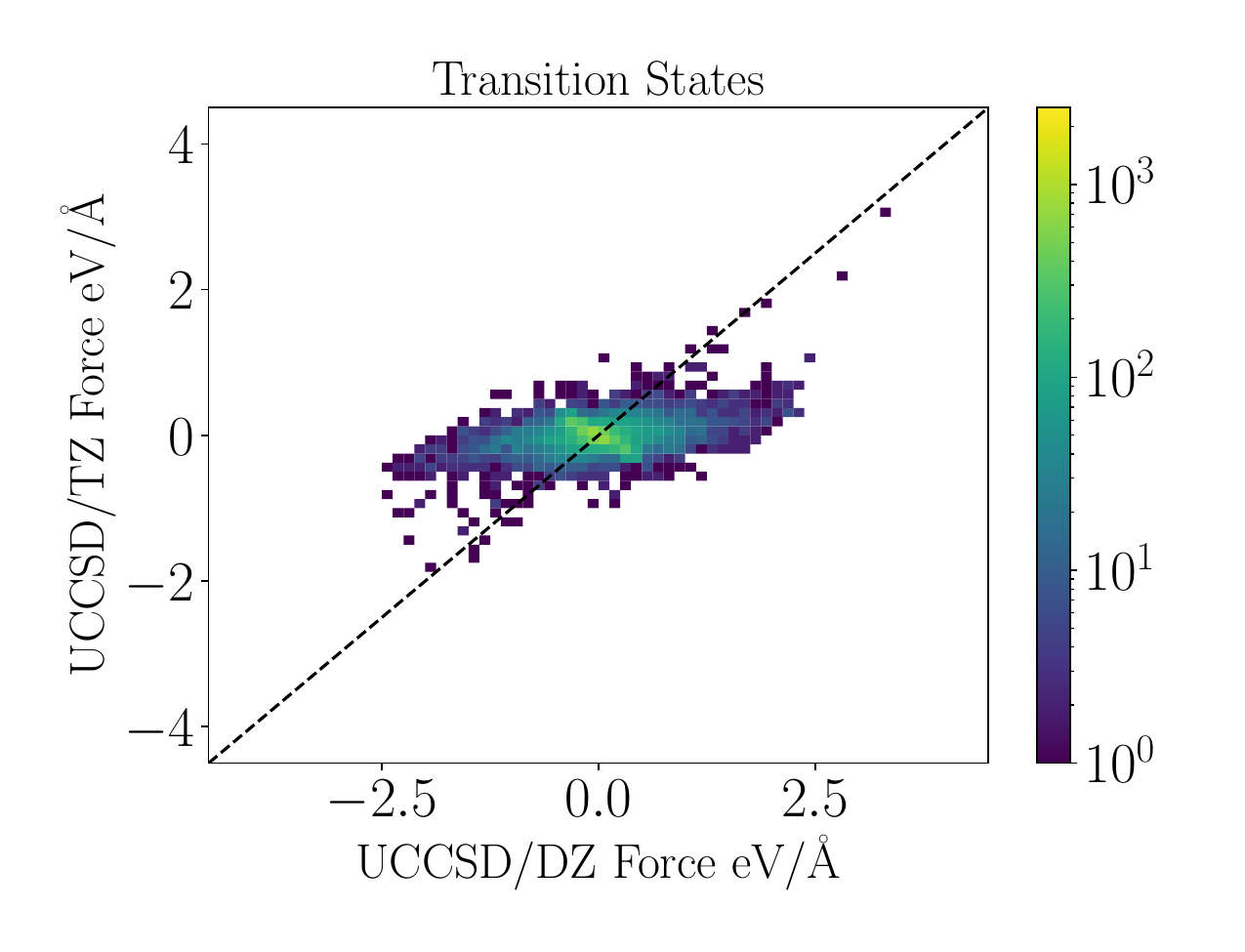}
    \caption{  A comparison between the unrestricted CCSD level of theory for DZ and TZ basis sets. }
    \label{fig:UCCSD_DZ_TZ}
\end{figure*}

An analysis of the forces for the UCCSD level of theory is shown. Large differences in the TZ and DZ forces are seen in Fig.~\ref{fig:UCCSD_DZ_TZ}. The advantages of using UCCSD(T) level of theory over UCCSD are seen in Fig.~\ref{fig:U_R_forces_2}.

\begin{figure*}[htb]
    \centering
    \includegraphics[width=6.85in]{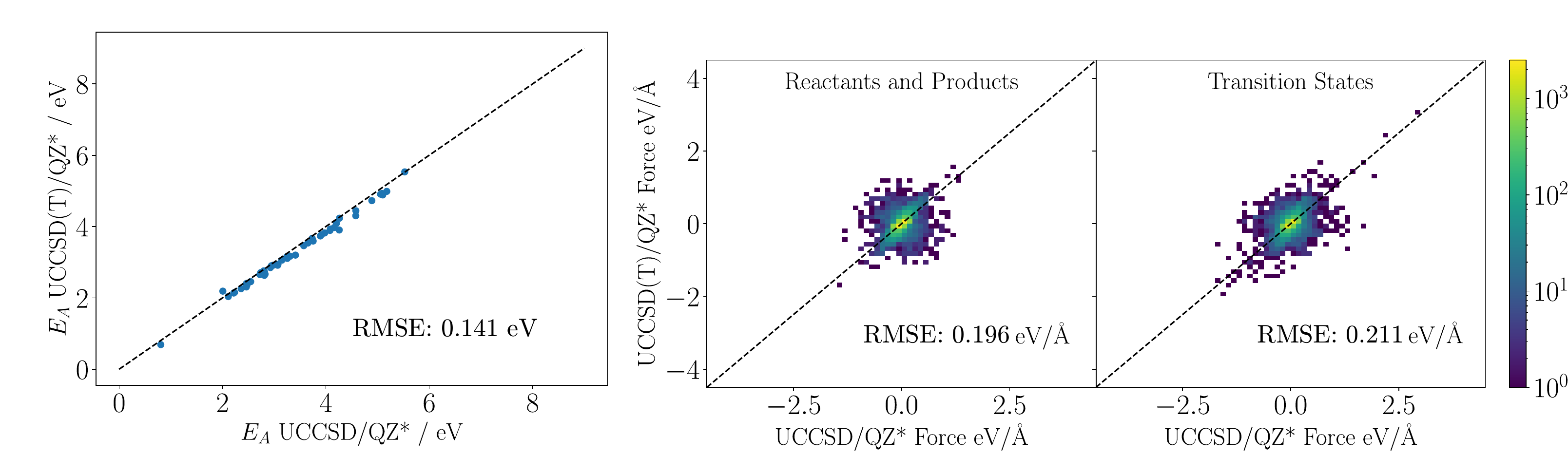}
    \caption{  A comparison of the unrestricted CCSD(T)/QZ* level of theory and unrestricted CCSD/QZ* for the $E_A$ and forces.  }
    \label{fig:U_R_forces_2}
\end{figure*}

\clearpage

\subsection{UMP2 Dataset}

A comparison of the forces for the UMP2 dataset with basis set is given in Fig.~\ref{fig:MP2_forces_BS}. The difference between the UMP2/QZ $E_A$ and forces and UCCSD(T)/QZ* $E_A$ and forces is shown in Fig.~\ref{fig:U_R_forces_2}.

\begin{figure*}[htb]
    \centering
    \includegraphics[width=2.10in]{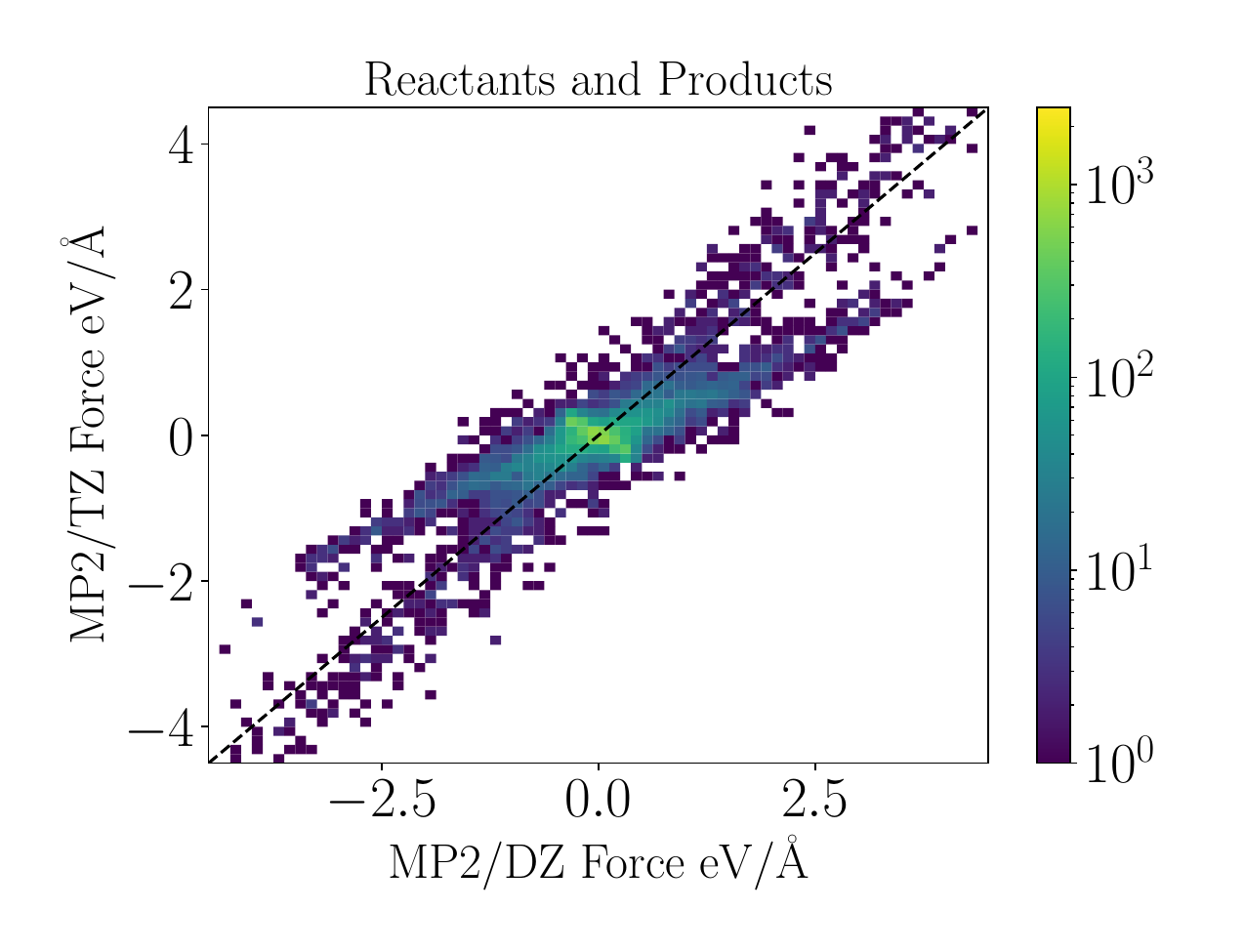}
    \includegraphics[width=2.10in]{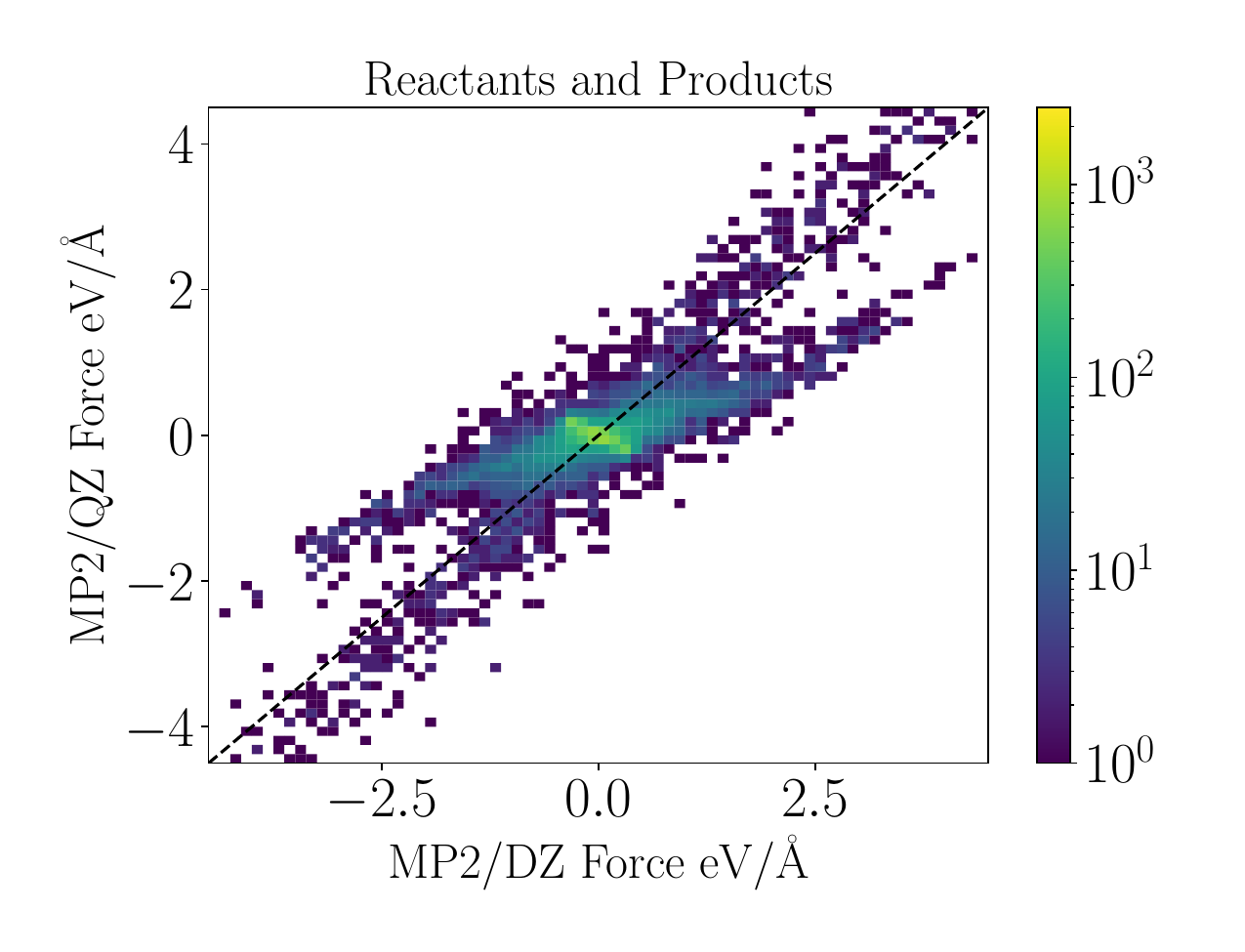}
    \includegraphics[width=2.10in]{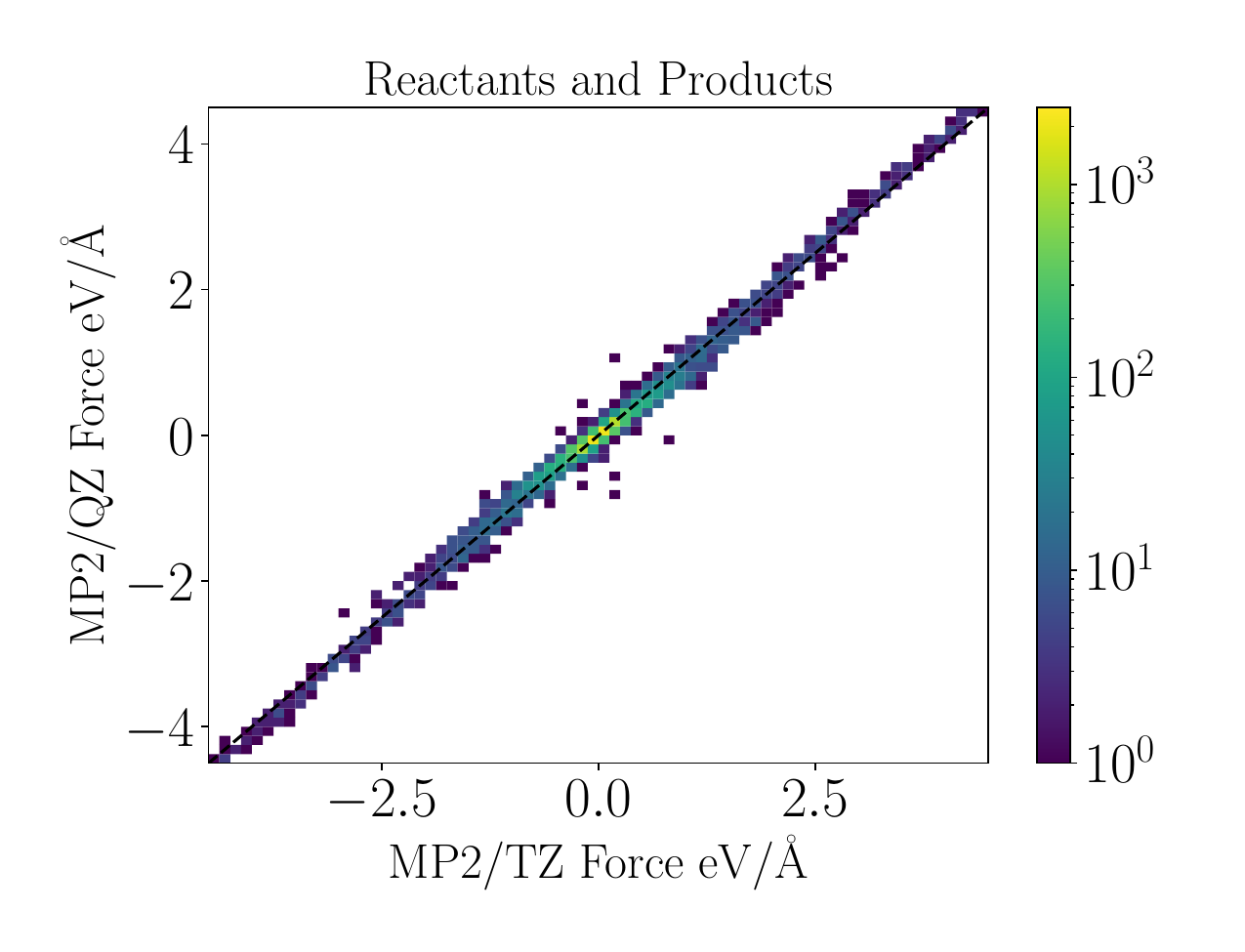}
    \includegraphics[width=2.10in]{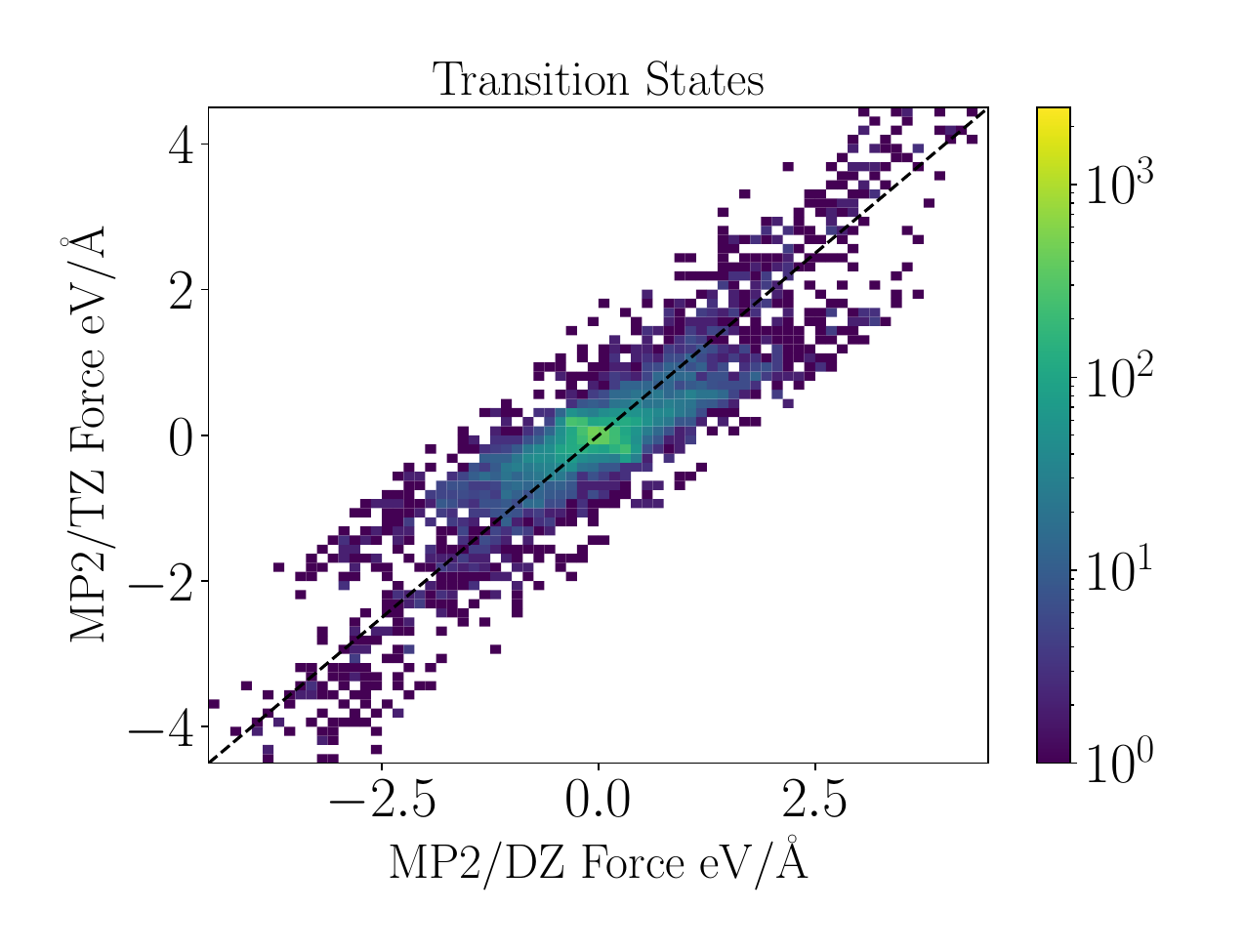}
    \includegraphics[width=2.10in]{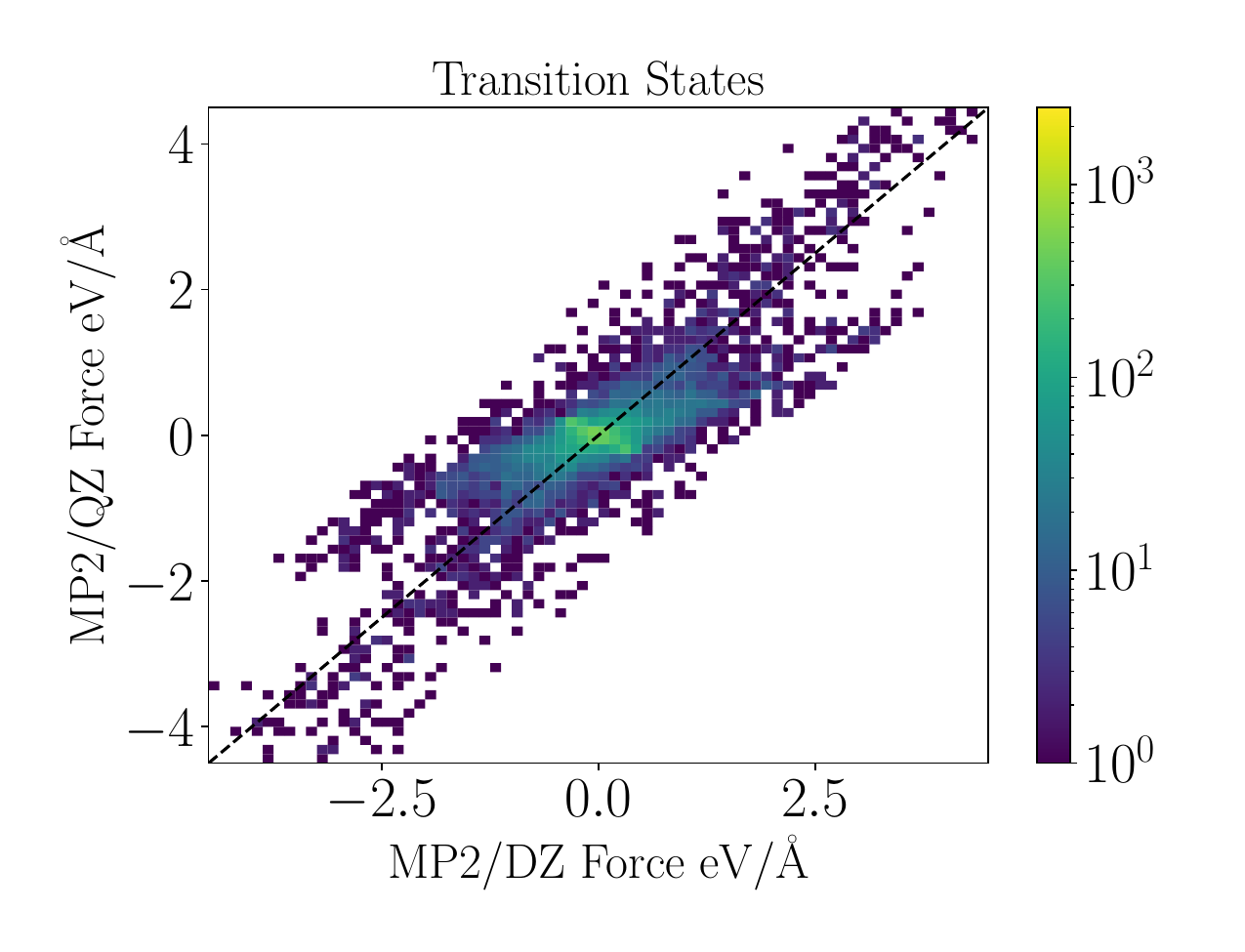}
    \includegraphics[width=2.10in]{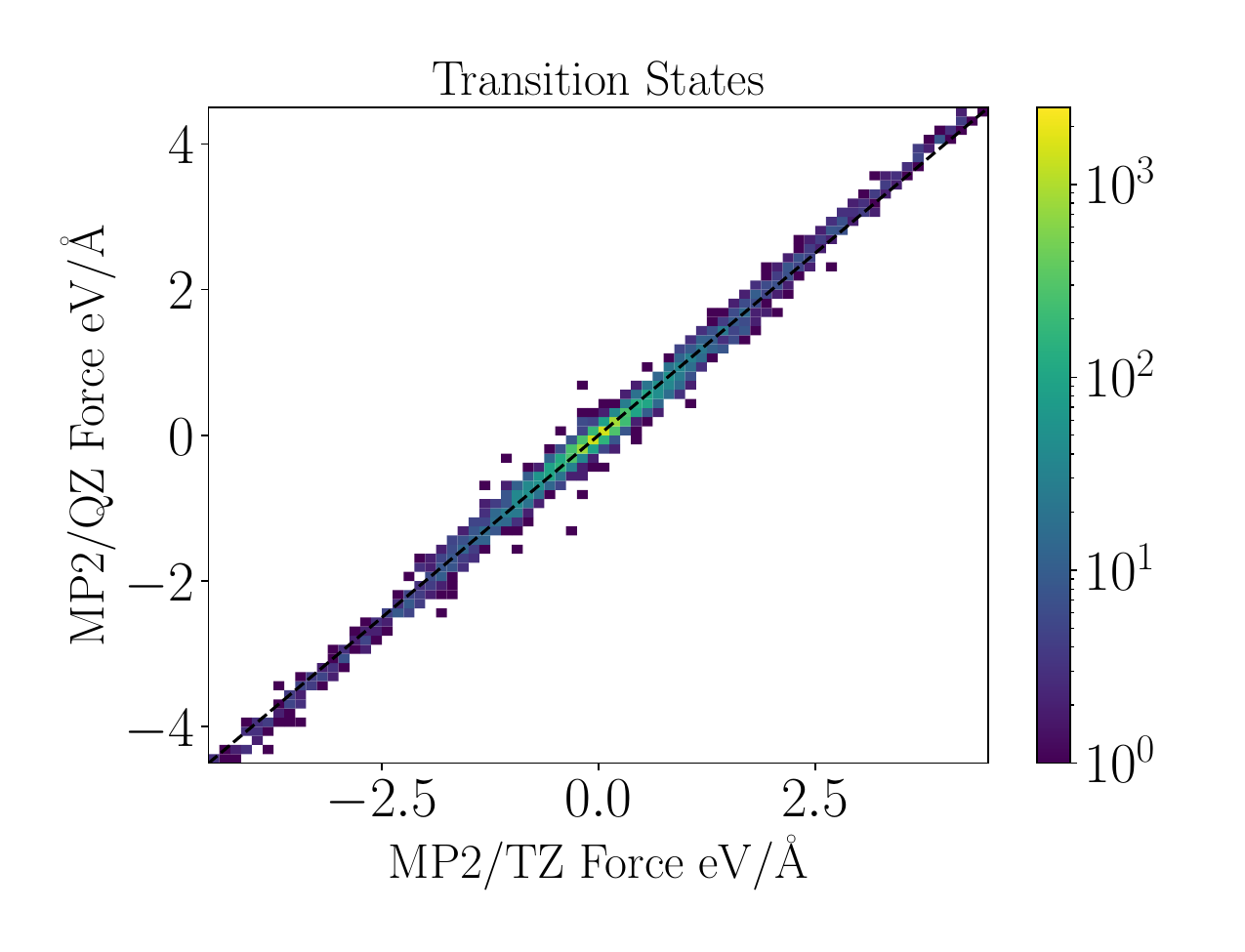}
    \caption{  A comparison between the unrestricted MP2 level of theory for DZ, TZ and QZ basis sets. }
    \label{fig:MP2_forces_BS}
\end{figure*}

\begin{figure*}[htb]
    \centering
    \includegraphics[width=6.85in]{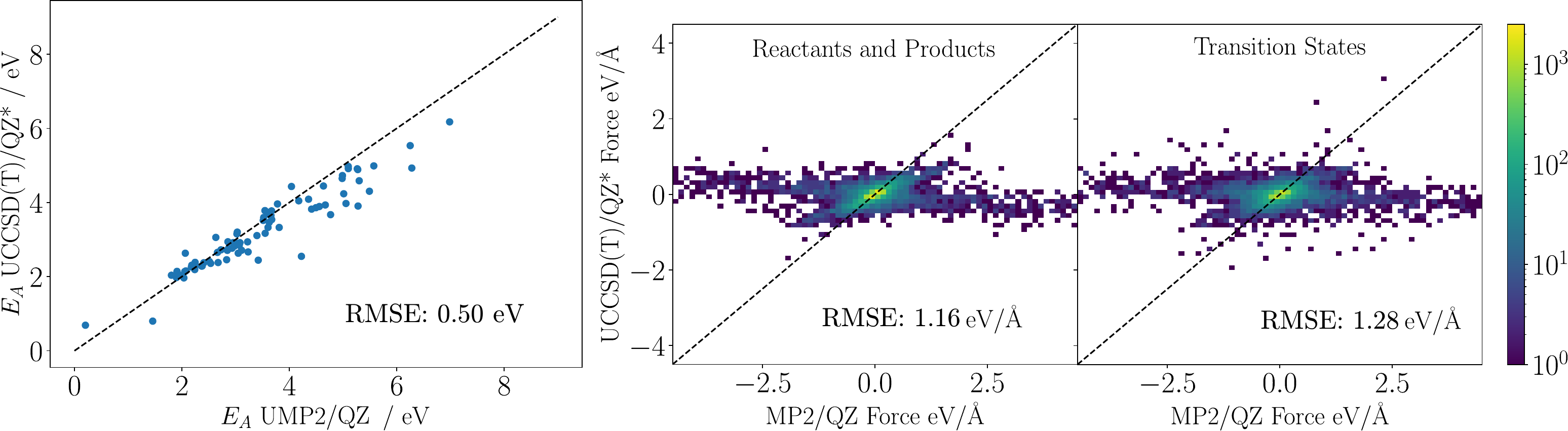}
    \caption{  A comparison between the unrestricted CCSD(T)/QZ* level of theory and unrestricted MP2/QZ for the $E_A$ and forces.  }
    \label{fig:U_R_forces_2}
\end{figure*}

\clearpage

\section{The NEB/Dimer/SEGS Datasets}

\begin{figure}[htb]
    \centering
    \includegraphics[width=3.25in]{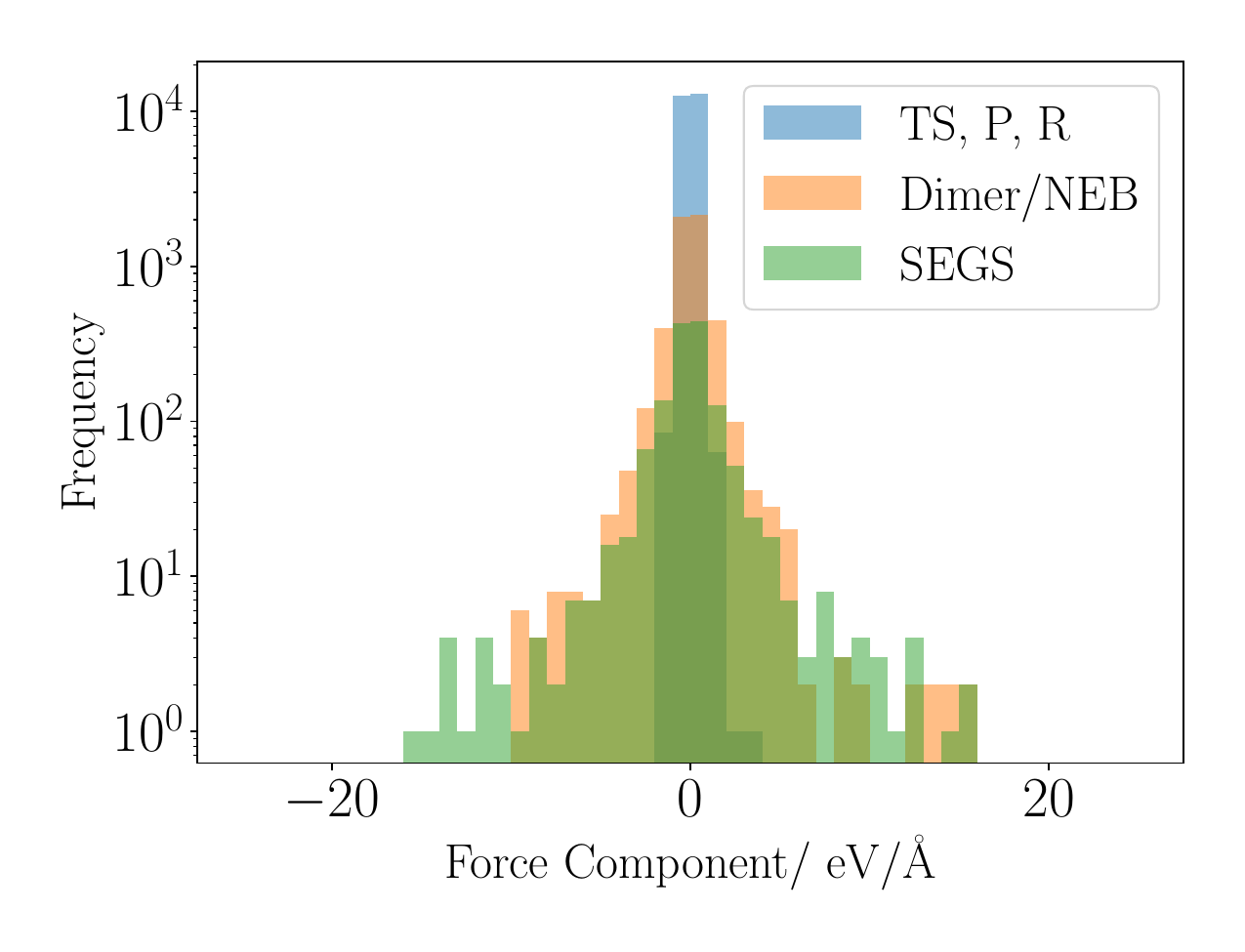}
    \caption{ The distribution of UCCSD(T)/QZ* force components for the transition states, products and reactants (TS, P, R), the dimer and NEB structures and SEGS. A wider distribution of forces is present for the dimer, NEB and SEGS structures. }
    \label{fig:Dis_Forces_NEB_SEGS_DFT_uccsdt}

\end{figure}

\begin{figure}[htb]
    \centering
    \includegraphics[width=6.85in]{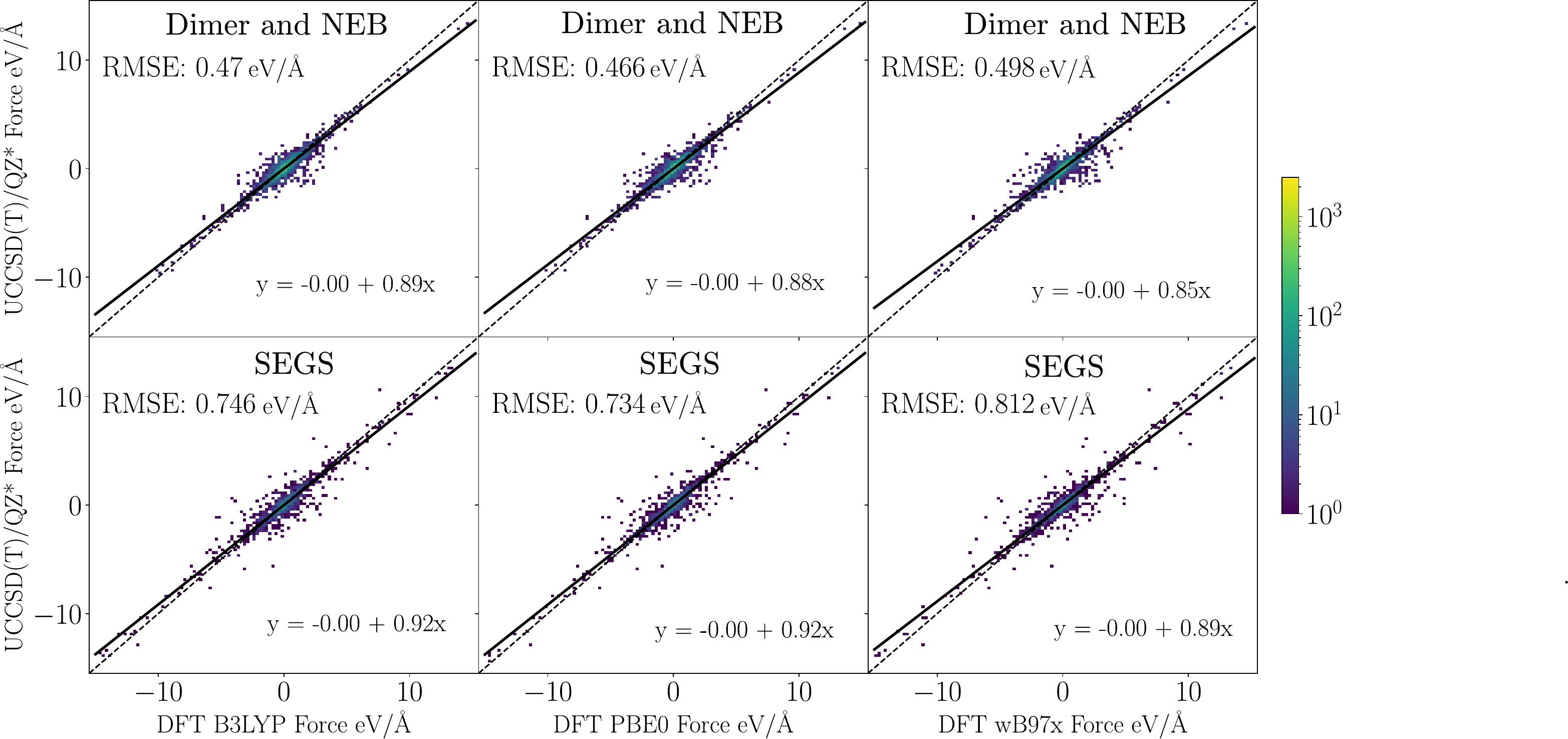}
        \caption{ A comparison in forces between the DFT functionals  calculated in the singlet state and UCCSD(T)/QZ* for NEB, the dimer method and the SEGS. }
    \label{fig:Forces_NEB_SEGS_DFT_uccsdt}

\end{figure}

Additional analysis of the NEB/Dimer/SEGS dataset for the DFT functionals and UCCSD(T)/QZ* is given in Fig.~\ref{fig:Forces_NEB_SEGS_DFT_uccsdt} and Fig.~\ref{fig:Dis_Forces_NEB_SEGS_DFT_uccsdt}.

\clearpage

\section{Dataset Sizes}
A summary of the size of each dataset created is presented in this section. First, an analysis by the level of theory used is provided, followed by a categorization based on the sampling technique for the UCCSD(T)/QZ* data.

\vspace{0.5cm}

\begin{table}[h]
\centering

\begin{tabular}{|r|c|}
\hline
\textbf{QM Method} & \textbf{\begin{tabular}[c]{@{}c@{}}Number of \\ Structures\end{tabular}} \\ \hline
\textit{$\omega$B97X-D3/def2-TZVP} & 270720 \\ \hline
\textit{UCCSD/DZ} & 30154 \\ \hline
\textit{UCCSD(T)/QZ*} & 3119 \\ \hline
\end{tabular}
\caption{The number of calculations performed and provided at DFT/TZ, UCCSD/DZ and UCCSD(T)/QZ* levels of theory.}
\label{tab:dft_uccsdt_no}
\end{table}

\vspace{0.5cm}

Furthermore, a breakdown of the sampling techniques for the UCCSD(T)/QZ* is given. It should be noted that the number of structures alone can give a misleading indication of the importance of the sampling technique when fitting MLIPs. The bond dissociation structures are much smaller than the other sampling techniques and therefore provide a smaller fraction of the forces than otherwise indicated. 

\vspace{0.5cm}

\begin{table}[]
\begin{tabular}{|r|c|}
\hline
\textbf{Sampling Method} & \textbf{\begin{tabular}[c]{@{}c@{}}Number of \\ Structures\end{tabular}} \\ \hline
\textit{\begin{tabular}[c]{@{}r@{}}Transition States,\\ Reactants, Products\end{tabular}} & 950 \\ \hline
\textit{NEB/Dimer} & 167 \\ \hline
\textit{SEGS} & 49 \\ \hline
\textit{Bond Dissociation} & 1953 \\ \hline
\end{tabular}
\caption{The number of calculations performed for each sampling method for the UCCSD(T)/QZ* dataset.}
\label{tab:uccsdt_no}
\end{table}

\clearpage

\section{MACE Fine-tuning}

The fine-tuning of the MACE-MP medium potential to the UCCSD(T) dataset used the $\mathrm{run\_train.py}$ python command with the following settings:

\begin{itemize}[noitemsep,topsep=0pt]
\item[] foundation$\_$model=medium
\item[] multiheads$\_$finetuning=False                          \item[] train$\_$file=``All$\_$UCCSDT$\_$train.xyz"  
\item[] valid$\_$file=``All$\_$UCCSDT$\_$valid.xyz"             \item[] test$\_$file=``All$\_$UCCSDT$\_$test.xyz" 
\item[]E0s=``{1:-13.60421,6:-1028.91941,7:-1484.46528,8:-2041.48145}"\
\item[]lr=0.001\
\item[]batch$\_$size=10\
\item[]max$\_$num$\_$epochs=150\
\item[]device=cuda\
\item[]default$\_$dtype=``float64"\
\item[]energy$\_$key=`energy'\
\item[]forces$\_$key=`forces'\
\item[]clip$\_$grad=10000\
\item[]loss=``weighted"\
\item[]forces$\_$weight=1000\
\item[]energy$\_$weight=10\
\item[]swa\
\item[]start$\_$swa=100\
\item[]swa$\_$lr=0.000001
\end{itemize}

\section{Machine Learning Potential Performance}

Additional results for the machine learning potential trained to the UCCSD(T)/QZ* data is given. 

\subsection{Bond Dissociation Curves}

The bond dissociation curves for hydrogen bond dissociation for QM9 molecules are shown in Fig.~\ref{fig:MP2_forces_BS}.

\begin{figure*}[htb]
    \centering
    \includegraphics[width=5.95in]{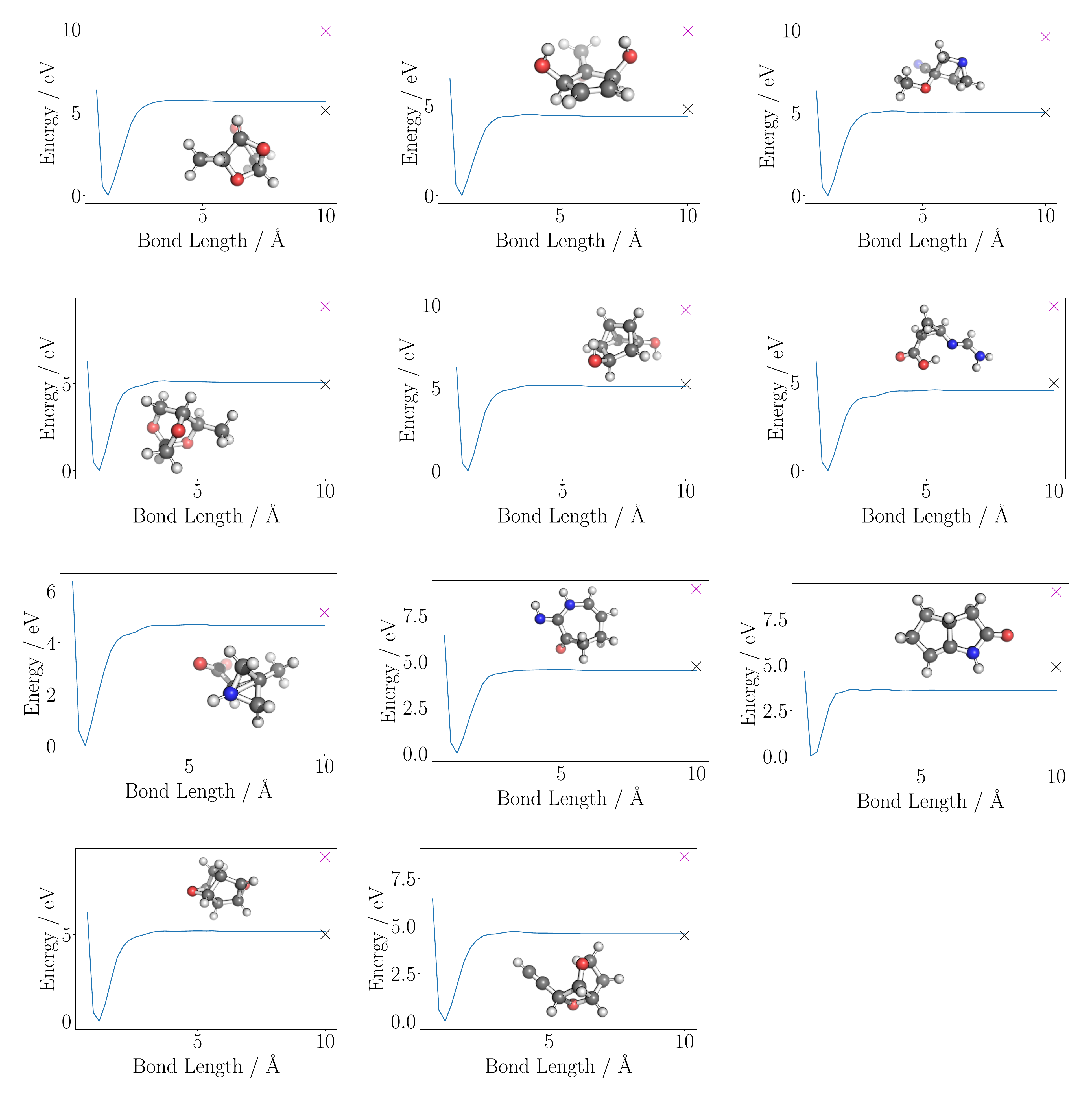}
    \caption{ The hydrogen bond dissociation energies given by the MLIP for randomly chosen molecules from the QM9. The black cross shows the UCCSD(T) bond dissociation energy relative to the equilibrium value. The magenta cross shows the DFT singlet bond dissociation energy. All of the molecules chosen are larger than the largest molecule used in the training dataset and rigid scans are performed. }
    \label{fig:MP2_forces_BS}
\end{figure*}

\clearpage

\subsection{UCCSD(T) Forces on Transition States Found with MLIPs}

\begin{figure}[htb]
    \centering
    \includegraphics[width=4.5in]{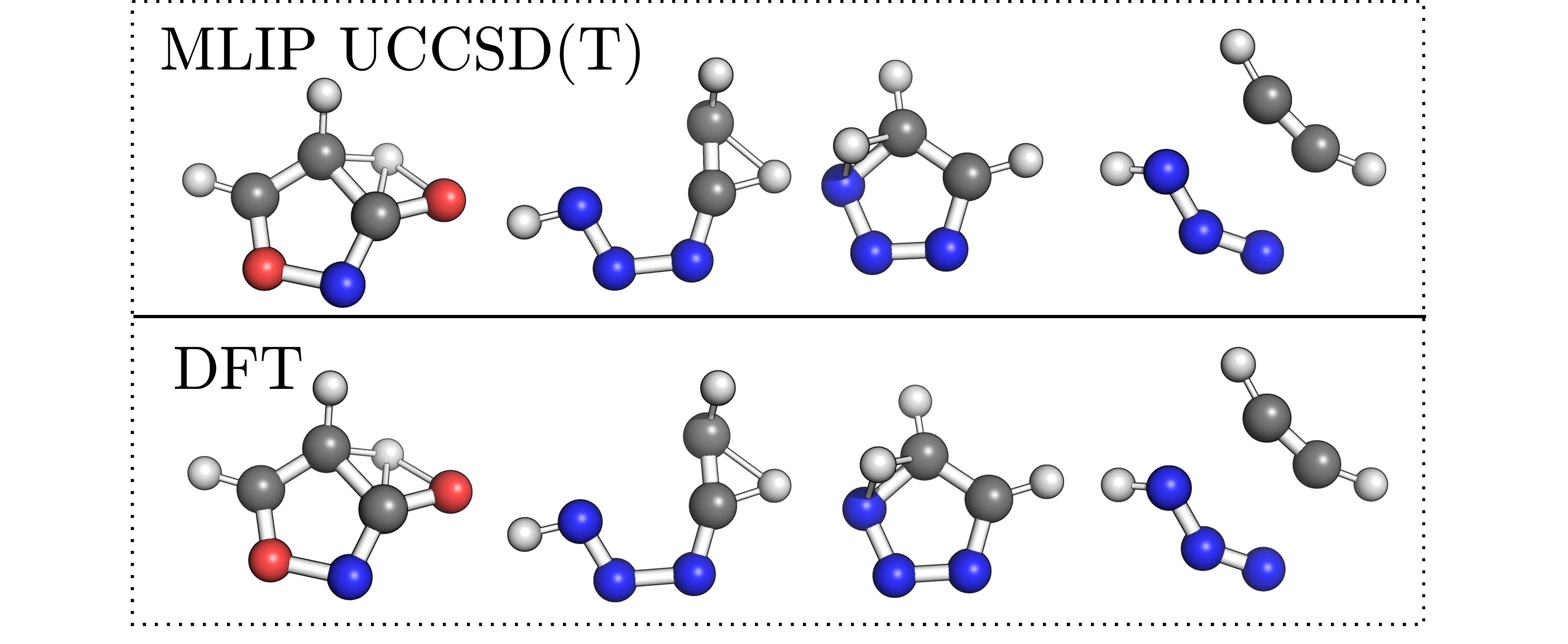}
    \caption{Transition state structures found with the dimer method using DFT and the MLIP trained to UCCSD(T) data are shown. }
    \label{fig:Dimer_ML}
\end{figure}

The UCCSD(T)/QZ* forces for the TS found with the MLIP trained to the UCCSD(T)/QZ* dataset are shown in Fig.~\ref{fig:Forces_TS_MLIP}. Whilst DFT recreates UCCSD(T) transition states forces better for this subset, it is orders of magnitude slower. A comparison of the DFT and MLIP TS found are shown in Fig.~\ref{fig:Dimer_ML}.

\iffalse
\begin{figure*}[htb]
    \centering
    \includegraphics[width=4.85in]{./SI_Graphs/Forces_dis_All.pdf}
    \caption{The distribution of forces for transition states found using the dimer method using different potentials. DFT recreates UCCSD(T) transition states best, however it is orders of magnitude slower. }
    \label{fig:Forces_TS_MLIP}
\end{figure*}
\fi 

\begin{figure}[htb]
    \centering
    \includegraphics[width=3.5in]{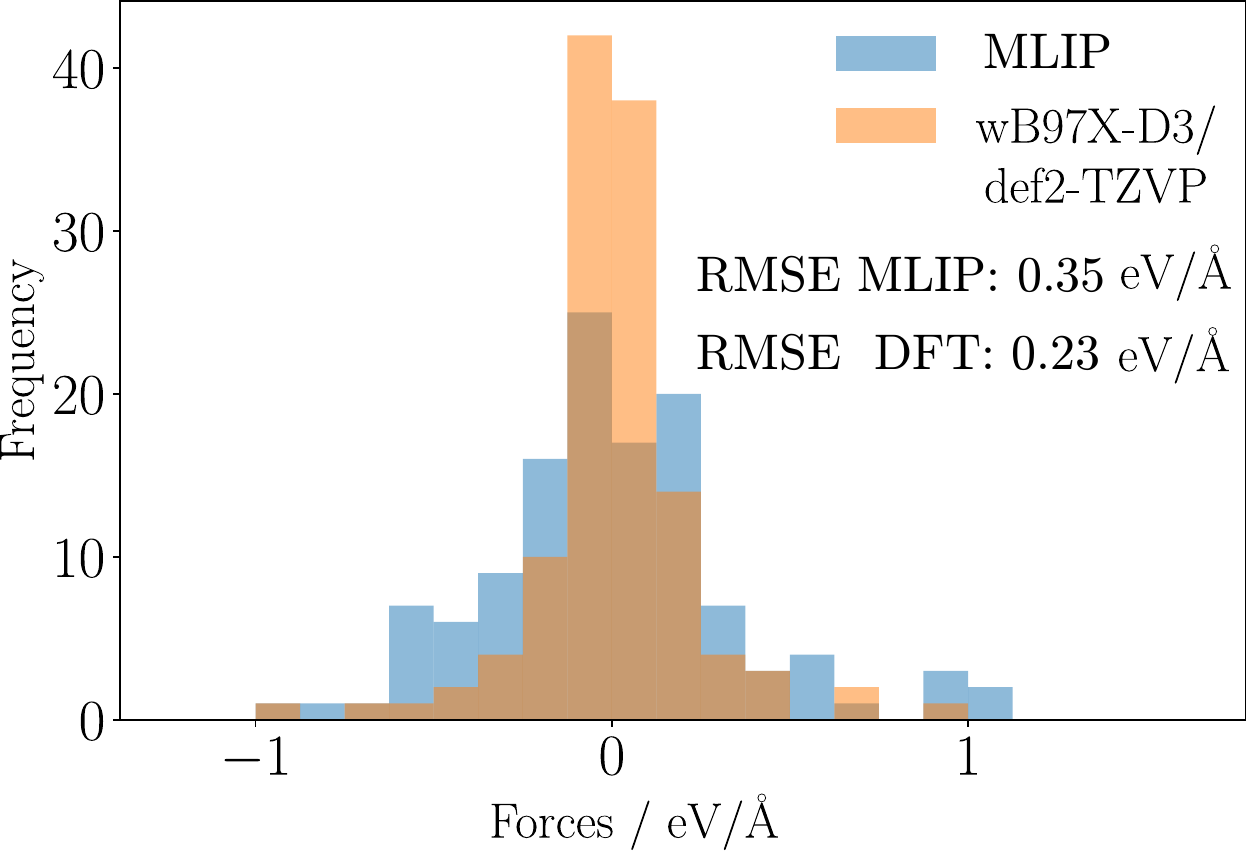}
    \caption{The distribution of forces for transition states found using the dimer method using different potentials. }
    \label{fig:Forces_TS_MLIP}
\end{figure}

\clearpage

\subsection{Isomerization Reactions}

\begin{figure}[htb]
    \centering
    \includegraphics[width=6.5in]{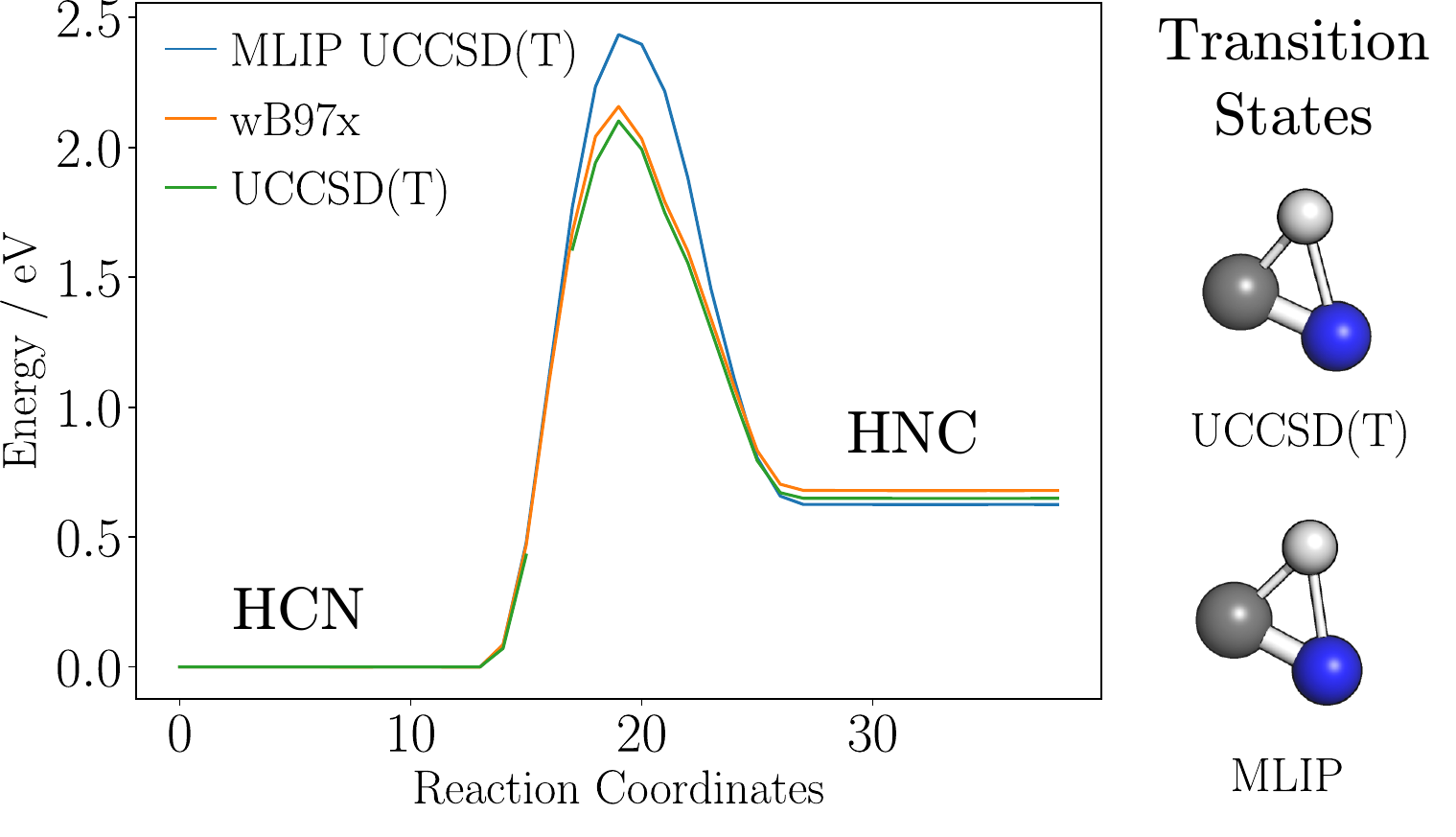}
    \caption{ The minimum energy path found with the nudged elastic band (NEB) technique using the HIP-HOP MLIP for HCN to HNC. The transition states for the MLIP and for UCCSD(T) found with the dimer method are also shown.}
    \label{fig:HCN}
\end{figure}

\begin{figure}[htb]
    \centering
    \includegraphics[width=3.25in]{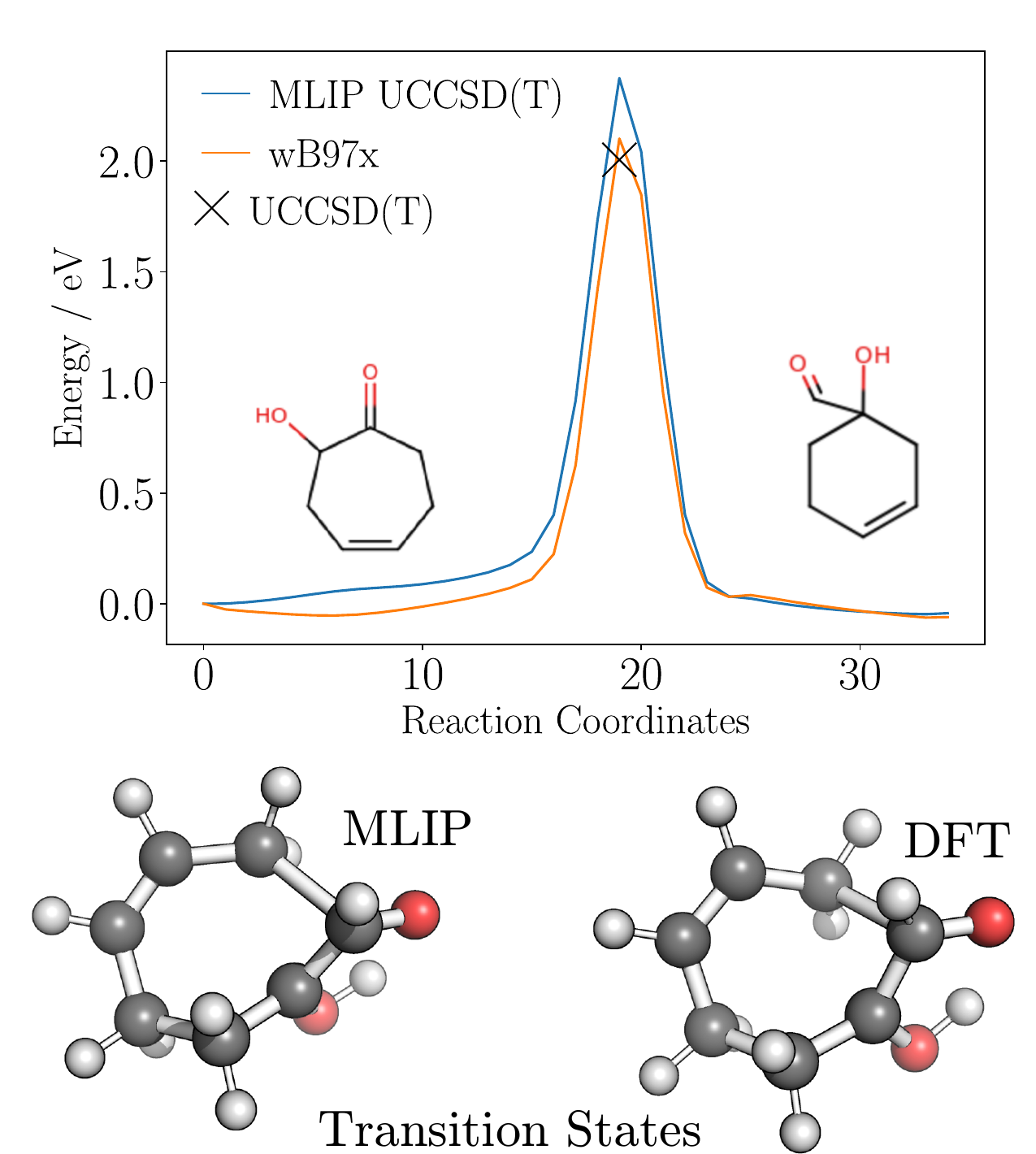}
    \caption{The minimum energy path found with the NEB technique using the HIP-HOP MLIP for the cyclic isomerization reaction from Ref. 53. The black cross shows the energy difference between the product and transition state for UCCSD(T).  The transition states with the MLIP trained at UCCSD(T) is compared to the DFT structure from Ref. 53. }
    \label{fig:Cyclic_MEP}
\end{figure}

The minimum energy path for isomerization reactions are shown in Fig.~\ref{fig:HCN} and in Fig.~\ref{fig:Cyclic_MEP}. 

\clearpage

\bibliography{refs}